\documentclass[showpacs,twocolumn,preprintnumbers,amsmath,amssymb,amsfonts,prb,floatfix, superscriptaddress]{revtex4}

\usepackage{graphicx, color,rotating, textcomp} 
\usepackage{dcolumn} 
\usepackage{bm} 

\begin{document}

\title{Theory of superconductor-ferromagnet point contact spectra:\\ the case of
strong spin polarization}

\author{Roland~Grein}
\affiliation{Institut f\"ur Theoretische Festk\"orperphysik
and DFG-Center for Functional Nanostructures,
Karlsruhe Institute of Technology, D-76128 Karlsruhe, Germany}
\author{Tomas~L\"ofwander}
\affiliation{Department of Microtechnology and Nanoscience—MC2,
Chalmers University of Technology, S-412 96 G\"oteborg, Sweden}
\author{Georgo~Metalidis}
\affiliation{Institut f\"ur Theoretische Festk\"orperphysik
and DFG-Center for Functional Nanostructures,
Karlsruhe Institute of Technology, D-76128 Karlsruhe, Germany}
\author{Matthias~Eschrig}
\affiliation{Institut f\"ur Theoretische Festk\"orperphysik
and DFG-Center for Functional Nanostructures,
Karlsruhe Institute of Technology, D-76128 Karlsruhe, Germany}
\affiliation{Fachbereich Physik, Universit\"at Konstanz, D-78457 Konstanz, Germany}
\date{\today}

\begin{abstract}
We study the impact of spin-active scattering on Andreev spectra of point contacts
between superconductors(SCs) and strongly spin-polarized ferromagnets(FMs) using
recently derived boundary conditions for the Quasiclassical Theory of Superconductivity.
We describe the interface region by a microscopic model for the interface
scattering matrix. Our model includes both spin-filtering and spin-mixing and is
non-perturbative in both transmission and spin polarization.
We emphasize the importance of spin-mixing caused by interface scattering,
which has been shown to be crucial for the creation of exotic pairing correlations
in such structures. We provide estimates for the possible magnitude of this effect
in different scenarios and discuss its dependence on various physical parameters.
Our main finding is that the shape of the interface potential has a tremendous impact
on the magnitude of the spin-mixing effect. Thus, all previous calculations, being
based on delta-function or box-shaped interface potentials,
underestimate this effect gravely. As a consequence, we find that with
realistic interface potentials the spin-mixing effect can easily be large enough
to cause spin-polarized sub-gap Andreev bound states in SC/sFM point contacts. In addition, we show
that our theory generalizes earlier models based on the Blonder-Tinkham-Klapwijk approach.
\end{abstract}

\maketitle

\renewcommand{\phi}{\varphi}
\newcommand{\eps}{\varepsilon}
\newcommand{\ud}{\uparrow,\downarrow}
\renewcommand{\u}{\uparrow}
\renewcommand{\d}{\downarrow}
\newcommand{\ket}[1]{| {#1}\rangle}
\newcommand{\bra}[1]{\langle {#1}|}
\newcommand{\barlambda}{{\lambda \!\!\!^{-}\,\!}}
\newcommand{\blFe}{{\lambda \!\!\!^{-}\,\!}_{\mathrm{F}1}}
\newcommand{\blFeta}{{\lambda \!\!\!^{-}\,\!}_{\mathrm{F}\eta}}
\newcommand{\blF}{{\lambda \!\!\!^{-}\,\!}_{\mathrm{F}}}
\newcommand{\blJ}{{\lambda \!\!\!^{-}\,\!}_{J}}
\newcommand{\EF}{E_{\mathrm{F}}}
\newcommand{\vpfe}{\vec{p}_{\mathrm{F}1}}
\newcommand{\vpfz}{\vec{p}_{\mathrm{F}2}}
\newcommand{\vpfd}{\vec{p}_{\mathrm{F}3}}
\newcommand{\vpfeta}{\vec{p}_{\mathrm{F}\eta }}
\newcommand{\vvfe}{\vec{v}_{\mathrm{F}1}}
\newcommand{\vvfz}{\vec{v}_{\mathrm{F}2}}
\newcommand{\vvfd}{\vec{v}_{\mathrm{F}3}}
\newcommand{\vvfzd}{\vec{v}_{\mathrm{F}2,3}}
\newcommand{\vvfeta}{\vec{v}_{\mathrm{F}\eta }}
\newcommand{\pfe}{p_{\mathrm{F}1}}
\newcommand{\pfz}{p_{\mathrm{F}2}}
\newcommand{\pfd}{p_{\mathrm{F}3}}
\newcommand{\pfeta}{p_{\mathrm{F}\eta }}
\newcommand{\vfe}{v_{\mathrm{F}1}}
\newcommand{\Nfe}{N_{\mathrm{F}1}}
\newcommand{\vfz}{v_{\mathrm{F}2}}
\newcommand{\Nfz}{N_{\mathrm{F}2}}
\newcommand{\vfd}{v_{\mathrm{F}3}}
\newcommand{\Nfd}{N_{\mathrm{F}3}}
\newcommand{\vfzd}{v_{\mathrm{F}2,3}}
\newcommand{\vfeta}{v_{\mathrm{F}eta}}
\newcommand{\JFM}{J_{\mathrm{FM}}}
\newcommand{\vJFM}{\vec{J}_{\mathrm{FM}}}
\newcommand{\JI}{J_{\mathrm{I}}}
\newcommand{\vJI}{\vec{J}_{\mathrm{I}}}
\newcommand{\VI}{V_{\mathrm{I}}}
\newcommand{\gr}{\gamma^R}
\newcommand{\ga}{\gamma^A}
\newcommand{\grt}{\tilde{\gamma}^R}
\newcommand{\gat}{\tilde{\gamma}^A}
\newcommand{\gra}{\gamma^{R,A}}
\newcommand{\grat}{\tilde{\gamma}^{R,A}}
\newcommand{\Rs}{{R}_1}
\newcommand{\Tsn}{{T}_{sn}}
\newcommand{\Tns}{{T}_{ns}}
\newcommand{\Rn}{{R}_n}
\newcommand{\ru}{r_{2}}
\newcommand{\rd}{r_{3}}
\newcommand{\Tsu}{{T}_{12}}
\newcommand{\Tsd}{{T}_{13}}
\newcommand{\Tus}{{T}_{21}}
\newcommand{\Tds}{{T}_{31}}
\newcommand{\rud}{r_{23}}
\newcommand{\rdu}{r_{32}}

\section{Introduction}
The proximity effect near interfaces between superconductors and
ferromagnetic materials has been a field of intense research in
recent years.
\cite{bergeret05,buzdin05,eschrig08,cuoco08,galak08,haltermann08,volkov08,brydon08,zhao08,linder09,grein09,brydon09,kalenkov09,beri09,Barsic,eschrig09}
This interest is mainly triggered by the observation that exotic types of pairing
symmetries that are difficult (or impossible) to be observed in bulk materials
can be created in such heterostructures.\cite{eschrig07,tanaka07,eschrig08}
Examples are the recent revival of pairing states that exhibit a sign change under
the exchange of the time coordinates of the particles that constitute a Cooper pair
(``odd-frequency pairing''),\cite{bergeret05}
or mechanisms for the creation of long-range equal-spin pairing components in
half-metallic ferromagnets.\cite{eschrig03,volkov03,kopu04,eschrig04}
Supercurrents in half-metals have subsequently been observed,\cite{keizer06} which
ignited a strong activity in further theoretical modeling of this effect.
\cite{eschrig07,braude07,asano07,linder,linder07,takahashi07,cuoco08,galak08,haltermann08,volkov08,kalenkov09,beri09,Barsic,eschrig09}
Spin triplet pairing has proven to be at the heart of new physical phenomena,
like $0$-$\pi $-transitions in Josephson junctions
with FM interlayers\cite{buzdin05,lofwander05,pajovic06,champel08,brydon08,brydon09}
or the interplay between magnons and triplet pairs.\cite{houzet1,takahashi07}

So far, transport calculations in SC/FM hybrids have mostly been
concentrated on either fully polarized FMs, so-called half metals (HM),
or on the opposite limit of weakly polarized systems.
However, most FMs have an intermediate exchange
splitting of the energy bands of the order of 0.2-0.8 times the Fermi
energy $\EF $, which we here refer to as strongly spin-polarized FMs (sFM).
As alternative to solving full Bogoliubov-de Gennes equations,\cite{linder, Valls, Barsic, cottet08, cottet08b} we have
recently presented a quasiclassical theory appropriate for this
intermediate range of spin-polarizations, which is of considerable importance
for applications.\cite{grein09}

For such strongly spin-polarized materials, it has been argued that Andreev point contact
spectra can be used to obtain
the spin-polarization of the FM,\cite{Upad,soulen98,Beenakker,Mazin} which is an important
information for spintronics applications.
Experimental studies of point contact spectra with strongly spin-polarized systems
have been performed for a number of systems.\cite{desisto00,ji01,angu01,parker02,woods04,Perez,dyachenko06,yates07,krivoruchko08,bocklage07}
However, Xia \emph{et al.}\cite{Xia}
have objected rightfully, that without taking into account a realistic description of the
interface region, the results obtained with this method are questionable.

In the quasiclassical approach to superconducting hybrid structures, interfacial
scattering is taken into account by the interface scattering matrix $S$ of the structure
in its normal state. This is ideal for discussing microscopic models of interfacial scattering which go well beyond the standard Blonder-Tinkham-Klapwijk (BTK) approach.\cite{blonder82}
The latter has been employed to fit experimental data of SC/FM point contact spectra,\cite{Beenakker} with the interface being described by a single parameter $Z$ related to its transparency and the ferromagnet by its spin-polarization $P$.
The modification of the Andreev point contact spectrum compared to a normal metal contact is then uniquely related to the spin-dependent density of states (DOS) in the FM bulk. This model allows for good fits to experimental data, however, comparing different probes with varying interface transparency, a systematic dependence $P(Z)$ was found by Woods \emph{et al.}\cite{woods04} This shows that the extracted spin polarization is not
a bulk property, as was originally assumed, but at least partially an interface property.
This important difference has been emphasized also in Ref.~\onlinecite{Perez}.

From the theoretical point of view,
it is obvious that if scattering is spin-active, i.e. the scattering event is sensitive to the spin of the incident electron, this may not only imply a spin-dependent transmission probability (spin-filtering)\cite{meservey94} but also a spin-dependent phase shift of the wavefunction.\cite{tokuyasu88} The latter is called the \emph{spin-mixing effect} and it has been shown to be of crucial importance for the creation of exotic pairing correlations.\cite{tokuyasu88,fogelstrom00,huertas02,eschrig03,eschrig08,cottet05,bobkova07}

So far, estimates of the magnitude of this effect and its dependence on physical parameters including not only the structure of the interface but also the Fermi surface geometry of the adjacent materials and the FM exchange splitting are still lacking. Instead, phenomenological models have been adopted that introduce a free parameter to account for it.\cite{fogelstrom00,eschrig03,eschrig08,Zaikin}

The main point of this paper is to provide a microscopic analysis of the characteristic
interface parameters.
In the following we adopt a simple model of the interface region consisting of a spin-dependent scattering potential whose quantization axis may be misaligned with that of the adjacent FM. We allow for an arbitrary shape of this scattering potential and illustrate that this may enhance the spin-mixing effect considerably compared to the previously
used box-shaped or delta-function potentials.
We also study in detail the relation between spin-mixing angle and impact angle of the quasiparticle, showing that this relation can be non-trivial for transparent interfaces. Furthermore, we provide a very general mathematical discussion of suitable parameterizations and representations of the scattering matrix in this context.

Andreev bound states have proven invaluable for studying the internal structure
of the superconducting order parameter.\cite{saint64,deutscher05}
Andreev states are also induced at spin-polarized interfaces by
the spin-mixing effect.\cite{fogelstrom00} In fact, the measurement of such
bound states at spin-active interfaces would be an elegant method do
determine the spin-mixing angle of the interface. To date this quantity has
never been determined in experiment. Our results show, that a measurable effect
is more likely to appear when leakage of
spin polarization into the superconductor takes place, for example due to
diffusion of magnetic atoms.
Our theory can discriminate between conventional Andreev reflection processes
(AR) and spin-flip Andreev reflection (SAR), the latter being responsible for the
long-range triplet proximity effect. We discuss the Andreev bound state associated
to the spin-mixing effect and show that it may be observable in experiment.
Furthermore we show that for highly polarized FMs, spin-flip scattering can
bias the spectra considerably, proving that such processes must be precluded if
one wishes to extract the FM spin-polarization from such spectra.

The paper is organized as follows. In Section~\ref{QCT}, we discuss quasiclassical theory to describe transport through a point contact. In Section~\ref{IM} we turn to interface models and discuss the spin-mixing effect and the scattering matrix. In Section~\ref{ACS} we present results for Andreev conductance spectra of SC/FM point contacts. We dicuss analytical results,
focusing on the Andreev bound state spectrum, as well as numerical results.
In Subsection~\ref{BTK} we establish the connection to earlier transport theories for such systems which are based on the BTK approach. We prove analytically that they are contained as limiting cases in our formalism.
Eventually, in Section~\ref{Conc}, we conclude on our results.

\section{Quasiclassical theory}
\label{QCT}

We make use of the quasiclassical theory of
superconductivity\cite{larkin68,eilen,Serene,schmid75,schmid81,rammer86,Larkin86,FLT}
to calculate electronic transport
across the SC/FM interface. This method is based on the observation that, in most situations, the superconducting state
varies on the length scale of the superconducting coherence length $\xi_0=\hbar |\vec{v}_{\rm{F}}| /2\pi k_{\rm B} T_{\rm c}$,
with the normal state Fermi velocity $\vec{v}_{\rm F}$.
The appropriate many-body Green's function for describing
the superconducting state has been introduced by Gor'kov,\cite{gorkov58} and
the Gor'kov Green's function
can then be decomposed in a fast oscillating component, varying on the scale of
the Fermi wave length $2\pi/k_F$,
and an envelop function varying on the scale of $\xi_0$.
The quasiclassical approximation consists of integrating out the fast oscillating component:
\begin{equation} \check{g}(\vec{p}_{\rm{F}} , \vec{R}, \eps, t)=\frac{1}{a(\vec{p}_{\rm F})}\int {\rm d}\xi_p \hat{\tau}_3 \check{G}(\vec{p}, \vec{R}, \eps, t)\end{equation}
where $a(\vec{p}_{\rm F})$ is the inverse quasiparticle renormalization factor (due to
self-energy effects from high-energy processes),\cite{Serene}
a ``check''
denotes a matrix in Keldysh-Nambu-Gor'kov space,\cite{keldysh64}
a ``hat'' denotes
a matrix in Nambu-Gor'kov particle-hole space (with $\hat{\tau}_3$ the third Pauli matrix),
$\vec{p}_{\rm F}$ is the Fermi momentum, $\vec{R}$ the spatial coordinate,
$\eps$ the quasiparticle energy, $t$ the time,
and $\xi_p=\vec{v}_{\rm{F}} (\vec{p}-\vec{p}_{\rm{F}} )$.
The quasiclassical Green's function obeys the transport equation\cite{larkin68,eilen}
\begin{equation} \label{eilen1} i \hbar \vec{v}_{\rm{F}} \cdot \nabla_{\vec{R}}\check{g}+[\eps\hat{\tau}_3-\check{\Delta}-\check{h}, \check{g}]_\otimes =\check{0}.\end{equation}
Here,
$\check{\Delta}$ is the superconducting order parameter, $\check{h}$ contains
external fields and self-energies due to impurities etc,
and $[\bullet,\bullet]_\otimes$ denotes the commutator with respect to a time convolution
product (for details see Ref.\onlinecite{Serene}). Eq.~\eqref{eilen1} must be supplemented by a normalization condition\cite{larkin68,shelankov85} $\check{g}\otimes \check{g}=-\check{1}\pi^2$.
The current density is related to the Keldysh component of the Green's function via:
\begin{equation}\vec{j}(\vec{R},t)=e N_{\rm F}\int \frac{{\rm d}\eps}{8\pi i}
\mathrm{Tr}
\Big\langle \vec{v}_{\rm{F}}(\vec{p}_{\rm F})
\hat{\tau}_3\hat{g}^K (\vec{p}_{\rm{F}},\vec{R}, \eps, t)\Big\rangle, \end{equation}
where $N_{\rm F}$ is the density of states at the Fermi level in the normal state, and
$\langle \bullet \rangle$ denotes a Fermi surface average
which is defined as follows:
\begin{eqnarray}
\langle \bullet\rangle&=&\frac{1}{N_{\mathrm{F}}}\int_{FS} \frac{{\rm d}^2 p_{\mathrm{F}}}{(2\pi\hbar)^3|\vec{v}_{\mathrm{F}}(\vec{p}_{\mathrm{F}})|} \; (\bullet )\; ,\\
N_{\rm F}&=&\int_{FS} \frac{{\rm d}^2 p_{\mathrm{F}}}{(2\pi\hbar)^3|\vec{v}_{\mathrm{F}}(\vec{p}_{\mathrm{F}})|}.
\end{eqnarray}

\begin{figure*}[t]
\includegraphics[width = 1.99\columnwidth]{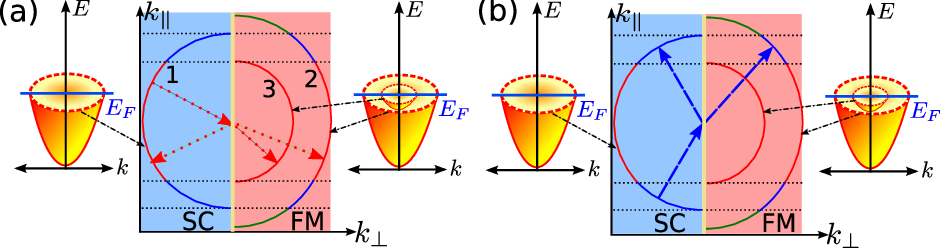}
\caption{\label{fig1}(Color online)
SC/sFM interface, showing the Fermi surfaces on either side (thick lines).
Assuming momentum conservation parallel to the interface ($\vec{k}_{||}$),
a quasiparticle incident from the SC can either
scatter into two (a), or into
only one (b) spin band of the FM.
}
\end{figure*}

The direct inclusion of an exchange energy $\JFM $ of order of 0.1 $\EF$ or larger
in the quasiclassical scheme
violates the underlying assumptions of quasiclassical theory. As
we aim to describe a strongly spin-polarized FM, which means that its exchange field $\JFM $ will be of the order of the Fermi energy,
we cannot
include it as a source term $-\frac{1}{2} \vJFM \cdot \vec{\sigma }$ (with
$\vec{\sigma }$ the vector of Pauli spin matrices) in the quasiclassical equation of
motion. Such an approach would neglect terms of order of $\JFM^2 /\EF $ compared to $\Delta $. The resulting condition
$\JFM \ll \sqrt{\EF \Delta}$, assuming e.g. a gap of 1 meV and $\EF\sim 1$ eV,
would imply $\JFM \ll 30$ meV. In general, the condition for the possibility to
include $\JFM$ in the quasiclassical low energy scale is violated for
most SCs if $\JFM >0.1\EF $.

To deal with the strong exchange splitting, we make use of the fact that it results in a rapid suppression of superconducting correlations between quasiparticle states with opposite spin, i.e. singlet ($\ket{\u\d}-\ket{\d\u}$) or $S_{\rm{z}}=0$ triplet ($\ket{\u\d}+\ket{\d\u}$) correlations. They decay on the short length scale
$\blJ=\hbar /(\pfz  -\pfd ) \ll \hbar \vfzd /\Delta \equiv \xi_{0\eta }$. Here $\pfz$, $\pfd$ are the Fermi-momenta of the two spin-bands (2 and 3) in the sFM and $\xi_{0\eta }$ with $\eta=2,3$ the coherence length in the respective band. Consequently, only equal-spin triplet correlations can penetrate the FM-bulk. Hence we pursue the following approach to model a strongly polarized FM in the frame of QC theory. We define independent QC Green's functions (QCGF) for each spin-band which are scalar in spin-space, i.e. describe correlations with $\ket{\u\u}$, respectively $\ket{\d\d}$ spin-wavefunction. The boundary conditions must now match three QC propagators at the SC/FM interface, which we label $\check{g}_{\eta}$ with $\eta=1\equiv \rm{SC}$, $\eta=2\equiv$ $\u$-band and $\eta=3\equiv$ $\d$-band (see Fig.~\ref{fig1}). These three QCGFs are formally obtained from:
\begin{equation} \check{g}(\vec{p}_{\rm{F}\eta} , \vec{R}, \eps, t)=\frac{1}{a(\vec{p}_{\rm F\eta })}\int {\rm d}\xi_{p\eta} \hat{\tau}_3 \check{G}(\vec{p}, \vec{R}, \eps, t)\end{equation}
with $\xi_{p\eta}=\vec{v}_{\rm{F}\eta} (\vec{p}-\vec{p}_{\rm{F}\eta} )$, $\vec{p}_{\rm{F}\eta}$ and $\vec{v}_{\rm{F}\eta}$ being the respective Fermi-momenta/velocities of the bands. Consequently, the current must then be evaluated for each band separately
\begin{equation}\vec{j}_{\eta}(\vec{R},t)=e N_{{\rm F}\eta}\int \frac{{\rm d}\eps}{8\pi i }
\mathrm{Tr}
\Big\langle \vec{v}_{\rm{F}\eta}(\vec{p}_{{\rm F}\eta} )
\hat{\tau}_3\hat{g}_{\eta}^K (\vec{p}_{\rm{F}\eta}, \vec{R}, \eps, t)\Big\rangle_{\eta}. \end{equation}
Here, $N_{{\rm F}\eta }$ is the partial density of states at the Fermi level
in band $\eta $, and
$\langle \bullet \rangle_\eta $ denotes the corresponding Fermi surface average
\begin{eqnarray}
\langle \bullet\rangle_\eta &=&\frac{1}{N_{{\rm F}\eta }}\int_{FS\eta } \frac{{\rm d}^2 p_{\mathrm{F}\eta }}{(2\pi\hbar)^3|\vec{v}_{\mathrm{F}\eta}(\vec{p}_{\mathrm{F}\eta })|} \; (\bullet )\; ,\\
N_{{\rm F}\eta }&=&\int_{FS\eta } \frac{{\rm d}^2 p_{\mathrm{F}\eta }}{(2\pi\hbar)^3|\vec{v}_{\mathrm{F}\eta }(\vec{p}_{\mathrm{F}\eta })|}.
\end{eqnarray}

In addition,
the system's properties vary on the atomic length scale in the interface region between the two materials. Thus the QC theory is also not applicable in the
immediate
proximity to the interface (on the scale of the Fermi wavelength).
This is a general problem in the quasiclassical description of heterostructures, which can be circumvented by deriving appropriate boundary conditions for matching the QC propagators on both sides of the
interface.\cite{zaitsev84}
The full boundary conditions for the present problem have been developed only recently.\cite{eschrig09}
Earlier works on Andreev spectra using QC theory were restraint to either SC/normal metal contacts with spin-active interfaces,\cite{Perez,fogelstrom00,barash02,zhao04} or contacts with weak ferromagnets.
We refer to Ref.\onlinecite{eschrig09} and references therein for a detailed discussion of this problem.
In the following subsection we
discuss a parameterization of the QC propagator, and
return to the problem of boundary conditions at the interface in Subsection~\ref{BC}.

\subsection{Riccatti parameterization}
\label{RP}

For our calculations we choose a representation of the quasiclassical Green's function
(QCGF) that has proven very useful in the past and is standard by now.
In this representation, the Keldysh QCGF is determined by six parameters in
particle-hole space, $\gamma^{\rm R,A}, \tilde{\gamma}^{\rm R,A}, x^{\rm K}, \tilde{x}^{\rm K}$,
of which $\gamma^{\rm R,A}(\vec{p}_{\rm F},\vec{R}, \eps, t)$ and $\tilde{\gamma}^{\rm R,A}(\vec{p}_{\rm F},\vec{R}, \eps, t)$ are the retarded (${\rm R}$) and
advanced (${\rm A}$) coherence functions,
describing the coherence between particle-like and hole-like states,
whereas $x^{\rm K}(\vec{p}_{\rm F},\vec{R}, \eps, t)$ and $\tilde{x}^{\rm K}(\vec{p}_{\rm F},\vec{R}, \eps, t)$ are distribution functions, describing the
occupation of quasiparticle states.\cite{eschrig00,cuevas06}
The coherence functions are a generalization of the so-called
Riccatti amplitudes\cite{nagato93,schopohl95} to non-equilibrium situations.
All six parameters are 2$\times $2 spin-matrix functions of Fermi momentum, position, energy, and time.
The parameterization is simplified by the fact that, due to symmetry relations,
only two functions of the six are independent. The particle-hole symmetry is
expressed by the operation $\tilde{(\bullet )}$, which is defined for any
function of the phase space variables by
\begin{equation}
\label{tilde}
\tilde Q(\vec{p}_{\rm F},\vec{R},z,t)=Q(-\vec{p}_{\rm F},\vec{R},-z^\ast,t)^{\ast },
\end{equation}
where $z=\epsilon $ is real for the Keldysh components  and $z$ is situated in
the upper (lower) complex energy half plane for retarded (advanced) quantities.
Furthermore, the symmetry relations
\begin{equation}
\gamma^{\rm A}=(\tilde \gamma^{\rm R})^\dagger , \quad
\tilde \gamma^{\rm A}=(\gamma^{\rm R})^\dagger , \quad
x^{\rm K}=(x^{\rm K})^\dagger
\end{equation}
hold. As a consequence, it suffices to
determine fully the parameters $\gamma^{\rm R}$ and $x^{\rm K}$.

The QCGF is related to these amplitudes in the following way
[here the upper (lower) sign corresponds to retarded (advanced)]:\cite{eschrig09}
\newcommand{\plus}{\;\;\,}
\newcommand{\mat}{\left( \begin{array}{cc} }
\newcommand{\matend}{\end{array}\right)}
\begin{equation}
\label{cgretav}
\hat g^{\rm R,A} =
\mp \, 2\pi i\,
\mat \plus {\cal G} & \plus {\cal F} \\ -\tilde{\cal F} &
-\tilde{\cal G} \matend^{\!\!\! \rm R,A }
\pm i\pi \hat \tau_3
,
\end{equation}
with the abbreviations
${\cal G}=({\it 1}-\gamma \tilde\gamma )^{-1}$ and ${\cal F}={\cal G} \gamma $, and
\begin{equation}
\label{ckelgf2}
\hat g^{\rm K} =
-2\pi i
\mat \plus {\cal G} & \plus {\cal F} \\ -\tilde{\cal F} &
-\tilde{\cal G} \matend^{\!\!\! \rm R }
\mat x^{\rm K} & 0 \\ 0 & \tilde x^{\rm K} \matend
\mat \plus {\cal G} & \plus {\cal F} \\ -\tilde{\cal F} &
-\tilde{\cal G} \matend^{\!\!\! \rm A }.
\end{equation}
Note that
all multiplication and inversion operations include 2$\times $2 matrix algebra
(and, more general, for time-dependent cases also a time convolution).


From the transport equation for the quasiclassical Green's functions
one obtains a set of 2$\times $2 matrix equations of motion for the
six parameters above.\cite{eschrig99,eschrig00} For the coherence amplitudes this
leads to Riccatti differential equations,\cite{schopohl95} hence the name Riccatti parameterization.
As we are interested in this paper only in the interface problem in relation to
a point contact, the transport equations are not relevant for the problem at hand.
For a point contact, the superconductivity is modified only in a very small spatial
region, and this modification can be neglected consistent with quasiclassical
approximation. We assume that the half-space problem is solved and calculate
the conductance across the point contact. For this, we turn now to
the problem of solving the boundary conditions for the point contact.

\subsection{Boundary conditions}
\label{BC}

\subsubsection{General case}

The QCGF mixes particlelike and holelike amplitudes, and as a result the transport
equations are numerically stiff, with exponentially growing solutions in
both positive and negative directions along each trajectory, which must be projected out.
A particular advantage of the coherence and distribution functions is
that, in contrast to the QCGF, they have a stable integration direction for each trajectory. This direction coincides with their
propagation direction, and is opposite for
holelike and particlelike amplitudes as well as advanced and retarded ones.
This allows to distinguish between incoming and outgoing amplitudes at the interface.
We adopt the notation\cite{eschrig00} that incoming amplitudes are denoted by small
case letters and outgoing ones by capital case letters.
Boundary conditions express outgoing amplitudes as functions of incoming ones and
as functions of the parameters of the normal-state scattering matrix.
They are formulated in terms of the solution of the equation\cite{eschrig09}
\begin{equation}
\label{GR}
[\Gamma_{k\leftarrow k'}]^{\rm R} = \big[ \gamma'_{kk'}
+\sum_{k_1\ne k}
\Gamma_{k\leftarrow k_1} \tilde \gamma_{k_1} \gamma'_{k_1k'}
\big]^{\rm R}
\end{equation}
for $[\Gamma_{k\leftarrow k'}]^{\rm R}$,
where the trajectory indices $k,k',k_1$ run over outgoing trajectories involved in the
interface scattering process, and the scattering matrix parameters enter only via
the ``elementary scattering event''
\begin{equation}
\label{gp}
[\gamma'_{kk'}]^{\rm R } = \sum_{p} S^{\rm R}_{kp} \gamma^{\rm R}_{p} \tilde S^{\rm R}_{pk'}
\end{equation}
(the trajectory index $p$ runs over all incoming trajectories).
It is useful to split the quantity
$[\Gamma_{k\leftarrow k'}]^{\rm R}$ into its forward scattering contribution, which
determines the quasiclassical coherence amplitude,
\begin{equation}
\label{GRA}
\Gamma_{k}^{\rm R }= \Gamma_{k \leftarrow k}^{\rm R},
\end{equation}
and the remaining part
\begin{equation}
\label{GbarR}
\,[\overline \Gamma_{k \leftarrow k'}]^{\rm R } =
[\Gamma_{k \leftarrow k'} - \Gamma_{k} \delta_{kk'} ]^{\rm R},
\end{equation}
which is relevant only for the Keldysh components.
Analogous equations\cite{eschrig09} hold for the advanced and particle-hole conjugated components,
$[\tilde \Gamma_{p\leftarrow p'}]^{\rm R}$, $[\Gamma_{p'\rightarrow p}]^{\rm A}$,
and $[\tilde \Gamma_{k'\rightarrow k}]^{\rm A}$.
The boundary conditions for the distribution functions read\cite{eschrig09}
\begin{eqnarray}
\label{X}
X^{\rm K}_{k} \! \!&= & \!\!
\sum_{k_1,k_2}
[\delta_{kk_1} + \overline \Gamma_{k \leftarrow k_1} \tilde \gamma_{k_1}]^{\rm R}
[x'_{k_1k_2}]^{\rm K}
[\delta_{k_2k} + \gamma_{k_2} \overline{\tilde \Gamma}_{k_2 \rightarrow k} ]^{\rm A}
\nonumber \\
&&\qquad \qquad -\sum_{k_1}
[\overline \Gamma_{k \leftarrow k_1}]^{\rm R} \tilde x^{\rm K}_{k_1}
[\overline{\tilde \Gamma}_{k_1 \rightarrow k} ]^{\rm A},
\end{eqnarray}
which depend on the scattering matrix parameters only via the elementary scattering event
\begin{equation}
\label{xp}
[x'_{kk'}]^{\rm K} = \sum_{p} S^{\rm R}_{kp}  x^{\rm K}_{p}  S^{\rm A}_{pk'}.
\end{equation}
Analogous relations hold for $\tilde X^{\rm K}_{p}$.

\subsubsection{Special case for point contact}

In the case under consideration the trajectory labels $k$ and $p$ run from 1 to 3, with 1 denoting
(spin-degenerate) trajectories on the superconducting side, and 2 and 3 trajectories for
the two spin directions on the ferromagnetic side.
We use the following notation for the (unitary) scattering matrix:
\begin{equation}\label{S}
S=
\left(\!\! \begin{array}{c|cc} \Rs & \Tsu &
\Tsd \\ \hline \Tus & r_{2} & r_{23}\\
\Tds & r_{32}  & r_{3}\end{array}\right).
\end{equation}
The current across the interface is conserved (this is ensured by our boundary conditions), so that it suffices to calculate the current density at the FM side of the
interface.
We proceed with expressing the outgoing amplitudes for bands $2$ and $3$ in terms
of the incoming amplitudes and the scattering matrix.

For a point contact with semi-infinite SC and FM regions (assuming that
the Thouless energy related to the geometry of the system is negligibly small),
there are no incoming correlation function from the FM side,
$\gamma_{2,3}^{\rm R,A}=\tilde{\gamma}_{2,3}^{\rm R,A}=0$,
whereas on the SC side we can use the bulk solutions.
For a singlet order parameter the bulk solutions of the coherence functions read
\begin{align}\label{gbulk} \gamma^{R,A}_1=-\; \frac{\Delta_s i\sigma_2}{\eps\pm i\sqrt{|\Delta_s|^2-\eps^2}},\;
\tilde{\gamma}^{R,A}_1=\frac{\Delta_s^\ast i\sigma_2}{\eps\pm i\sqrt{|\Delta_s|^2-\eps^2}} ,
\end{align}
with the singlet superconducting order parameter $\Delta_s$.
Taking into account these facts, we obtain from Eq.~\eqref{GR}
\begin{eqnarray}
\label{G21}
\Gamma_{2\leftarrow 1}^{\rm R} &=& [\gamma'_{21}]^{\rm R} +
\Gamma_{2\leftarrow 1}^{\rm R} \tilde\gamma_1^{\rm R} [\gamma'_{11}]^{\rm R},\\
\label{G22}
\Gamma_{2}^{\rm R} &=& [\gamma'_{22}]^{\rm R} +
\Gamma_{2\leftarrow 1}^{\rm R} \tilde\gamma_1^{\rm R} [\gamma'_{12}]^{\rm R},\\
\label{G23}
\Gamma_{2\leftarrow 3}^{\rm R} &=& [\gamma'_{23}]^{\rm R} +
\Gamma_{2\leftarrow 1}^{\rm R} \tilde\gamma_1^{\rm R} [\gamma'_{13}]^{\rm R},
\end{eqnarray}
with $[\gamma'_{ij}]^{\rm R} = S_{i1} \gamma_1^{\rm R} S^\ast_{1j}$ for
$i,j=1,2,3$.
The first equation, Eq.~\eqref{G21}, can be solved,
\begin{equation}
\label{Gamma21}
\Gamma_{2\leftarrow1}^{\rm R} = \Tus \gamma_1^{\rm R} \Rs^\ast \left(
1- \tilde\gamma_1^{\rm R} \Rs \gamma_1^{\rm R} \Rs^\ast \right)^{-1} .
\end{equation}
It appears useful to introduce the notation
\begin{equation}\label{A}
A= \Gamma_{2\leftarrow1}^{\rm R} \tilde \gamma_1^{\rm R}.
\end{equation}
From Eqs.~\eqref{G22}-\eqref{G23} we obtain
\begin{eqnarray}
\Gamma_{2}^{\rm R} &=& (\Tus + A \Rs ) \gamma_1^{\rm R} \Tsu^\ast,\\
\Gamma_{2\leftarrow 3}^{\rm R} &=& ( \Tus + A \Rs ) \gamma_1^{\rm R} \Tsd^\ast .
\end{eqnarray}
Note that the identity
$\Tus + A \Rs= \Tus
(1- \gamma_1^{\rm R} \Rs^\ast \tilde \gamma_1^{\rm R} \Rs )^{-1} $ holds.
The corresponding solutions for band 3 are simply obtained by replacing $2\leftrightarrow 3$. Amplitudes $\tilde\Gamma^R_{2}$ and $\tilde\Gamma^R_{2\leftarrow3}$ are obtained
using Eq.~\eqref{tilde}, with $\tilde S_{ij}=S_{ij}^\ast$.
The required advanced amplitudes can be obtained from the fundamental symmetry relations of this formalism, which imply $\tilde{\Gamma}^A_{2}=(\Gamma^R_{2})^{\dagger}$ and $\tilde{\Gamma}^A_{3\rightarrow 2}=(\Gamma^R_{2\leftarrow 3})^{\dagger}$.

For the distribution functions, we use a gauge in
terms of {\it anomalous} components.\cite{eschrig09}
Taking the electrochemical potential equal to zero in the SC, and equal
to $-{\rm eV}$ in the ferromagnet,
these are $x_1= \tilde{x}_1=0$ and
\begin{align}\label{dist} x_{2,3}&=\plus \tanh\left(\frac{\eps+\rm{eV}}{2 k_B T}\right)-\tanh\left(\frac{\eps}{2 k_B T}\right)\\\nonumber \tilde{x}_{2,3}&=-\tanh\left(\frac{\eps-\rm{eV}}{2 k_B T}\right)+\tanh\left(\frac{\eps}{2 k_B T}\right). \end{align}
Note that in our notation $\rm{e}=-|\rm{e}|$.
From Eq.~\eqref{X}
we arrive at the following expressions for the outgoing
Keldysh amplitudes for band $2$:
\begin{eqnarray}
\label{X2}
X_{2}&=& [x'_{22}]^{\rm K}+
\Gamma_{2\leftarrow 1}^{\rm R}\tilde \gamma_1^{\rm R} [x'_{12}]^{\rm K}+
[x'_{21}]^{\rm K} \gamma_1^{\rm A}\tilde \Gamma_{1\rightarrow 2}^{\rm A}
\nonumber \\
&&+
\Gamma_{2\leftarrow 1}^{\rm R}\tilde \gamma_1^{\rm R} [x'_{11}]^{\rm K}
\gamma_1^{\rm A}\tilde \Gamma_{1\rightarrow 2}^{\rm A}
-
\Gamma_{2\leftarrow 3}^{\rm R}\tilde x_3 \tilde \Gamma_{3\rightarrow 2}^{\rm A}
\end{eqnarray}
with $ [x'_{ij}]^{\rm K}= S_{i2} x_2 S_{j2}^\dagger +
S_{i3} x_3 S_{j3}^\dagger $ for $i,j=1,2,3$.
Introducing what has been obtained before, we arrive at
\begin{align}\label{boundsol}
X_{2}= (\ru +A \Tsu ) &x_{2} (\ru+A\Tsu )^\dagger
\nonumber \\
+(\rud +A \Tsd ) &x_{3} (\rud+A\Tsd )^\dagger
\nonumber \\
-
(\Tus + A \Rs ) (\gamma_1^{\rm R} \Tsd^\ast )
&\tilde{x}_{3}
(\gamma_1^{\rm R} \Tsd^\ast)^\dagger (\Tus + A \Rs )^\dagger
\end{align}
Again, the corresponding solution for band 3 is obtained by replacing $2\leftrightarrow 3$.

\section{Interface model}
\label{IM}

\begin{figure}[b]
\includegraphics[width = 0.70\columnwidth]{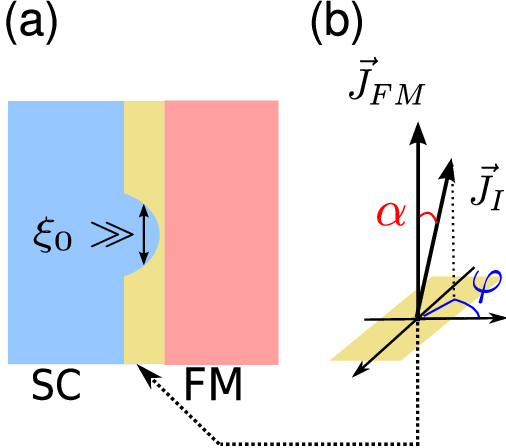}
\caption{\label{fig2}(Color online)
(a) The Andreev point contact with spin-active interface
(b) Interface with FM exchange field $\vJI$, $\alpha$ and $\phi$ characterize the orientation of $\vJI$ with respect to $\vJFM$.
The dashed arrow indicates the area where the misaligned interface magnetic moment resides.
}
\end{figure}

We consider a point contact with a diameter much smaller than the superconducting
coherence length but still larger than the Fermi-wavelength, as shown in Fig.~\ref{fig2} a.
A larger contact would result in a perturbation of the SC state, a smaller one would invoke
conductance quantization\cite{GolubovRMP}. This also allows for the decisive assumption
of translational invariance on the scale of $\lambda_{\rm{F}}$.
The region in the immediate vicinity of the interface (I) cannot be described
within QC theory. Instead, the normal state scattering matrix of the interface must
be obtained from microscopic calculations and then enters the QC theory through boundary
conditions as outlined above.

The mechanism giving rise to
spin-active scattering at the interface is the ferromagnetic exchange field
in both the adjacent ferromagnetic material and in the interface itself.
The interface will in general carry a magnetic moment, that in the simplest
case is induced by the
magnetization of the bulk ferromagnetic material; however, there might be cases
where an extra interface magnetic moment develops, either manufactured by using
a thin magnetic layer, or due to spin-orbit coupling, and related to that, magnetic
anisotropy.
The interface magnetic moment
can be misaligned with the one of the bulk sFM.
We characterize this misalignment by two spherical angles $\alpha$ and $\phi$,
as indicated in Fig.~\ref{fig2} b.
While the spin-activity of interfaces has been discussed extensively in the theory of
superconducting heterostructures, most of this work so far considered a set
of phenomenological parameters for characterizing the interfacial scattering.
Notably, one of these parameters, the so-called spin-mixing angle, or spin-dependent phase shift, turned out to be of decisive importance for the creation of unconventional
superconducting correlations in proximity to the interface.
The spin-mixing angle is essentially a relative phase difference between $\u$ and $\d$
electrons acquired upon scattering. Obviously, an exchange field in the interface
region will provide such an effect, but other mechanisms, like for instance spin-orbit
coupling are also candidates.

So far, estimates of the possible magnitude of this
effect based on a physical model of the interface region are still lacking. Here,
we will provide such an analysis based on wavefunction matching techniques.
In particular, we will discuss the dependence of the
spin-mixing effect on the shape of the barrier.
To this end, we consider a spin-split potential barrier which is assumed to conserve
the momentum component parallel to the interface upon scattering.
For the system we deal with, this gives rise to two types of transmission events
(see Fig.~\ref{fig1}). Depending on the impact angle the parallel momentum conservation
constraint either allows for or prohibits scattering into/from the minority
spin-band of the sFM.  For a half metal, where the $\d$-band is completely insulating,
only the latter case occurs.

\subsection{Interface scattering matrix}

At this point we mention some general considerations concerning the scattering matrix of
a spin-active interface. Such a matrix is unitary and of dimensions $4\times4$ in
the FM and $3\times3$ in the HM case.
The maximum number of free parameters is 16 or 9 respectively.
However, not all of these parameters will be relevant for the physical problem at hand.
For instance, spin-scalar phase factors do only matter for two or more interfaces.
Furthermore, since a singlet SC is spin-isotropic, one is free to choose the
spin-quantization axis in the SC conveniently.
To clearly identify these irrelevant parameters we use a special parameterization
of a general unitary matrix with the aforementioned dimensions,
as discussed in App.~\ref{app}. The most important result of these
considerations is that the spin-mixing effect can be fully described by only one
parameter in the HM case, but 3 are required in the FM case.

Neglecting irrelevant spin-scalar phases and using the gauge freedom in the SC
the scattering matrix reads for the first type of scattering
\begin{equation}\label{scatFM}
\hat{S}_{FM}=\left(\begin{array}{cc|cc}
r_{1\u} e^{i\vartheta/2} &
r_{1\u\d} & t_{2} e^{i\vartheta_2/2} & t'_{3} e^{i\vartheta_3/2} \\
r_{1\u\d} & r_{1\d} e^{-i\vartheta/2} & t'_{2} e^{-i\vartheta_2/2} &
t_3 e^{-i\vartheta_3/2}\\\hline
t_{2} e^{i\vartheta_2/2} & t'_{2} e^{-i\vartheta_2/2} & r_2 & r_{23} \\
t'_{3} e^{i\vartheta_3/2} & t_3 e^{-i\vartheta_3/2} & r_{23} & r_3
\end{array}\right).\end{equation}
The scattering matrix for the second, HM type, scattering is
\begin{equation}
\label{scatHM}
\hat{S}_{HM}=\left(\begin{array}{cc|c} r_{1\u} e^{i\vartheta/2} &
r_{1\u\d} & t_{2} e^{i\vartheta/4}  \\
r_{1\u\d} & r_{1\d} e^{-i\vartheta/2} & t'_{2} e^{-i\vartheta/4} \\\hline
t_{2} e^{i\vartheta/4} & t'_{2} e^{-i\vartheta/4} & r_2
\end{array}\right).\end{equation}
There is also the possibility of total reflection with no transmission on either side,
in which case the scattering matrix consists of the reflection parts only.
In writing the scattering matrices \eqref{scatFM} and \eqref{scatHM}
we have put the $\varphi$-phase that appears in Fig.~\ref{fig2}(b) to
zero, since the problem we consider is invariant with respect to rotation of the
interface magnetic moment around the bulk magnetization;
the scattering matrix is symmetric in this case, $S=S^T$.
We also omitted possible complex phases in the reflection part on the FM-side,
i.e. $r_2$, $r_3$ and $r_{23}$, as they are irrelevant to the problem at hand.
The requirement of unitarity leads to additional relations between the reflection
and transmission parameters.
The phases that we wrote explicitly in Eqs.~\eqref{scatFM} and ~\eqref{scatHM}
are crucial, since they account for the spin-mixing effect.
In the following section, we will discuss their magnitude
as a function of various interface parameters.

Using the set of independent
parameters described in the appendix we have:
\begin{align}\label{alphaU} r_{1\u}&=r_{\u}\cos(\alpha_Y/2)^2+r_{\d}\sin(\alpha_Y/2)^2\\\nonumber
 r_{1\d}&=r_{\u}\sin(\alpha_Y/2)^2+r_{\d}\cos(\alpha_Y/2)^2\\\nonumber
r_{1\u\d}&=-(r_{\u}-r_{\d})\frac{\sin(\alpha_Y)}{2}.
\end{align}
The angle $\alpha_Y$ defines a rotation in spin-space to the interface eigenstates,
characterized by transmission and reflection eigenvalues. Its precise definition is given
in the appendix. Most importantly, it is in general not
identical to the interface misalignment angle $\alpha $, however approaches
it in the limit of thick interfaces.  For thin interfaces it is renormalized
by the influence of the exchange field of the adjacent FM.
$r_{\u}$ and $r_{\d}$ are the singular values of the reflection block $\hat{R}_S$.
In the tunneling limit, $r_{\u},\ r_{\d}\approx 1$,
and the off-diagonal elements vanish even for $\alpha_Y\neq 0$. This is easily
understood from a physical point of view, since spin-flip reflections on the SC
side requires that the reflected quasiparticles ``feel'' both misaligned
exchange fields and not just that of the interface.
It is possible to provide analogous expressions for the remaining
parameters of the scattering matrix, however in the sFM case
they are rather cumbersome and also not needed for the following analytical
discussion.
For the half-metallic case, the only relevant phase parameter is the
spin-mixing-angle $\vartheta$, and for the remaining parameters we have
$r_{\d}=1$  and
\begin{align} t_2=t_{\u}\cos\left(\frac{\alpha_Y}{2}\right),\;
 t'_2=-t_{\u}\sin\left(\frac{\alpha_Y}{2}\right),\; r_2=-r_{\u}.
\end{align}

In the following we will discuss the influence of the shape of the scattering
potential, and will show that the widely used box shaped or delta-function
shaped potentials gravely underestimate the magnitude of the spin-mixing effect.

\subsection{Box-shaped scattering potential}

\begin{figure}[b]
\includegraphics[width = 0.99\columnwidth]{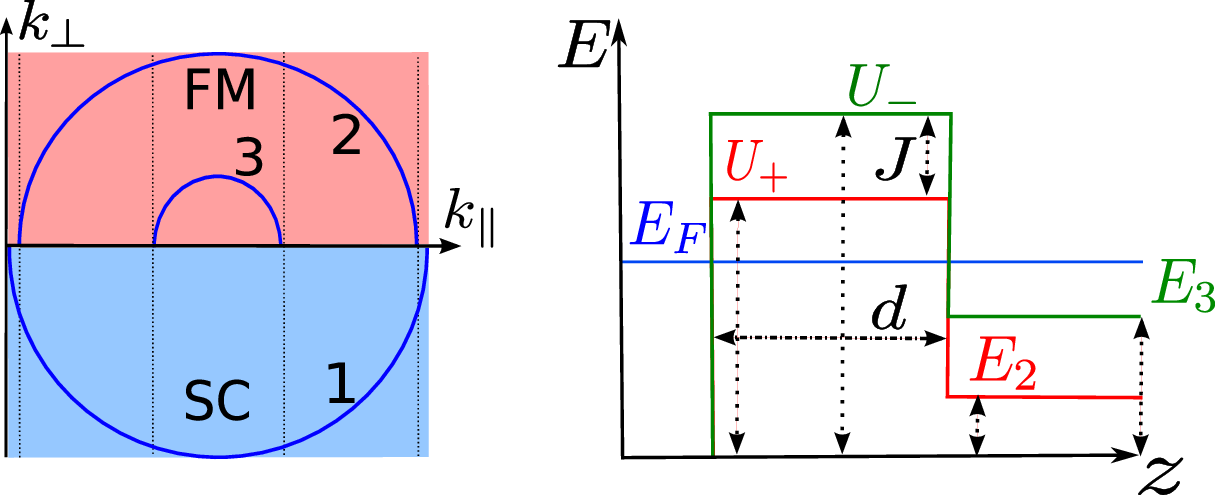}
\caption{\label{fig3}(Color online)
Sketch of the box-potential model that we consider in this section (right) and of the corresponding Fermi-surface geometry (left). The model parameters are indicated.
}
\end{figure}

In this section
we consider spin-dependent box potentials, for which analytical solutions can be obtained.
In particular, we discuss here the dependence of $\vartheta$ on the impact angle of the
incoming quasiparticle which is parameterized by the momentum component parallel to
the interface, $k_{||}$.
The model parameters are the misalignment angle $\alpha$ (see Fig.~\ref{fig2} b),
the energies of the band minima in the FM with respect to that in the SC ($E_2, \ E_3$),
the spin-dependent height of the potential ($U_{+}, \ U_{-}$),
and the width of the potential $d$ (see Fig.~\ref{fig3}).
All energies are given in units of $E_F$ and $d$ in units of $\lambda_{\rm{F}}/2\pi$.

The scattering matrix is defined with respect to the chosen spin-quantization axes
on both sides of the interface.
Naturally, on the FM side we use the bulk sFM magnetization axis.  On the SC side we use that of the interface magnetic moment.
To obtain an S-matrix with the structure defined above, one must subsequently calculate and apply a rotation of the quantization axis in the SC:
\begin{equation} \left(\begin{array}{cc} Q^{\dagger} & 0 \\ 0 & 1 \end{array}\right)\hat{S}\left(\begin{array}{cc} Q & 0 \\ 0 & 1 \end{array}\right),\end{equation}
where $Q$ is a spin rotation matrix acting on spins in the superconductor.
We describe this procedure in App.~\ref{appsing}.
All the quantities plotted are calculated in this rotated frame, the point being that otherwise one does not have an unambiguous definition of the mixing-phases.
Naturally, the Andreev spectra are invariant under these transformations.
We obtain the scattering matrix by matching wave functions as described
in App.~\ref{appbox}.

\begin{figure}[b]
\includegraphics[width = 0.49\columnwidth]{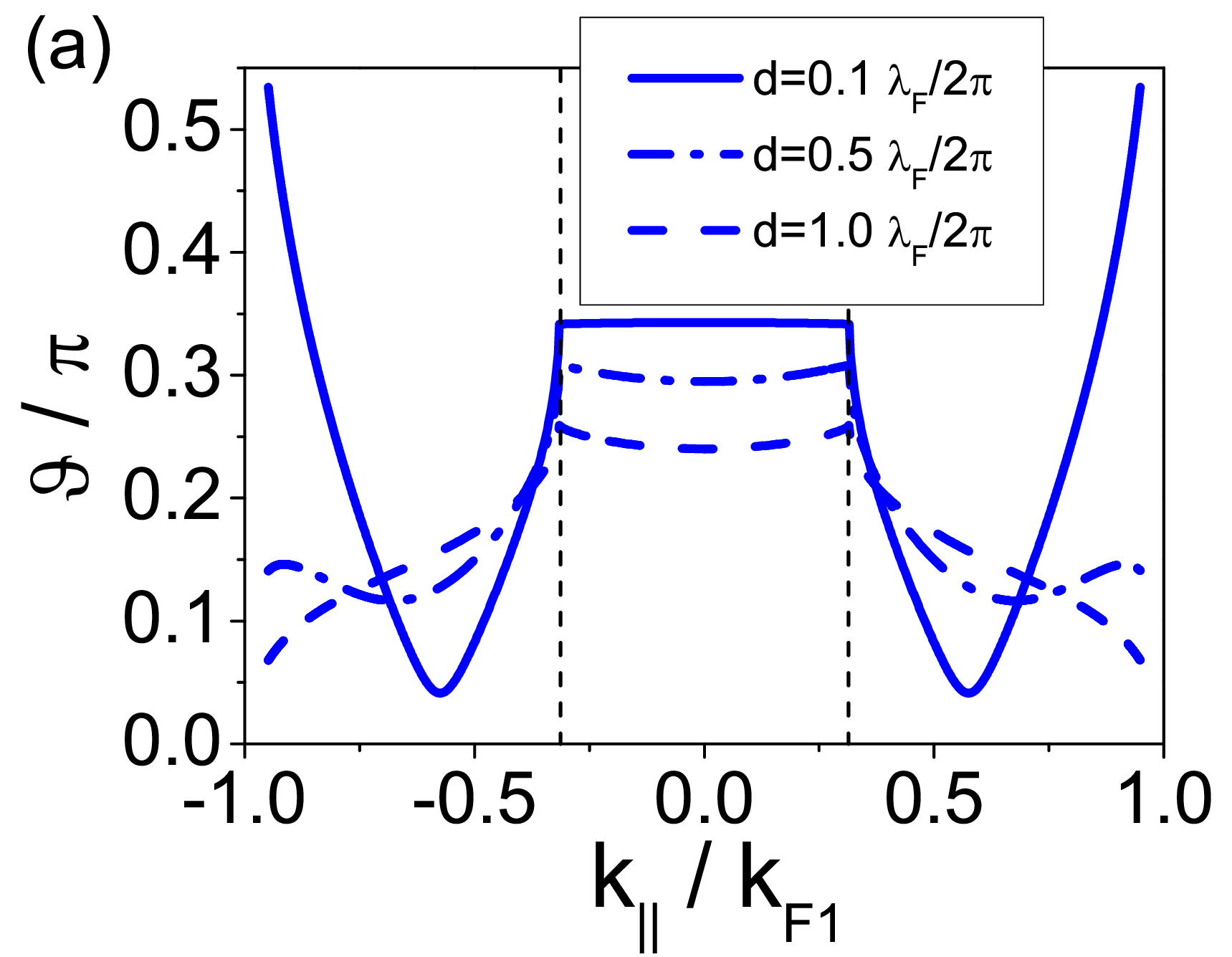}\includegraphics[width = 0.49\columnwidth]{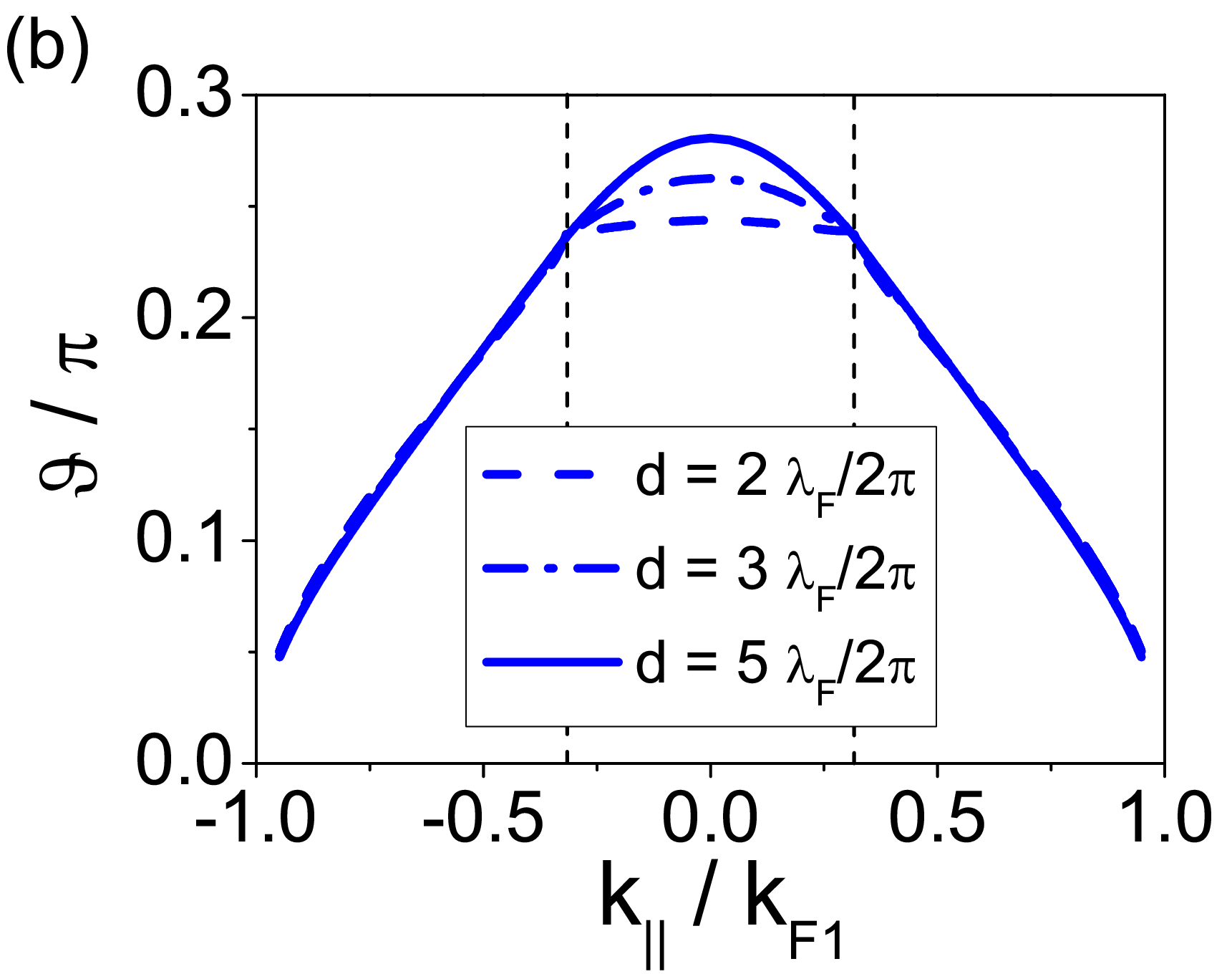}\\
\caption{\label{fig4}(Color online)
The spin-mixing angles $\vartheta $ as function of the
momentum component parallel to the interface, shown for different barrier thickness.
(a) $d=0.1,\ 0.5,\ 1.0\ \lambda_{\rm{F}}/2\pi$,
(b) $d=2.0,\ 3.0,\ 5.0\ \lambda_{\rm{F}}/2\pi$.
The remaining parameters for all plots are
$E_2=0.1 E_{\rm{F}}$, $E_3=0.9\ E_{\rm{F}}$,
$U_+=1.1\ E_{\rm{F}}$, $U_-=1.9\ E_{\rm{F}}$, $\alpha=0.5\ \pi$ (see text).
}
\end{figure}
In Fig.~\ref{fig4} a,b we show the spin-mixing angle for different values of the
interface potential width $d$. The band minima in the FM are $E_2=0.1\ E_{\rm{F}}$ and
$E_3=0.9\ E_{\rm{F}}$, which implies that at $k_{||}\ge 0.31\ k_{\rm{F}1}$
the minority band becomes insulating and the scattering matrix reduces to a $3\times3$ matrix.
In the tunneling limit ($d \gg \lambda_{\rm{F}}/2\pi$)
the spin-mixing angle behaves as expected:
it is approximately given by the value (see App.~\ref{appbox})
\begin{equation}
\vartheta =2 \left[ \mathrm{arctan}\left(\frac{k_1}{\kappa_{+}} \right)
-\mathrm{arctan}\left(\frac{k_1}{\kappa_{-}} \right)\right],
\end{equation}
which appoaches zero for grazing impact ($k_1\approx 0$), and
$2[\arctan\sqrt{E_F/(U_+-E_F)}-\arctan\sqrt{E_F/(U_--E_F)}]$ for
normal impact ($\approx 0.29\pi $ for Fig.~\ref{fig4}).
Here, $k_1$ is the component of the wavevector perpendicular to the interface
in the superconductor, and $\kappa_\pm $ are the exponential decay factors for the spin-up/down wave function in the barrier.
For thin (highly transparent) interfaces the mixing-angle $\vartheta $ is a
more complicated function of the quasiparticle impact angle.
In this regime, $\vartheta $ is predominantly controlled by the Fermi-surface geometry indicated in Fig.~\ref{fig3}.
There is a local minimum at $k_{\parallel}>k_{\rm{F}3}$,
and for very thin interfaces $\vartheta $
is largely enhanced for grazing impact ($d=0.1\ \lambda_{F}/2\pi$ in Fig.~\ref{fig4}).
This
enhancement can be understood from the $d=0$ limit, i.e. the case where the interface barrier is absent. In this case,
\begin{equation}
\vartheta=\pi-2\ \mathrm{arctan} \left(\frac{k_1}{\kappa_{3}}\right) ,
\end{equation}
where, $\kappa_3$ corresponds to the imaginary
wave vector in the insulating band 3, which controls the
exponential decay of the spin-down wave function into the ferromagnet.
In the particular case we show here, see Fig.~\ref{fig3},
$k_1$ takes a finite value for all trajectories that contribute to the current,
while $\kappa_3$ \emph{increases} monotonously from $0$ at
$k_{||}=k_{\rm{F}3} \approx 0.31\ k_{\rm{F}1}$ to some finite value at $k_{||}=k_{\rm{F}2}$.
This is because the effective height of the potential for tunneling into the insulating band increases with $k_{||}$.
For Fermi-surface geometries with
$k_{\rm{F}1}<k_{\rm{F}2}$ (not shown here) the wave vector $k_1$ drops to zero
for grazing impact, and the spin-mixing angle approaches $\pi$.

\begin{figure}
\includegraphics[width = 0.49\columnwidth]{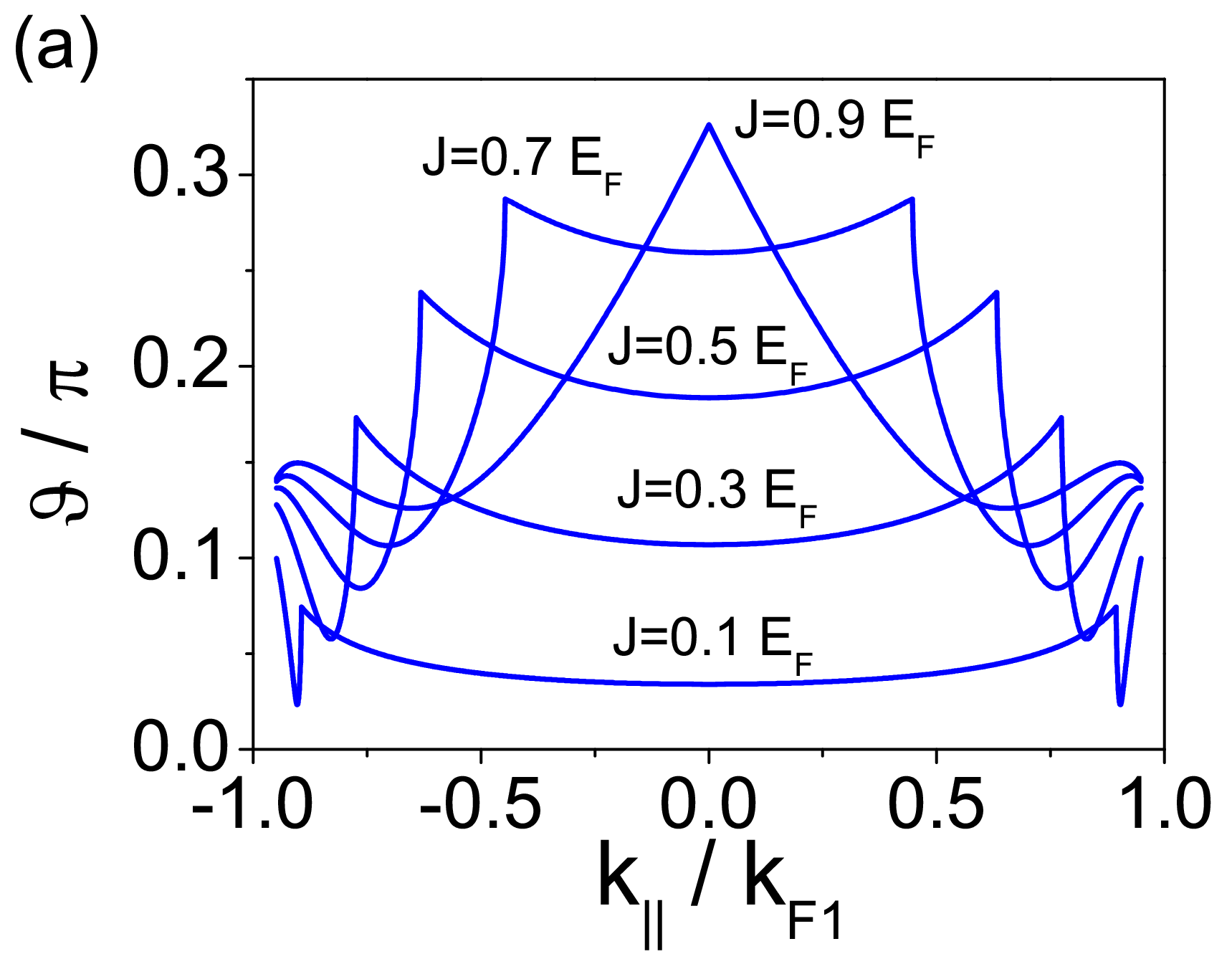}\includegraphics[width = 0.49\columnwidth]{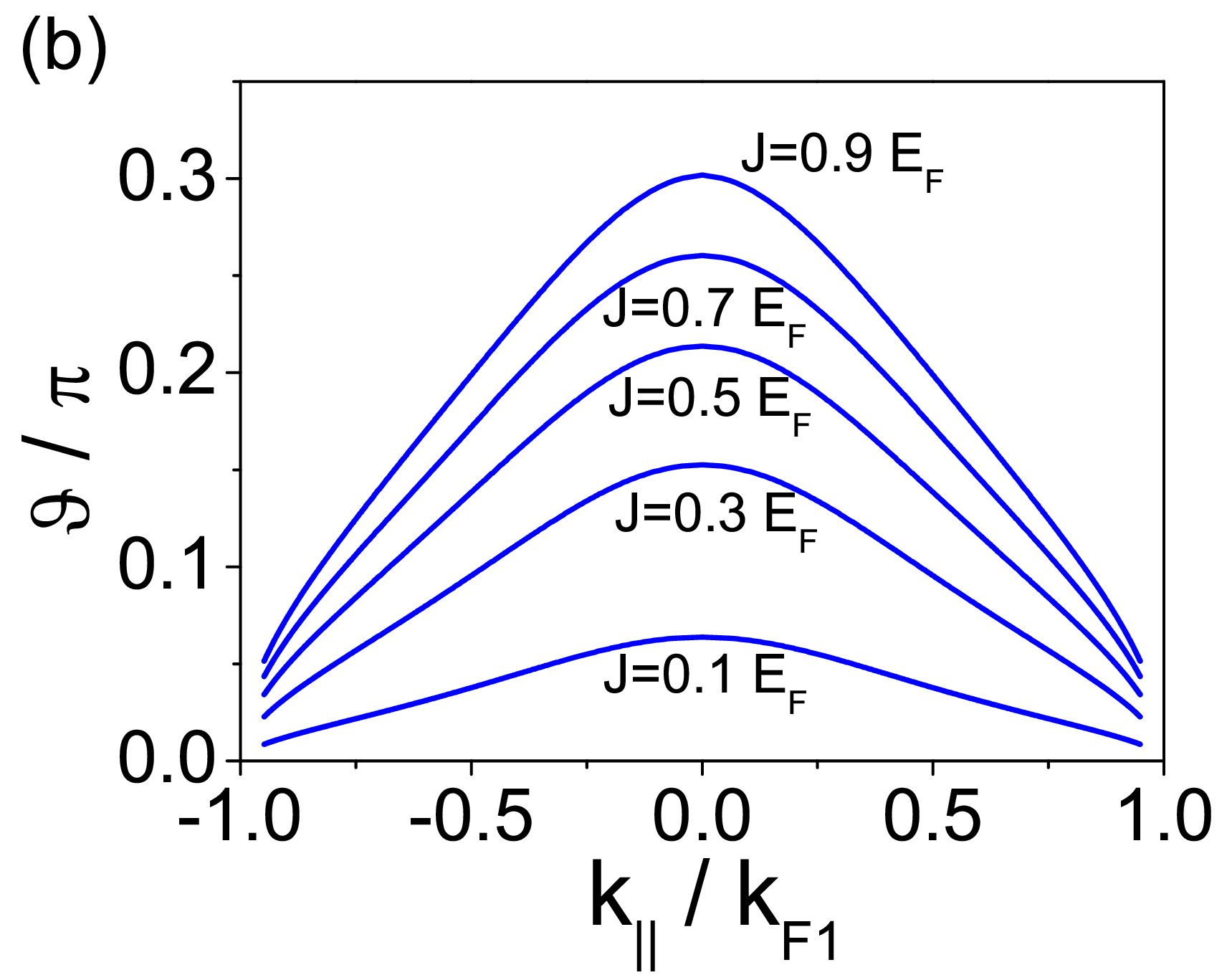}
\caption{\label{fig7}(Color online)
(a) Spin-mixing angle $\vartheta$ as a function of impact angle
for (a) $d=0.5\ \lambda_{\rm{F}}/\pi$, and (b) $d=5.0\ \lambda_{\rm{F}}/2\pi$.
In both plots, the curves are for $U_-=1.2...2.0\ E_{\rm{F}}$, $E_3=U_--1.0$,
the corresponding value of the exchange field $J$ is indicated.
The remaining parameters are
$E_2=0.1 E_{\rm{F}}$, $U_+=1.1\ E_{\rm{F}}$, and $\alpha=0.5\ \pi$.
}
\end{figure}

\begin{figure}[b]
\includegraphics[width = 0.49\columnwidth]{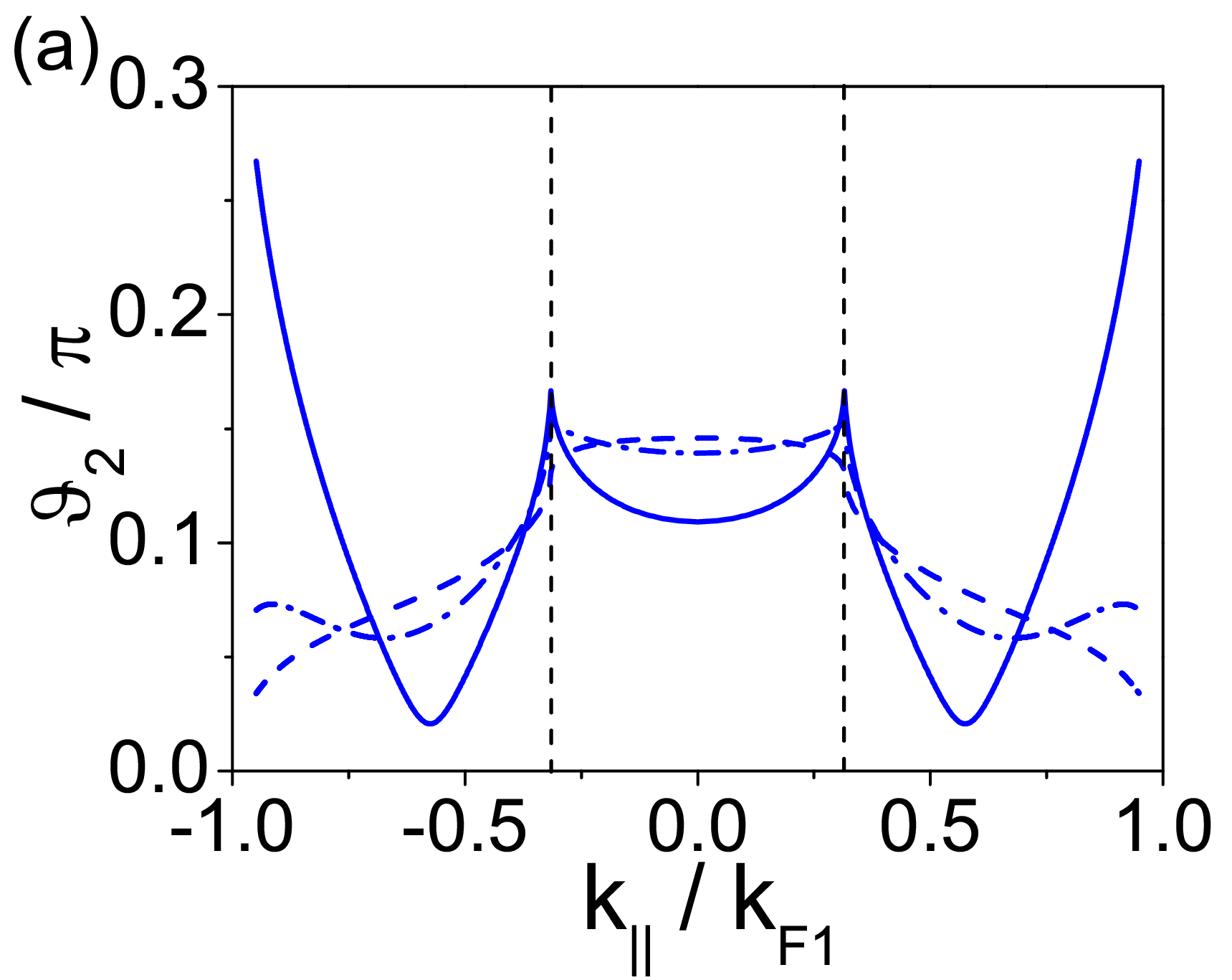}\includegraphics[width = 0.49\columnwidth]{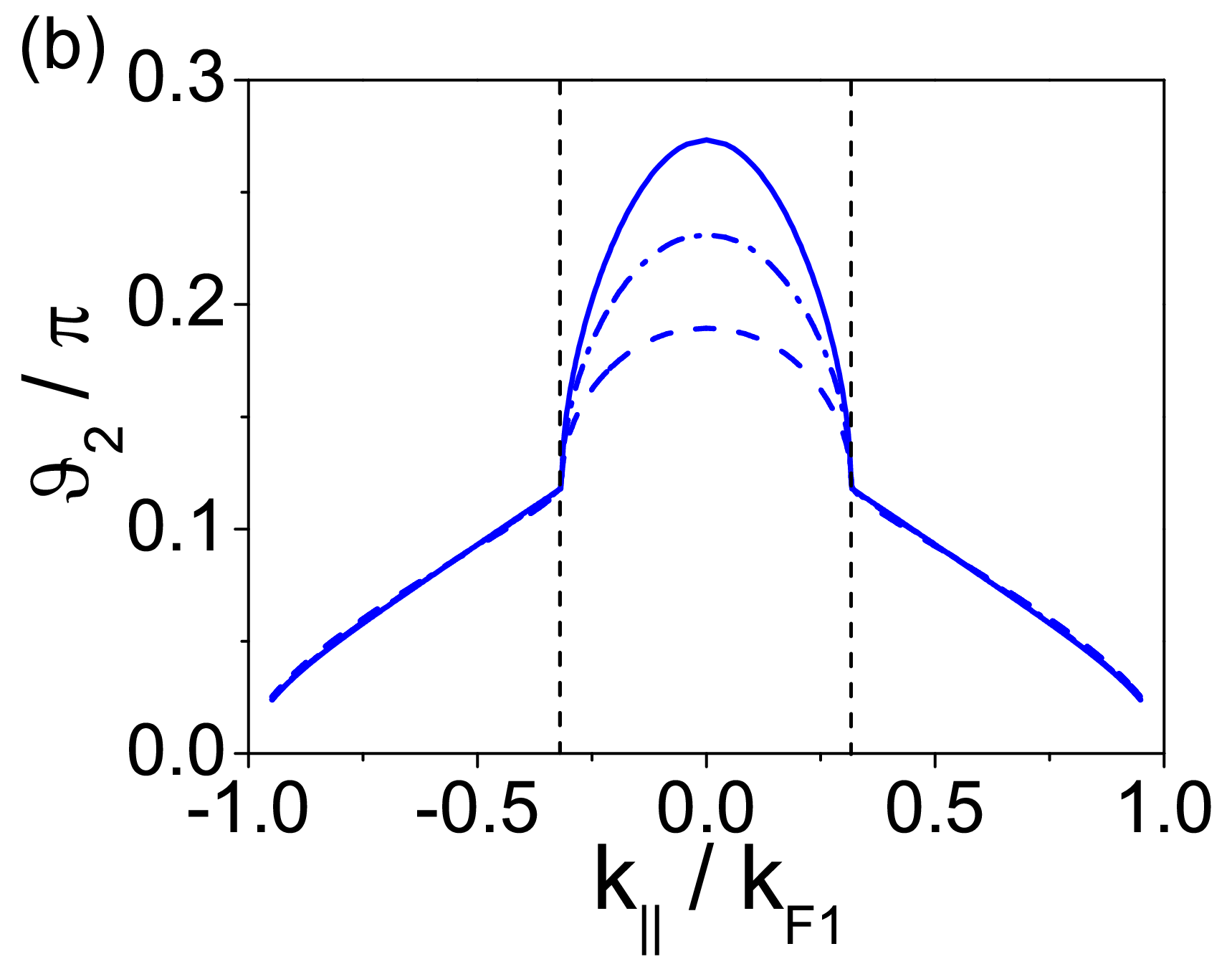}\\
\includegraphics[width = 0.49\columnwidth]{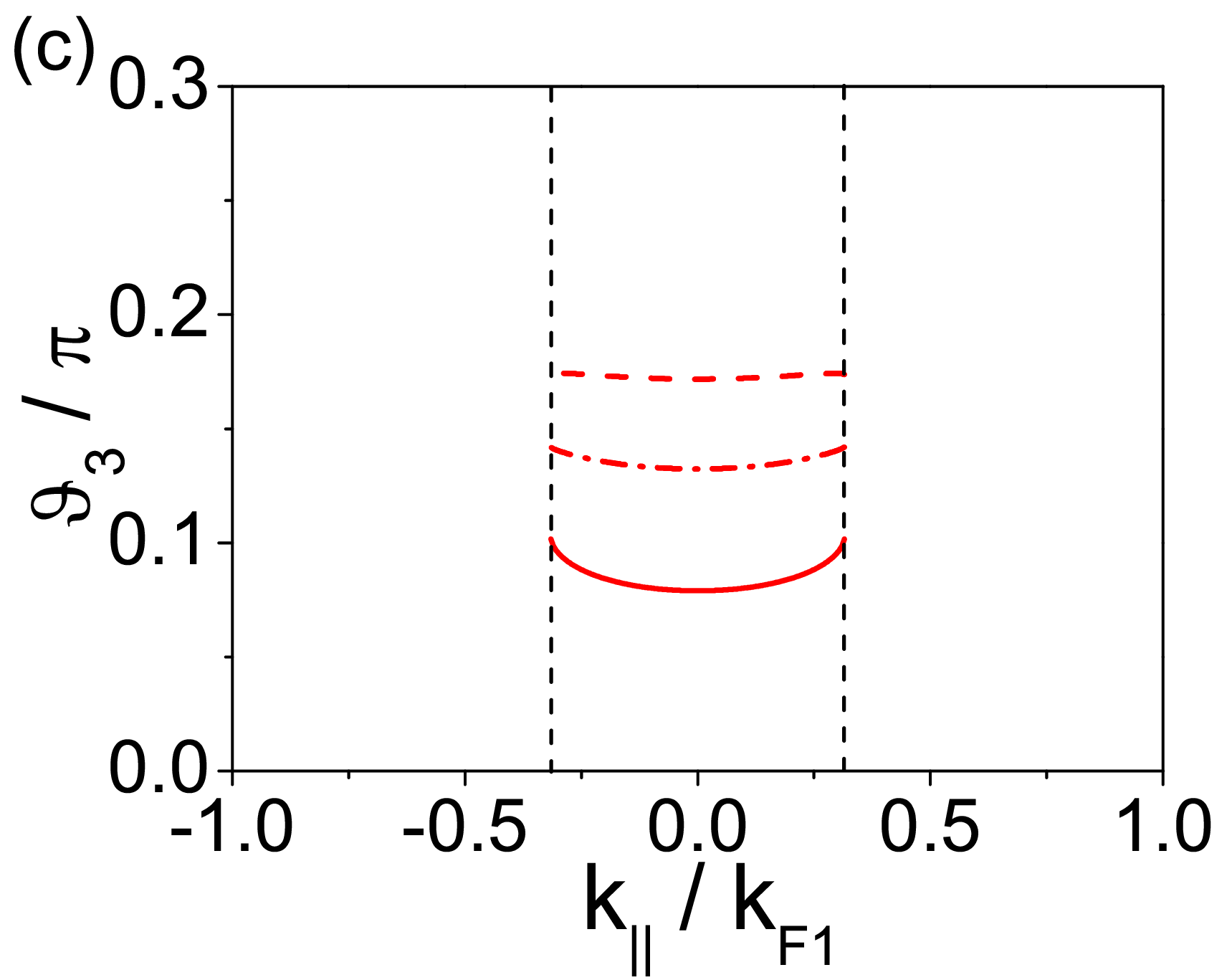}\includegraphics[width = 0.49\columnwidth]{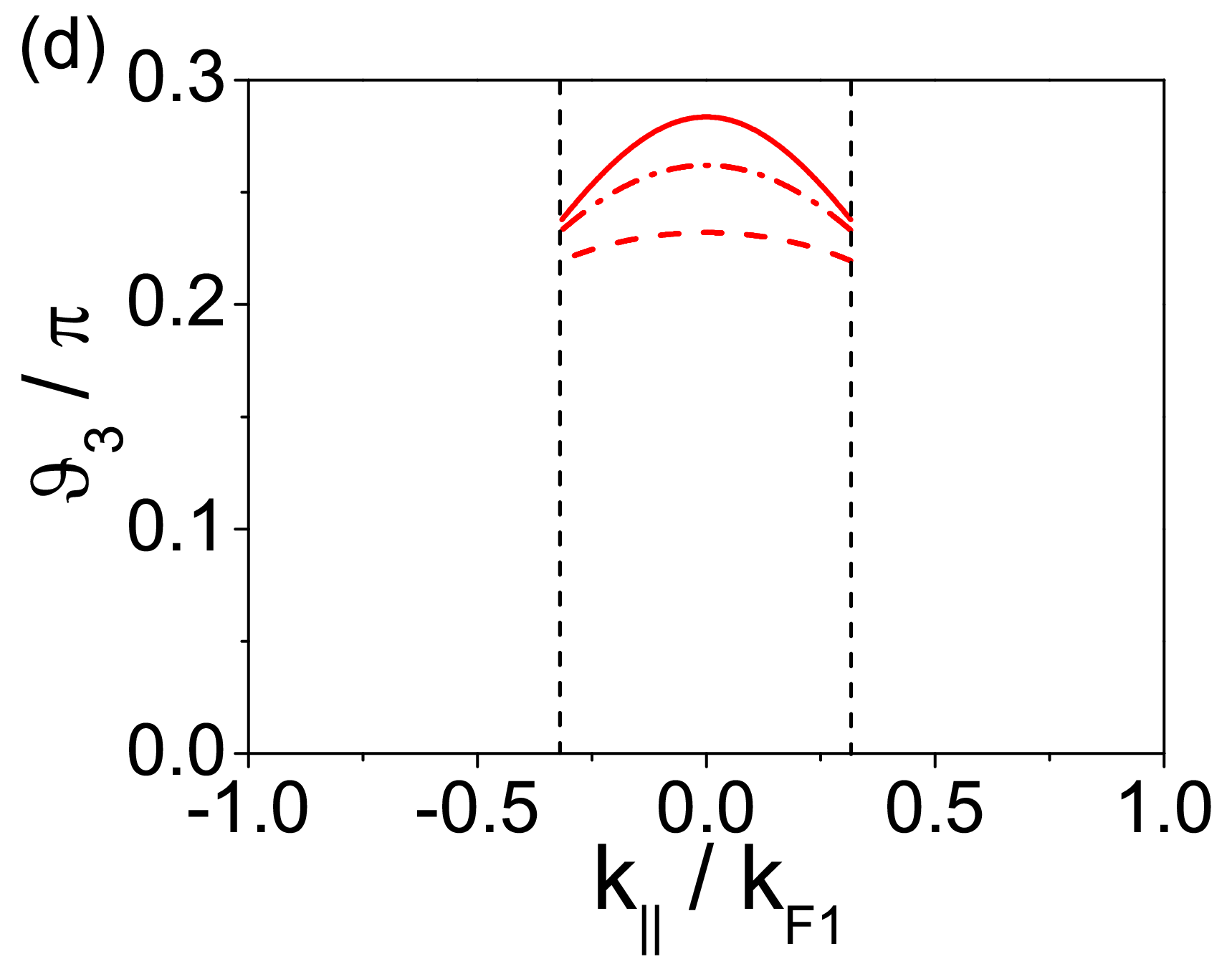}\\
\caption{\label{fig5}(Color online)
The spin-mixing angles $\vartheta_2$ (first row) and $\vartheta_3$ (second row) for thin (left column: $d=0.1\ \mathrm{(solid)}$, $0.5\ \mathrm{(dashed-dotted)}$, $1.0\ \mathrm{(dashed)}\ \lambda_{\rm{F}}/2\pi$) and thick (right column: $d=2\ \mathrm{(solid)}$, $3\ \mathrm{(dashed-dotted)}$, $5\ \mathrm{(dashed)}\ \lambda_{\rm{F}}/2\pi$) interfaces. All parameters are the same as in Fig.~\ref{fig3}.
(a) and (b) $\vartheta_2(k_{||})$;
(c) and (d) $\vartheta_3(k_{||})$.
}
\end{figure}

In the present case, the situation is complicated by the fact that
we consider both a finite interlayer and a broken spin-rotation symmetry.
This leads to a finite spin-mixing angle even for $k_{||}=k_{\rm{F}3}$ and below,
which leads to the non-trivial behavior with a minimum for intermediate impact
angles.
This illustrates that not only the scattering potential itself but also the Fermi-surface geometry is highly important for spin-active scattering beyond the
tunneling limit.

\begin{figure}[b]
\includegraphics[width = 0.49\columnwidth]{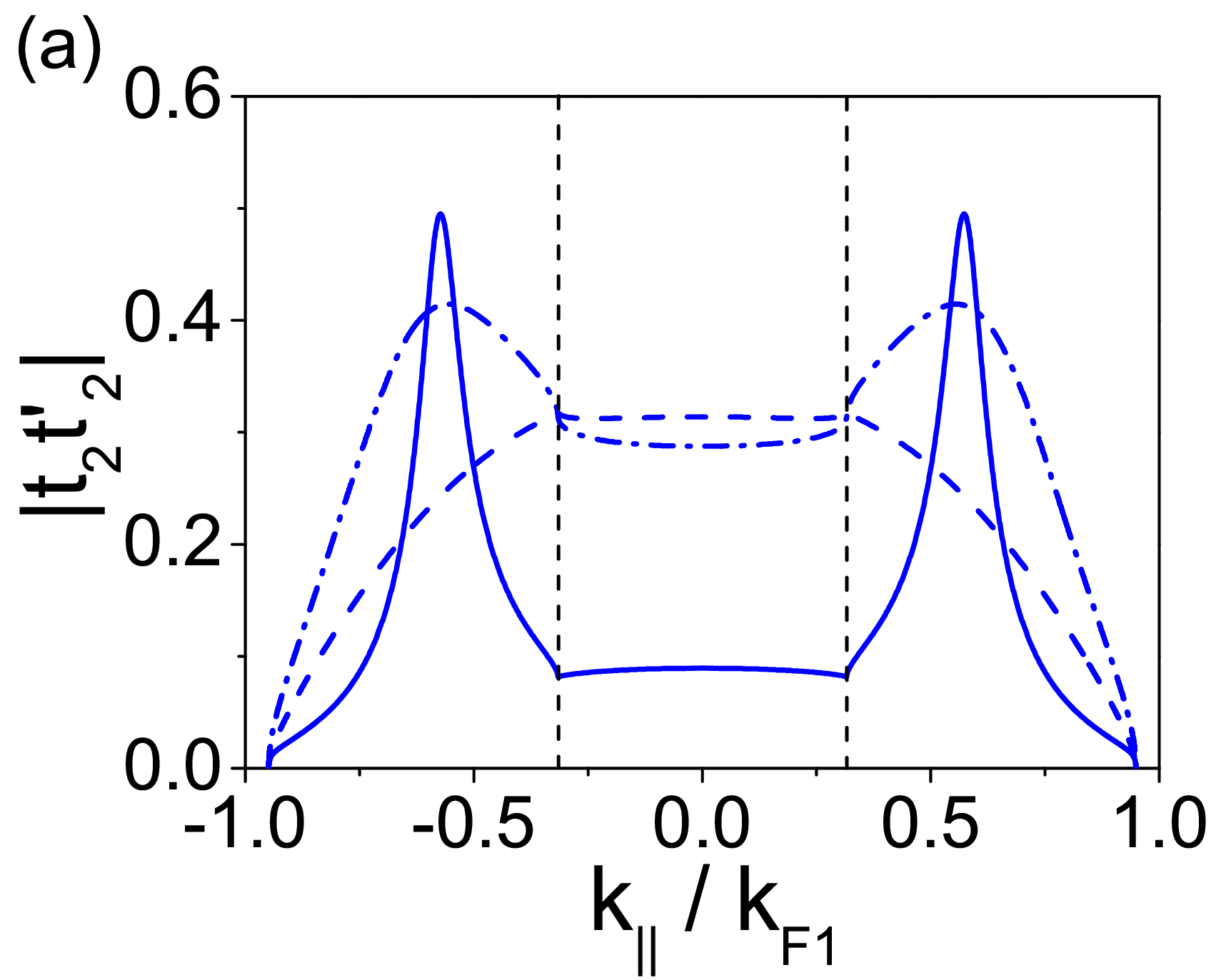}\includegraphics[width = 0.49\columnwidth]{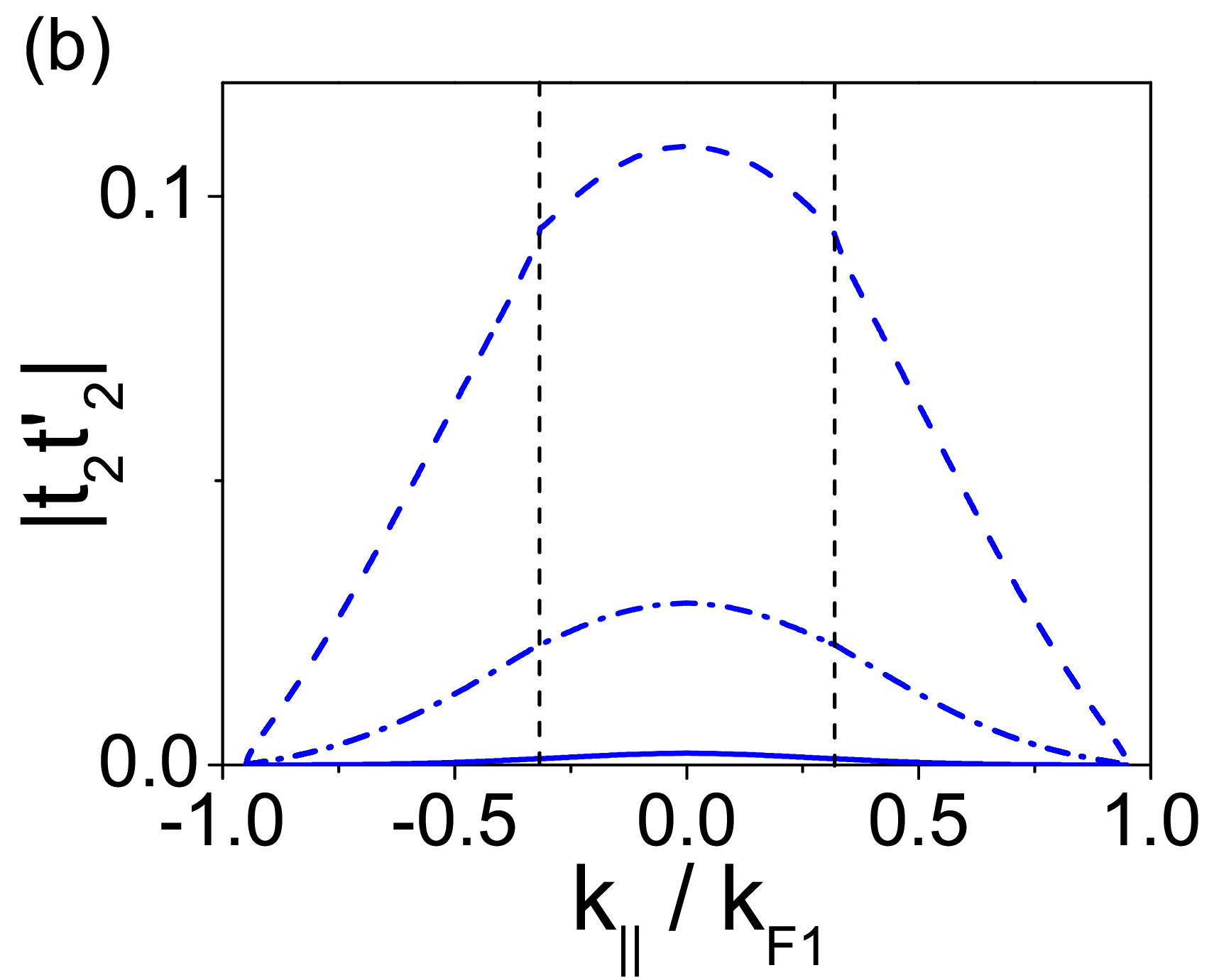}\\
\includegraphics[width = 0.49\columnwidth]{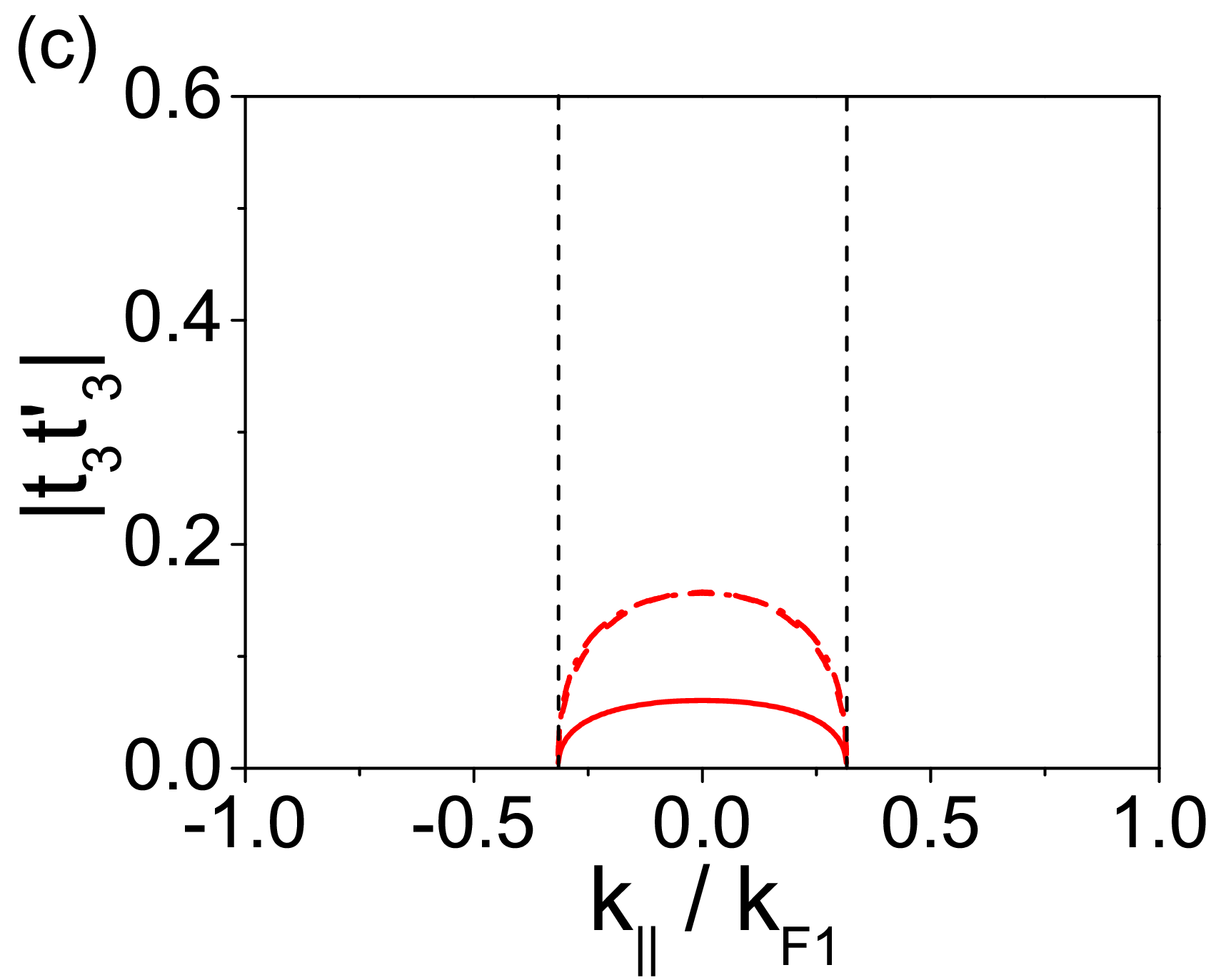}\includegraphics[width = 0.49\columnwidth]{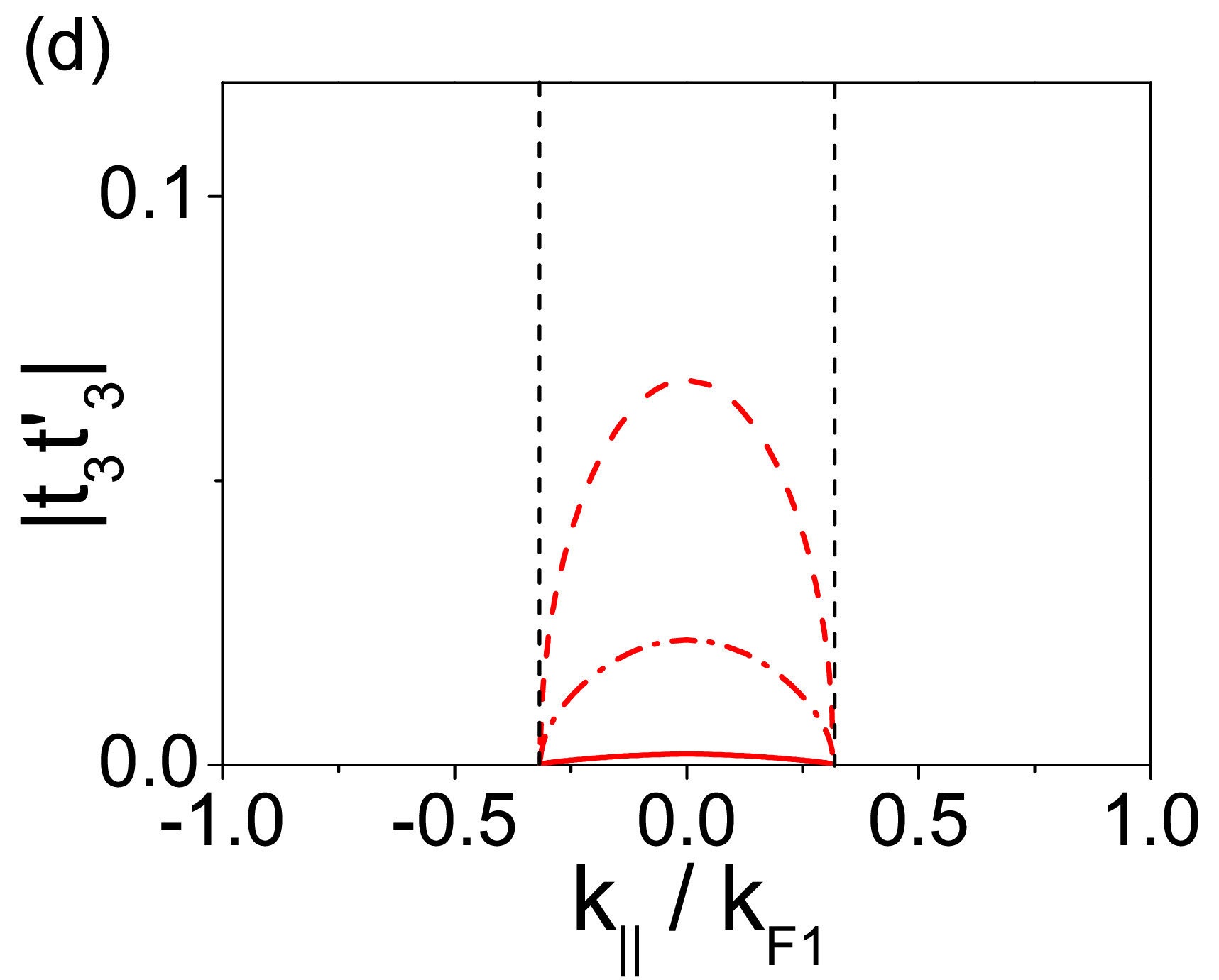}\\
\caption{\label{fig6}(Color online)
The spin-mixing angles $|t_2t'_2|$ (first row) and $|t_3t'_3|$ (second row) for thin (left column: $d=0.1\ \mathrm{(solid)}$, $0.5\ \mathrm{(dashed-dotted)}$, $1.0\ \mathrm{(dashed)}\ \lambda_{\rm{F}}/2\pi$) and thick (right column: $d=2\ \mathrm{(solid)}$, $3\ \mathrm{(dashed-dotted)}$, $5\ \mathrm{(dashed)}\ \lambda_{\rm{F}}/2\pi$) interfaces. All parameters are the same as in Fig.~\ref{fig3}.
(a) and (b) $|t_2t'_2|(k_{||})$;
(c) and (d) $|t_3t'_3|(k_{||})$.
}
\end{figure}

As for the magnitude of the mixing effect, we stress that for a realistic choice of parameters, it is hardly possible to
achieve mixing-phases above $0.5\pi$ in this model.
In Fig.~\ref{fig4} we use an exchange field of $J=0.8\ E_{\rm{F}}$,
which is close to the half-metallic limit. Using smaller exchange energies naturally
leads to a smaller effect, as can be seen in Fig.\ref{fig7} a,b, where we
plot $\vartheta$ for different values of the exchange field $J=E_3-E_2$.

In Fig.~\ref{fig5} we show the spin-mixing phases associated to transmission $\vartheta_2$ and $\vartheta_3$. One can see that $\vartheta_2=\vartheta/2$ for
$k_{||}>0.31\ k_{\rm{F}1}$. This relation one would expect for a SC contacted
with a half-metallic ferromagnet; the finding in Fig.~\ref{fig5}
is consistent with this and the discussion
presented above, since the trajectories under consideration
effectively correspond to the HM case.
For $k_{||}<0.31\ k_{\rm{F}1}$, the mixing phase is considerably enhanced above
the value of $\vartheta/2$. The plots also illustrate that $\vartheta_2$ and $\vartheta_3$ are different in magnitude and also vary differently with $k_{\parallel}$. As we show in the appendix, the mixing-phases $\vartheta_2$ and $\vartheta_3$
are correlated with $\vartheta$ but in general also depend on a number of other free parameters.
Their magnitude is decisive for the creation of triplet correlations in the corresponding band, as we will show below.

In Fig.~\ref{fig6} we present the product $|t_{\eta}t'_{\eta}|$ (which controls the
magnitude of long-range SAR).
We plot this quantity for both the
majority (upper row) and minority (lower row) band of the FM. Apparently there is a
non-monotonous dependence on the interface width $d$, which is related to the
fact that spin-flip scattering becomes more effective as the interface region
becomes larger. For even larger $d$ the global suppression of transmission
intervenes and we approach the tunneling limit. Again, we note that for thin
interfaces the dependence on trajectory impact angle is non-monotonous,
showing maxima for non-perpendicular impact. These maxima
coincide exactly with the minima of the spin-mixing angle.
Note that a nonzero $t'_{\eta}$ requires a non-vanishing misalignment angle $\alpha$.

To conclude on this section, we have shown that the magnitude of the spin-mixing
effect is limited to rather small values in the box potential case
if one assumes $J< E_{\rm{F}}$ and $d\approx \lambda_{\rm{F}}$.
Moreover, both spin-mixing effect and spin-flip scattering are very sensitive
to trajectory impact, interface thickness, exchange field of the interface and
the Fermi surface geometry of the adjacent materials.

\subsection{Delta-function scattering potential}
\begin{figure}[b]
\includegraphics[width = 0.49\columnwidth]{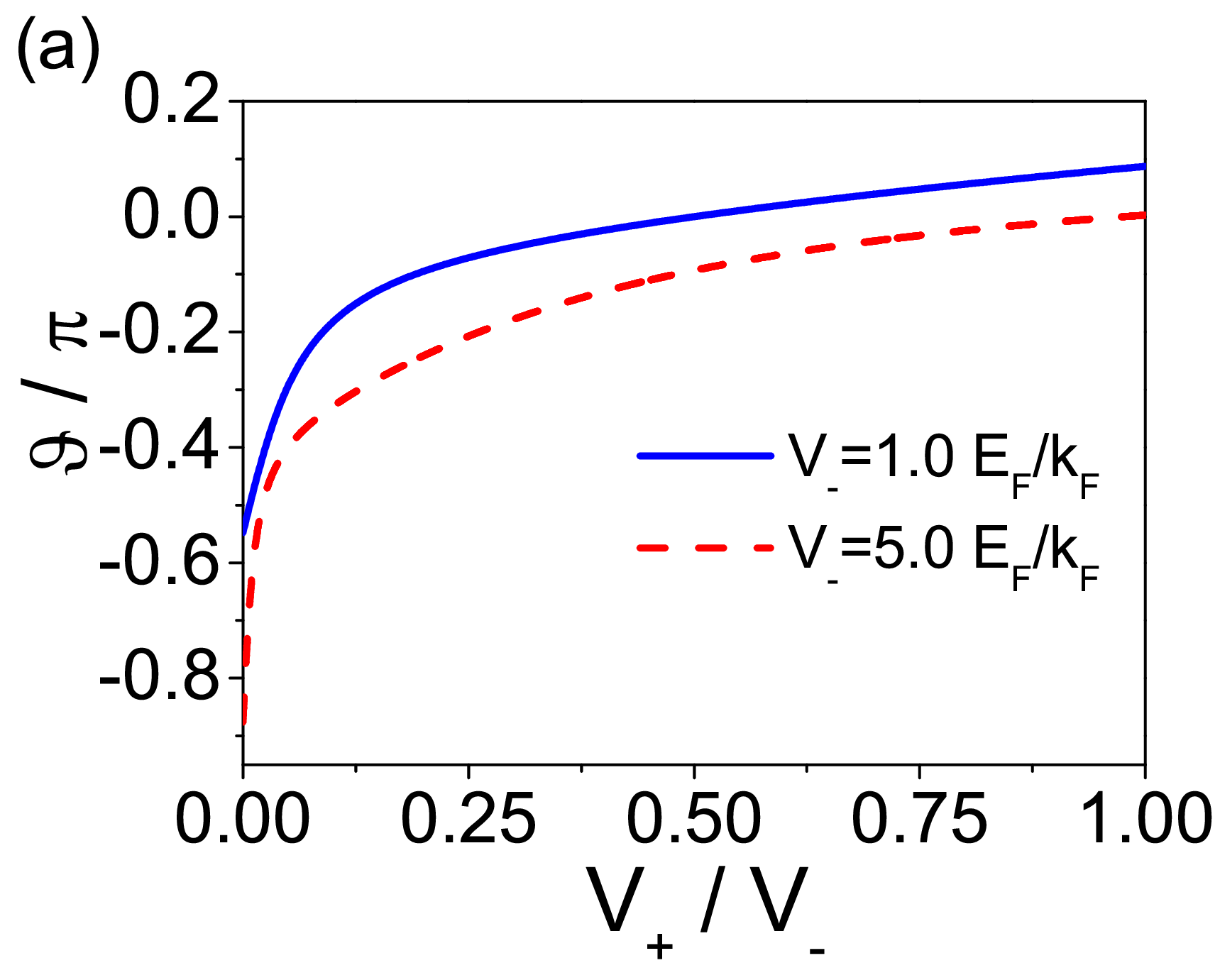}\includegraphics[width = 0.49\columnwidth]{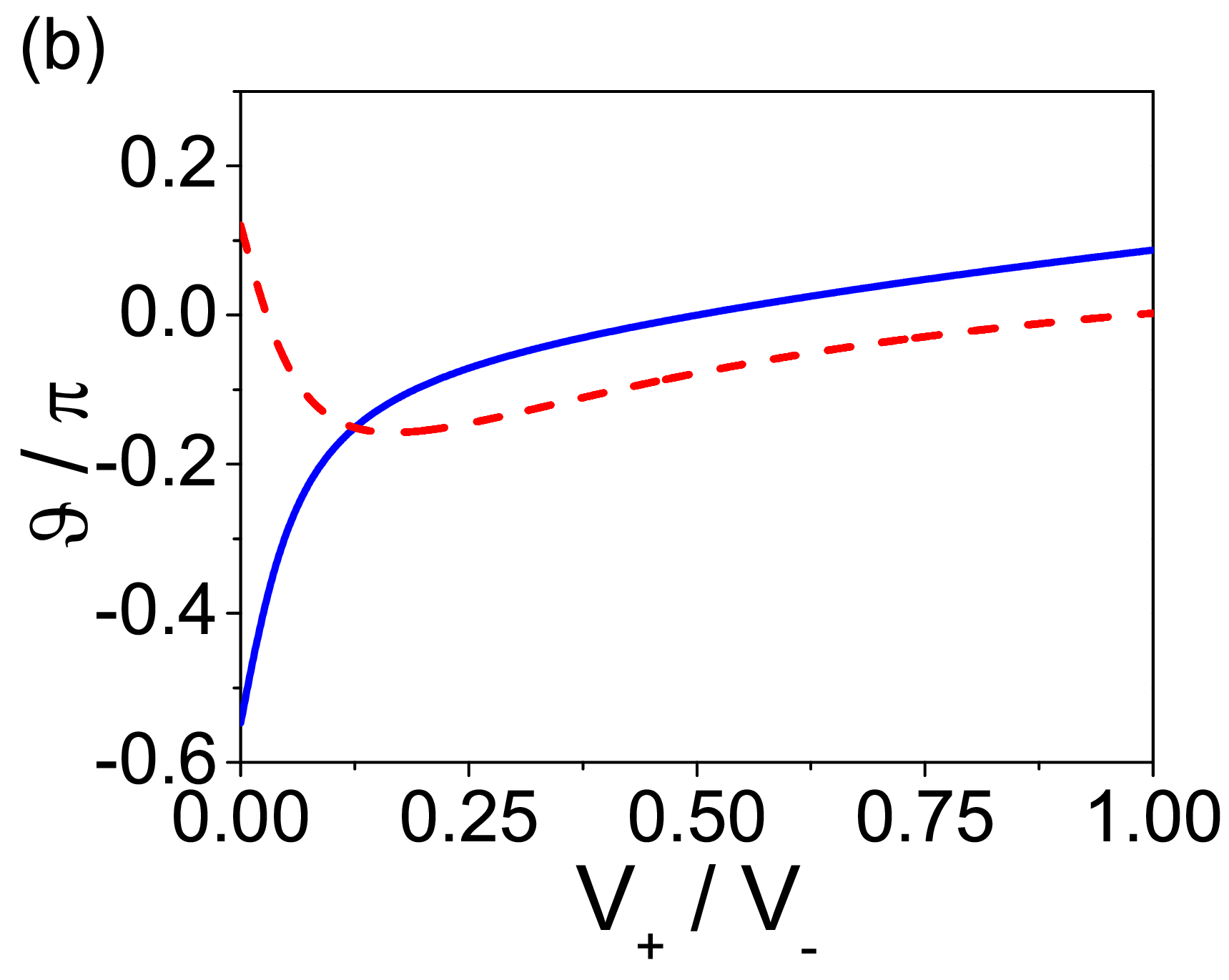}\\
\caption{\label{fig8}(Color online)
(a) Spin-mixing angle $\vartheta$ as a function of $V_+/V_-$ for the Delta-function potential. $E_2=0.1\ E_{\rm{F}}$,
 $E_{3}=0.7\  E_{\rm{F}}$.
(b) The same as (a) for $E_2=-0.7\ E_{\rm{F}}$,
 $E_{3}=-0.1\  E_{\rm{F}}$.
}
\end{figure}
A special case of the box-shaped potential is that of the delta-function potential,
that is widely used in describing interfaces within the BTK paradigm.
Here, we show that the situation is in this case comparable to that of the box potential.
Delta-function models introduce a weight factor $V_0$ of the Delta-function which enters the matching
condition for wavefunction derivatives:
\begin{equation} \frac{d}{dz}\Psi_1(z=0)-\frac{d}{dz}\Psi_2(z=0)=\frac{2 m V_0}{\hbar^2}\Psi_2(z=0).\end{equation}
A spin-dependent potential can simply be modeled by choosing a spin-dependent weight factor $V_{\pm}$.
This weight factor effectively corresponds to the area under the scattering potential,
i.e. we have $V_{\pm}=(U_{\pm}-E_F)\cdot d$, to connect with the notation above.
In Fig.~\ref{fig8} we plot $\vartheta$ as a function of $V_+/V_-$ for perpendicular impact and two different choices of
the Fermi-surface geometry. Since we do not calculate any spectra for this model, we choose $\alpha=0$ for simplicity.
Generically, spin-mixing angles $\vartheta>0.5\pi$ can only be reached for $V_+/V_- < 0.1$,
which requires either $V_+$ to be very small or an interface exchange field exceeding the Fermi-energy.

\subsection{Scattering potentials with arbitrary shape}

The box potential actually constitutes a high degree of idealization. The most obvious generalization is to consider a potential that varies smoothly on the scale of a few interatomic distances, or on the scale of the Fermi wavelength in metals.\cite{smooth} This is quite realistic taking into account that metallic screening of
charges takes place only on the Thomas-Fermi wavelength scale.
In addition, some magnetic ions might penetrate the superconductor from the
ferromagnet,
leading to a spin-dependent potential that decays in the bulk of the
superconductor. In the latter case a certain degree of disorder is introduced.
However, we will assume that any such disorder is weak, so that the momentum
component parallel to the interface is still a good quantum number.
A truly realistic description would have to drop the assumption of translational invariance and consider disorder on a microscopic level. In principle our theory can be extended to this regime, but this is beyond the scope of this paper. If one is only interested in transmission and reflection amplitudes, the difference between the box-shape and a smoothened potential is negligible. But when scattering phases are important, as in our case, this is not true, as we will show in the following.

\begin{figure}[b]
\includegraphics[width = 0.9\columnwidth]{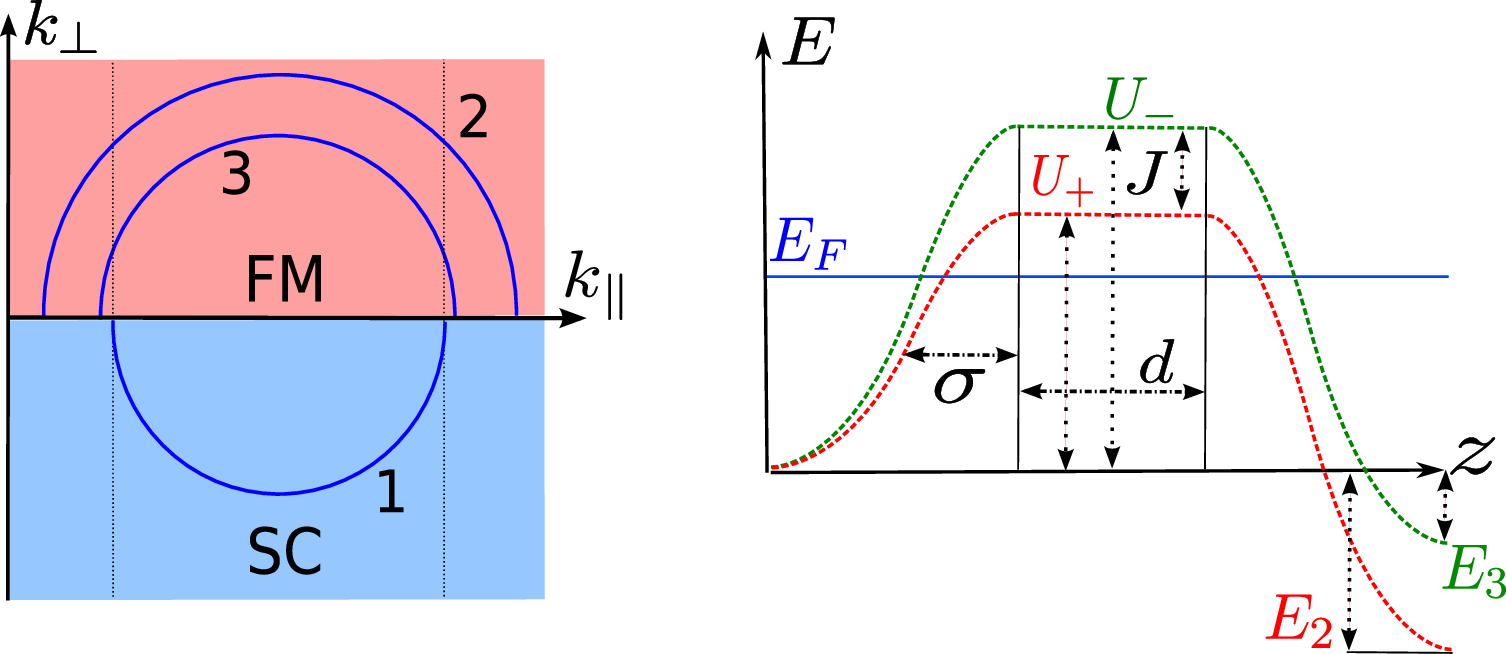}
\caption{\label{smooth}(Color online)
Sketch of the scattering potential for the smooth potential model (right) and the corresponding Fermi-surface geometry (left). The parameters introduced in Eq.~\ref{smootheq} are indicated.
}
\end{figure}
\begin{figure}
\includegraphics[width = 0.49\columnwidth]{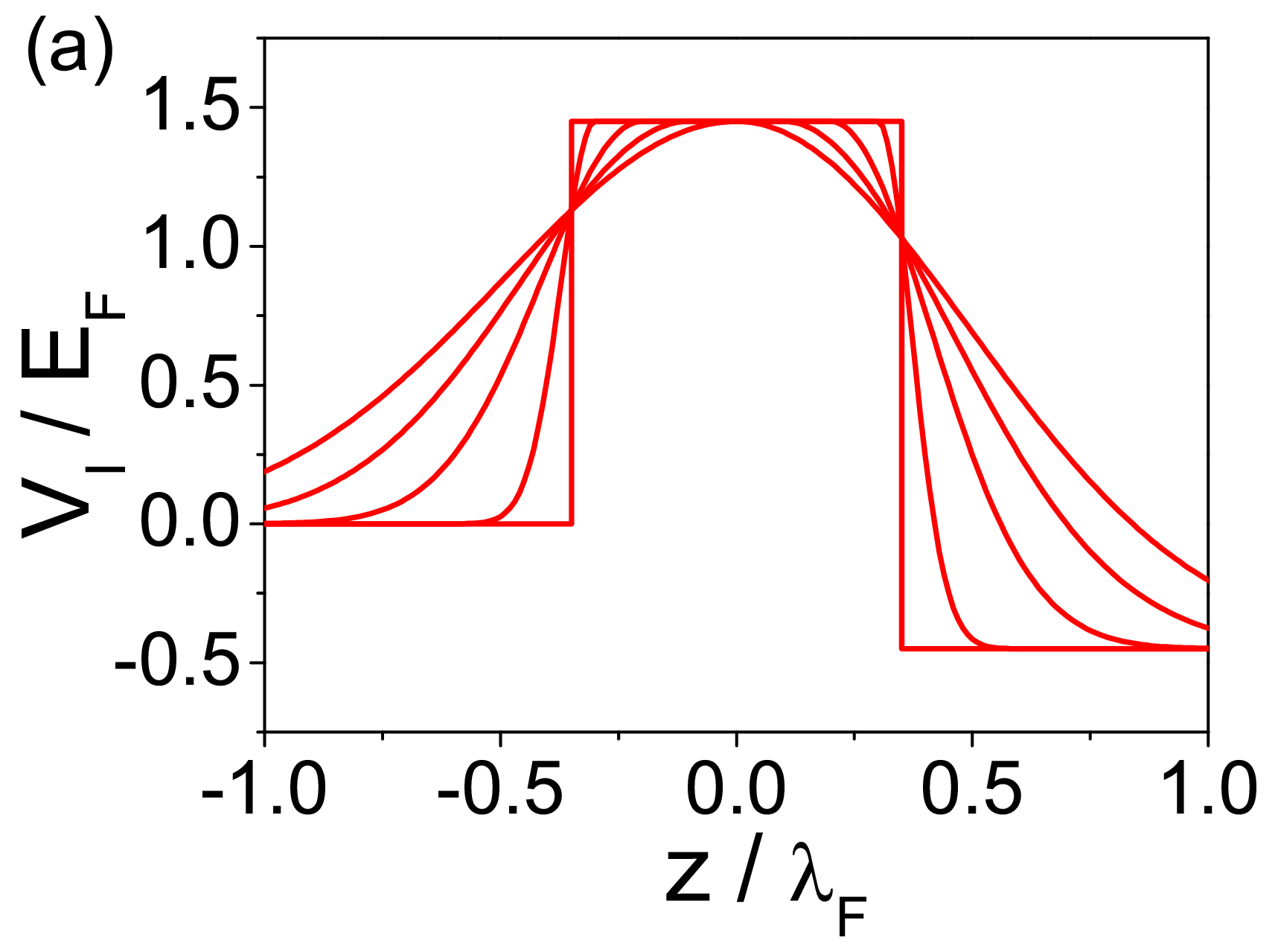}\includegraphics[width = 0.49\columnwidth]{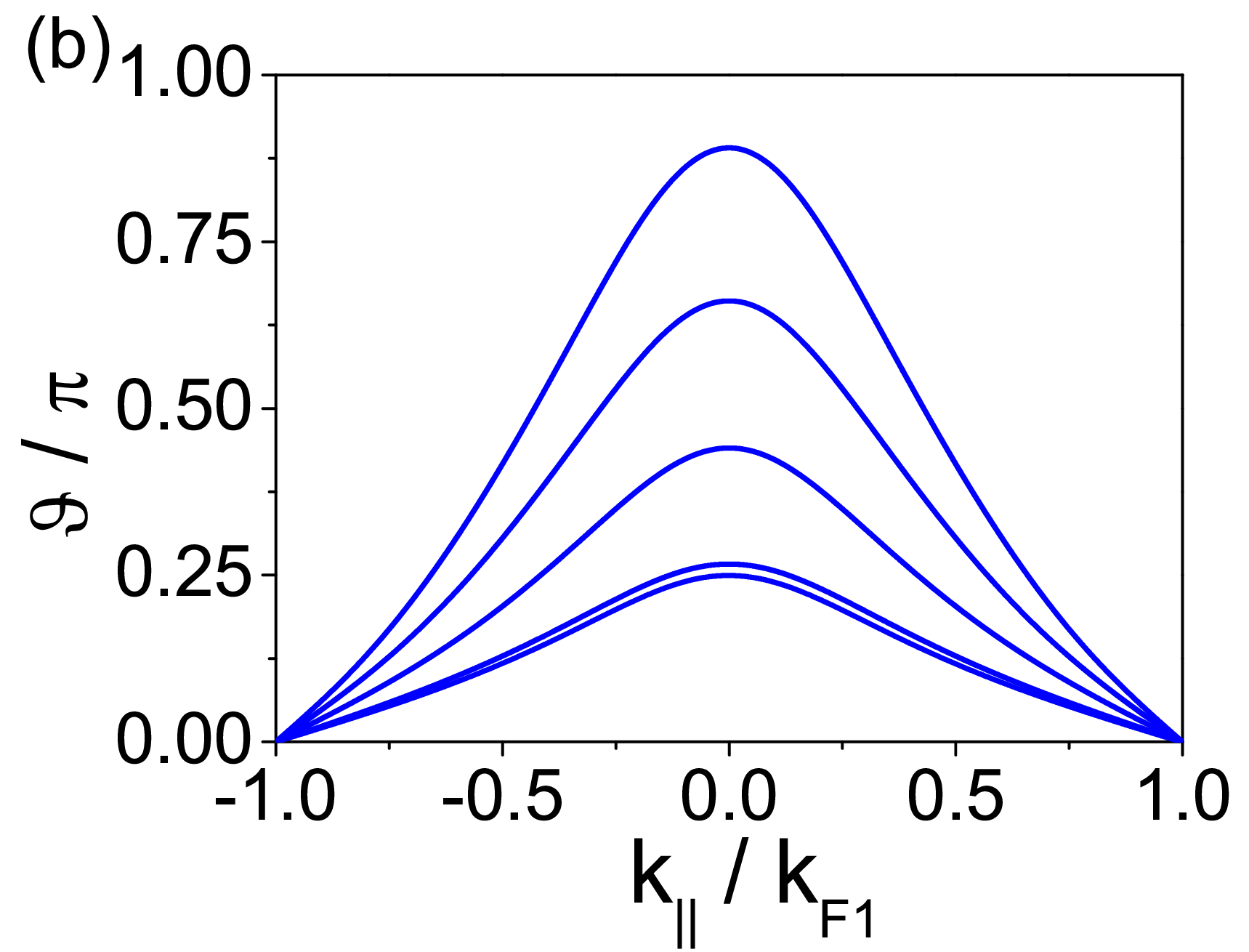}\\
\caption{\label{fig9}(Color online)
(a) Shape function of the scattering potential (average between both spin directions)
for $\sigma=0, 0.1, 0.3, 0.5, 0.7\ \lambda_{\rm{F}}$ and $\sigma+d=0.7\ \lambda_{\rm{F}}$. $E_2=-0.1\ E_{\rm{F}}$ and $E_3=-0.8\ E_{\rm{F}}$.
(b) The spin-mixing angle $\vartheta$ as a function of impact angle for the different potentials plotted in (a). $\sigma$ increases from bottom to top.
}
\end{figure}

For definiteness, we consider a potential shape as shown in Fig.~\ref{smooth},
with Gaussian ``slopes''. The `'smoothness'' of the interface barrier is then controlled by the standard deviation $\sigma$ of the Gaussian. Hence, we have the spin-dependent potential:
\begin{equation}\label{smootheq} U_{\pm}=\left\{ \begin{array}{ll}(V_I\mp J/2)\cdot e^{-(z+d)^2/\sigma^2} & z<-d \\ V_I\mp J/2 & -d<z<0 \\ E_{\eta}+(V_I\mp J/2-E_{\eta})e^{- z^2/\sigma^2} & z>0 \end{array}\right. .
\end{equation}
In the limit of a very smooth potential, one may resort to the
Wentzel-Kramers-Brillouin
(WKB) approximation\cite{WKB} to calculate the scattering problem. An interface that complies to the requirements of WKB would have to be much larger than the Fermi-wavelength however, which is unrealistic.
For this reason we resort to a numerical method for calculating the scattering problem. We use a recursive Green's function technique\cite{Georgo} to calculate the single particle Green's function of the interface Hamiltonian and obtain the scattering matrix from it using the Fisher-Lee relations.\cite{Fisher}
To study the effect of the potential shape on the spin-mixing angle, we plot
the angle $\vartheta$ in Fig.~\ref{fig9} b
for different values of $\sigma$.
To avoid a large variation of the interface transmission when varying $\sigma$, we keep $d+\sigma=0.7\ \lambda_{\rm{F}}$ (see Fig.~\ref{fig9} a).

Furthermore, we use $E_2,\ E_3<0$ here, i.e. both the FM-bands have a larger
Fermi-surface than the SC.
As we will see later on, this Fermi surface geometry and the
scattering constraints it implies can have an important effect on the shape of
the spectra, and in particular on features which are related to the spin-mixing effect.

The main result of considering a variation of the potential shape is however,
that it has a tremendous effect on the spin-mixing angle, as clearly
seen in Fig.~\ref{fig9} b. Its magnitude can exceed for a smooth potential
that for a box potential of similar transmission easily by a factor of 3-4 or more.
This is sufficient to observe some exotic features related to this effect in the
Andreev spectra of point contacts, as discussed in the next section. The physical reason
for this is that, unlike in the box potential case, electrons with opposite spin acquire a phase difference while
they are still propagating, which implies that a larger mixing phase is not inevitably tied to a strongly reduced
transmission. This can be best seen in the WKB limit, where the mixing angle is exclusively given by this dephasing:
\begin{equation}\vartheta=2 \left[\int_{-\infty}^{z_{\u}}{\rm d}z\; p_{\u}(z)-\int_{-\infty}^{z_{\d}}{\rm d}z\; p_{\d}(z)\right]. \end{equation}
Here $p_{\u,\d}=\sqrt{2m (E_F-U_{\pm})}$ and $z_{\u,\d}$ are the classical return points for the respective spin band (see Fig.~\ref{fig3} for the notation). In the intermediate
case, that we consider here both the different wavevector mismatches and the dephasing of propagating modes will add to the mixing effect.

The discussion in terms of scattering matrix parameters presented here is
flexible enough to be extended, e.g. to other Fermi surface geometries,
or adiabatic variation of the interface magnetization.
Furthermore,  instead of insulating interfaces one could consider interfaces
where one or even both channels are conducting.
The latter case has been considered by B{\'e}ri \emph{et al}.\cite{beri09}

\section{Andreev conductance spectra of SC/FM point contacts}
\label{ACS}

In the remaining part of the paper we discuss Andreev spectra that
result from our model. We use a definition
for the FM's spin-polarization given by
\begin{equation}\label{pol} P=\frac{N_{\rm F2}-N_{\rm F3}}{N_{\rm F2}+N_{\rm F3}}. \end{equation}
For parabolic bands, the density of states is proportional to the Fermi-momentum, $N_{{\rm F}\eta}\propto p_{\rm{F}\eta}\propto\sqrt{E_{\rm{F}}-E_{\eta}}$, assuming
equal effective masses.

The current density in terms of the distribution functions and coherence functions
is given by
\begin{align} \vec{j}_{\eta}&=-\frac{{\rm e} N_{{\rm F}\eta}}{2}\int {\rm d}\eps\ \langle \vec{v}_{\eta}\cdot j_{\eps,\eta}\rangle_{\eta+}, \label{currica}\\
j_{\eps,\eta}&=X_{\eta}-x_{\eta}
-\Gamma^R_{\eta} \tilde{x}_{\eta} \tilde{\Gamma}^A_{\eta},\label{currica2}
\end{align}
where the expression for $j_{\eps,\eta}$ is given by
\begin{align}
\label{currica3}
j_{\eps,2}&=x\Big\{|r_2+AT_{12}|^2+|r_{23}+AT_{13}|^2-1\Big\}-\\\notag
-&\tilde{x}\Big\{|(T_{21}+AR_1)(\gamma_1T_{13}^*)|^2+|(T_{21}+AR_1)(\gamma_1 T_{12}^*)|^2\Big\},\end{align}
and an analogous expression is obtained for $j_{\eps,3}$ by interchanging
2 with 3.
Here, $\langle \bullet \rangle_{\eta+}$ means a Fermi-surface average over
one half of the Fermi surface (positive momentum directions,
pointing into the FM, for the first and third
term of Eq.~\eqref{currica2}, negative directions for the second term).
To derive this expression, we used the universal symmetry relation \eqref{tilde}. Furthermore $x=x_2=x_3$ as defined in \eqref{dist}, $A$ is defined in \eqref{A} and the scattering matrix parameters in \eqref{S}.
Equations \eqref{currica}-\eqref{currica2} are the main result of this paper.

\begin{figure}[t]
\includegraphics[width = 0.5\columnwidth]{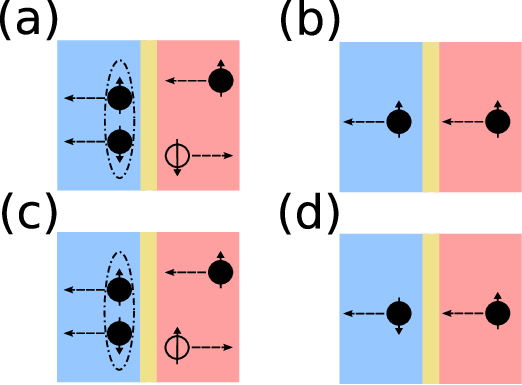}
\caption{\label{fig10}(Color online)
Transport process contributing to the current through the point-contact
(a) Normal Andreev reflection (AR)
(b) Normal transmission, requires $\eps>\Delta$
(c) Spin-flip Andreev reflection (SAR)
(d) Spin-flip transmission, requires $\eps>\Delta$
}
\end{figure}

The interpretation of Eqs.~\eqref{boundsol} and \eqref{currica} allows for identifying
two types of Andreev reflection, shown in Fig.~\ref{fig10},
one of them giving rise to a long-range proximity effect in the FM.
The terms $\Gamma^R_{2} x_{2} \tilde{\Gamma}^A_{2}$ in Eq.~\eqref{currica}
and $\Gamma^R_{2\leftarrow3}\tilde{x}_{3}\tilde{\Gamma}^A_{3\rightarrow2}$
entering $X_2$ in Eq.~\eqref{X2} both describe current contributions
from Andreev reflected holes to the current in band $2$.
The first term is proportional to the incoming distribution function in the same band.
Thus we refer to it as spin-flip Andreev reflection (SAR), as it requires a
spin-flip to transmit a singlet pair into the SC. The second term corresponds to
normal Andreev reflection, since it reflects a hole in the opposite band.
While SAR is related to the outgoing (equal-spin) triplet correlation function
in the respective band, AR is described as a term renormalizing the outgoing
distribution function.
Unlike SAR, AR does not contribute to the coherence functions in the FM spin-bands.

Using the scattering matrix parameterization introduced in Sec.~\ref{IM}, we can
obtain explicit analytical solutions for the coherence functions:
\begin{align}\label{Gfull} \Gamma^{R}_{\eta}=&\left[ \frac{T_{\eta1}\hat{N}i\hat{\sigma}_yT_{1\eta}^*}{D}\right]^{\rm R}\quad
\Gamma^{R}_{2\leftarrow3}=\left[ \frac{T_{21}\hat{N}i\hat{\sigma}_y T_{13}^*}{D}\right]^{\rm R}\\
\hat{N}=&\gamma(1+\gamma\tilde{\gamma}\left[e^{-i\vartheta\sigma_z}(\rho+\rho_{s})-\rho_s+2\sigma_y\hat{\rho}\sin (\vartheta/2)\right])\\
D=&1+2\gamma\tilde{\gamma}[\rho \cos \vartheta -\rho_{s}(1-\cos \vartheta )]+(\gamma\tilde{\gamma})^2\rho^2\\
\rho=&r_{\u}r_{\d}\qquad \rho_{s}=(r_{\u}-r_{\d})^2\frac{\sin^2(\alpha_Y)}{4} \\\notag
\hat{\rho}=&\sqrt{\rho_{s}}\cdot\mathrm{diag}[r_{\u}\cos^2(\alpha_Y/2)+r_{\d}\sin^2(\alpha_Y/2),\\\notag &r_{\d}\cos^2(\alpha_Y/2)+r_{\u}\sin^2(\alpha_Y/2)]
\end{align}
Here $\eta\in\{2,3\}$ and we omitted the index $1$ for the incoming coherence functions.
$\gamma^{R}$ is related to $\gamma^{R}_1$ in \eqref{gbulk} by $\gamma^{R}i\sigma_y=\gamma^{R}_1$.
The advanced component $\tilde\Gamma_\eta^{A}$ is obtained via
$\tilde\Gamma_\eta^{A}=(\Gamma_\eta^{R})^\ast $.

Note that the $\Gamma$-functions differ only by the transmission vectors $T$ but since the numerator is a matrix product, this still gives expressions that differ markedly.
In any case, we have
$\Gamma^{R}_{\eta}=0$ if $\alpha_Y=0$ or $\vartheta$ and $\vartheta_{\eta}=0$.
We focus on the  denominator which arises from the matrix inversion
in Eq.~\eqref{Gamma21} and is the same for all coherence functions. It is of
particular interest since it leads to the emergence of conductance peaks in the
Andreev spectrum.

\subsection{Andreev bound state spectrum}

The appearance of the Andreev conductance peaks
can be seen most clearly in the tunneling limit. Here $\rho_{s}=0$ and $\rho=1$ which simplifies the expressions above considerably. The full solutions read

\begin{align}\label{anaG} \Gamma^{R}_{\eta}&=\left[2i\ t_{\eta}t'_{\eta}\gamma\frac{\sin\vartheta_{\eta}-\sin(\vartheta-\vartheta_{\eta})\gamma\tilde{\gamma}}{1+(\gamma\tilde{\gamma})^2+2\cos(\vartheta)\gamma\tilde{\gamma}}\right]^{R} \\
\label{anaG23}
\Gamma^{R}_{2\leftarrow3}&=\left[\gamma\frac{\tau+\tau_{\vartheta}\gamma\tilde{\gamma}}{1+(\gamma\tilde{\gamma})^2+2\cos(\vartheta)\gamma\tilde{\gamma}}\right]^{R},\end{align}
with
\begin{align*} \tau^R &=t_2t_3e^{i\vartheta_{23}}-t_{23}t_{32}e^{-i\vartheta_{23}},\\
\tau^R_{\vartheta} &=t_2t_3e^{i(\vartheta_{23}-\vartheta)}-t_{23}t_{32}e^{-i(\vartheta_{23}-\vartheta)},\\
\vartheta_{23}&=(\vartheta_2+\vartheta_3)/2.
\end{align*}

\begin{figure}[b]
\includegraphics[width=0.8\columnwidth]{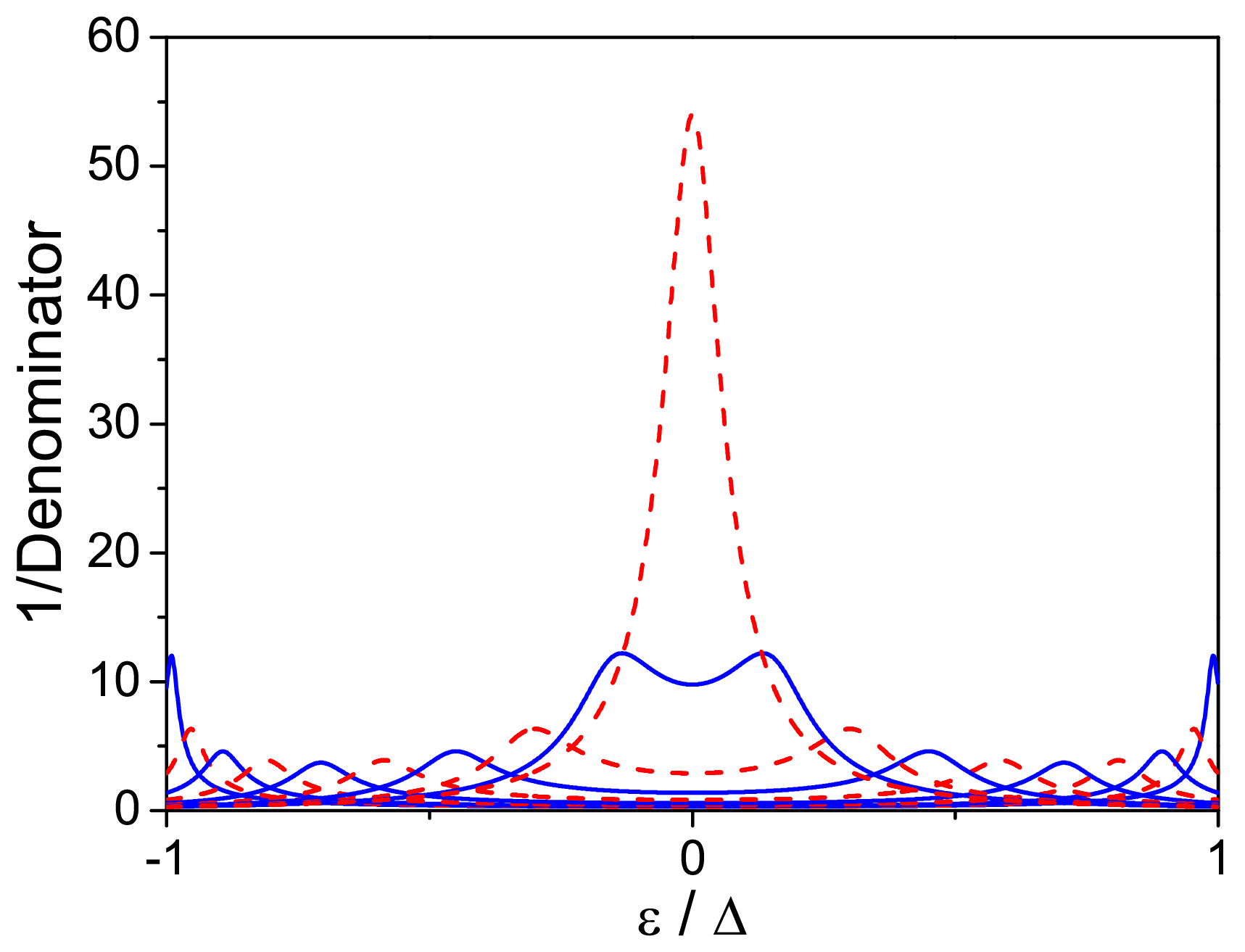}
\caption{\label{fig11} The denominator of equation \eqref{Gfull} as a function of the quasiparticle energy $\eps$. $r_{\u}=0.9$, $r_{\d}=0.95$, $\alpha_Y=0.5\ \pi$. Plots for $\vartheta=0.1 \ \pi...1.0\ \pi$ in steps of $0.1\ \pi$. The maximum moves to lower energies with increasing $\vartheta$.}
\end{figure}
For $\eps<\Delta$ we have $\tilde{\gamma}^{R}=-\gamma^{R}$ and $|\gamma^{R}|=1$ and we can easily show that \eqref{anaG} and \eqref{anaG23} both have a pole at\cite{fogelstrom00}
\begin{equation}\label{pole} \eps_{\rm{pole}}=\pm\Delta\cos(\vartheta/2). \end{equation}
This pole corresponds to an Andreev bound state induced by the spin-mixing effect at
the superconducting side of the sample. Following Fogelstr\"om,\cite{fogelstrom00}
one can show that these bound states appear in the DOS of the superconductor close to
the interface and are actually associated to different spins. The bound state
for $\eps>0$ appears in the DOS of $\u$-quasiparticles and that for $\eps<0$ in
that of $\d$-quasiparticles if $\pi\geq\vartheta\geq 0$. This is why the appearance of the sub-gap peak is only tied to
the spin-mixing angle $\vartheta$. It does not depend on spin-flip scattering
or the mixing phases associated to transmission. However, we shall see that a high
mixing angle of $\geq 0.5\ \pi$ is required to make this bound state appear in a finite
temperature spectrum. If we consider the full expression of the denominator, we find
that the pole is lifted, yet 2 local maxima remain (see Fig.~\ref{fig11}) which
decrease in magnitude with increasing transparency of the interface,
as the bound state acquires a finite lifetime.
Obviously, a tunneling barrier in addition to an appreciable mixing effect is required to
observe a sub-gap-peak in the Andreev spectrum.

In the half-metallic case, the full solution for the outgoing coherence function reads:
\begin{equation}\label{GammaHM}\Gamma^{R,A}_{2}=\frac{-i\mathcal{T}_{\alpha} t^2(1+r)\gamma^{R,A}(1-\gamma\tilde{\gamma}^{R,A})}{1+2\gamma\tilde{\gamma}^{R,A}[r\cos\vartheta -\mathcal{T}_{\alpha}^2t^4]+(\gamma\tilde{\gamma}^{R,A})^2r^2}\end{equation}
with $\mathcal{T}_{\alpha}=\frac{\sin(\vartheta/2)\sin\alpha_Y}{1+r}$. This solution is already discussed in
Ref.~\onlinecite{eschrig09}, but we state it here again to comment on a recent result
obtained by B\'eri \textit{et al}.\cite{beri09}
Using a different approach to calculate the conductance of a SC/HM
point-contact, they find that generically $G(\rm{eV}=0)=0$, at zero temperature.
This agrees perfectly with our results. One can show that below the gap, \begin{equation} G(\mathrm{eV})\propto|\Gamma^R_2(\eps)|^2\partial_Vx(\eps, \mathrm{eV}).\end{equation} For $T=0$, $\partial_Vx=2{\rm e}\delta(\eps+\mathrm{eV})$ holds and  since $(1-\gamma\tilde{\gamma}^{R,A})=0$
for $\eps=0$ we also find $G(\rm{eV}=0)=0$. Note that \eqref{GammaHM} holds for
arbitrary scattering matrices. Thus, this property is
universal with the exception of $\vartheta=\pi$, $\alpha_Y=0.5 \ \pi$ where the
denominator is zero for $\eps=0$.

\subsection{Andreev conductance spectra}

As we have pointed out above, two competing Andreev processes participate in the presence
of spin-flip scattering, shown in Fig.~\ref{fig10}.
Normal AR is suppressed as the polarization of the FM increases, since it requires
one quasiparticle from each spin-band. SAR on the other hand takes two quasiparticles
from the same band and thus dominates the spectrum for high polarization.
We can define the corresponding contributions to the differential conductance
for each spin-band by
\begin{align}\label{form13}\nonumber \frac{G_{\rm{AR},2}}{2}&=\partial_V\frac{{\rm e} N_{\rm F2}}{2}\int_{\eps}{\rm d}\eps\ \langle\Gamma^R_{2\leftarrow3}\tilde{x}_{3}\tilde{\Gamma}^A_{3\rightarrow2}\rangle_+\\
\frac{G_{\rm{SAR},2}}{2}&=\partial_V\frac{{\rm e} N_{\rm F2}}{2}\int_{\eps}{\rm d}\eps\ \langle \Gamma^R_{2} \tilde{x}_{2} \tilde{\Gamma}^A_{2}\rangle_+
\end{align}
and correspondingly ($2\leftrightarrow 3$) for band $3$. The factor $1/2$ appears because the expressions in the integrand describe only one of the two charges which are transferred by the process. The other charge is contained in $X_{\eta}-x_{\eta}$ and cannot be disentangled from the one-quasiparticle transmission processes. The total contribution to the conductance is given by the sum over both bands: $G_{\rm{AR}}=G_{\rm{AR},2}+G_{\rm{AR},3}$, $G_{\rm{SAR}}=G_{\rm{SAR},2}+G_{\rm{SAR},3}$.

\begin{figure}
\includegraphics[width=0.49\columnwidth]{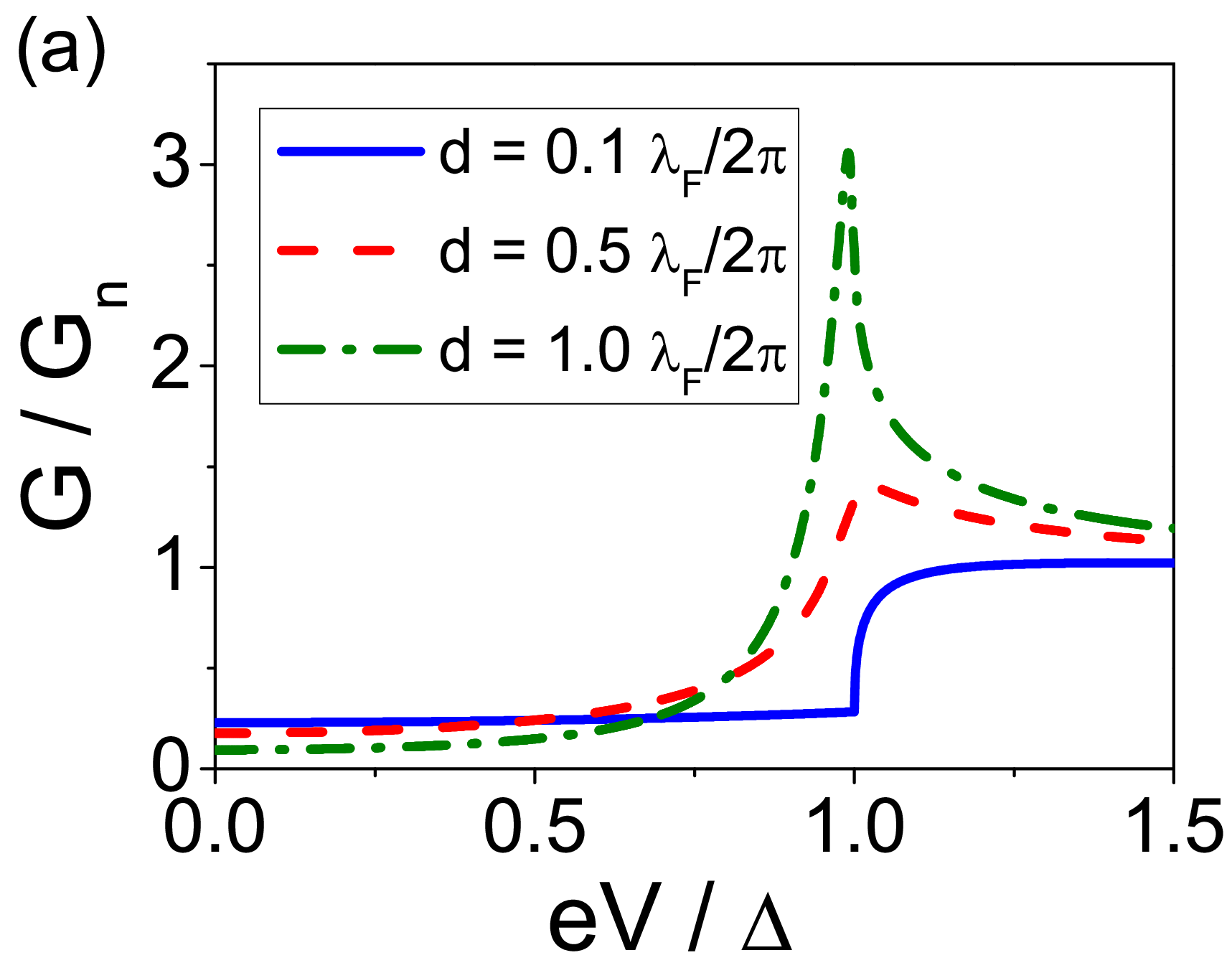}\includegraphics[width=0.49\columnwidth]{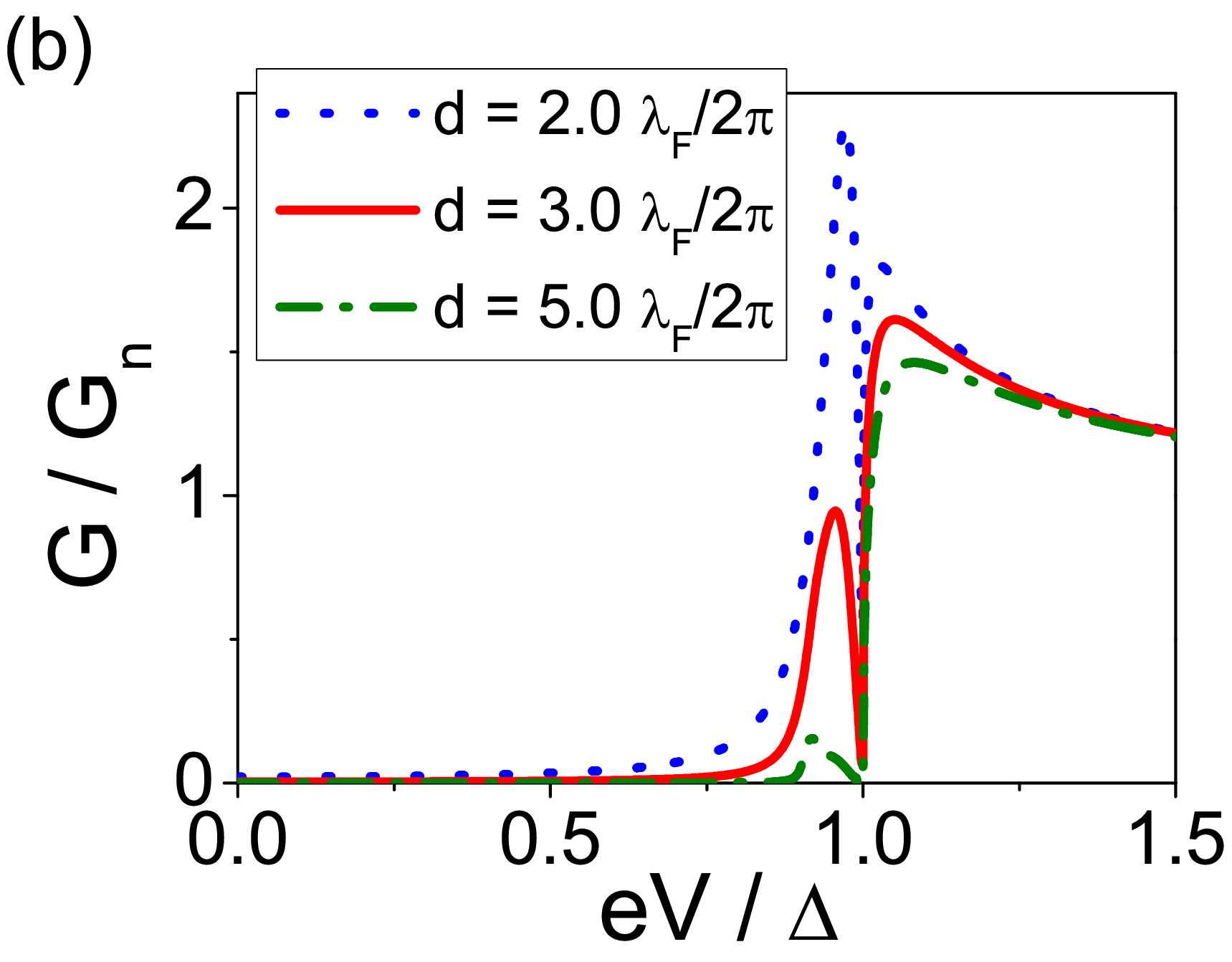}\\
\includegraphics[width=0.49\columnwidth]{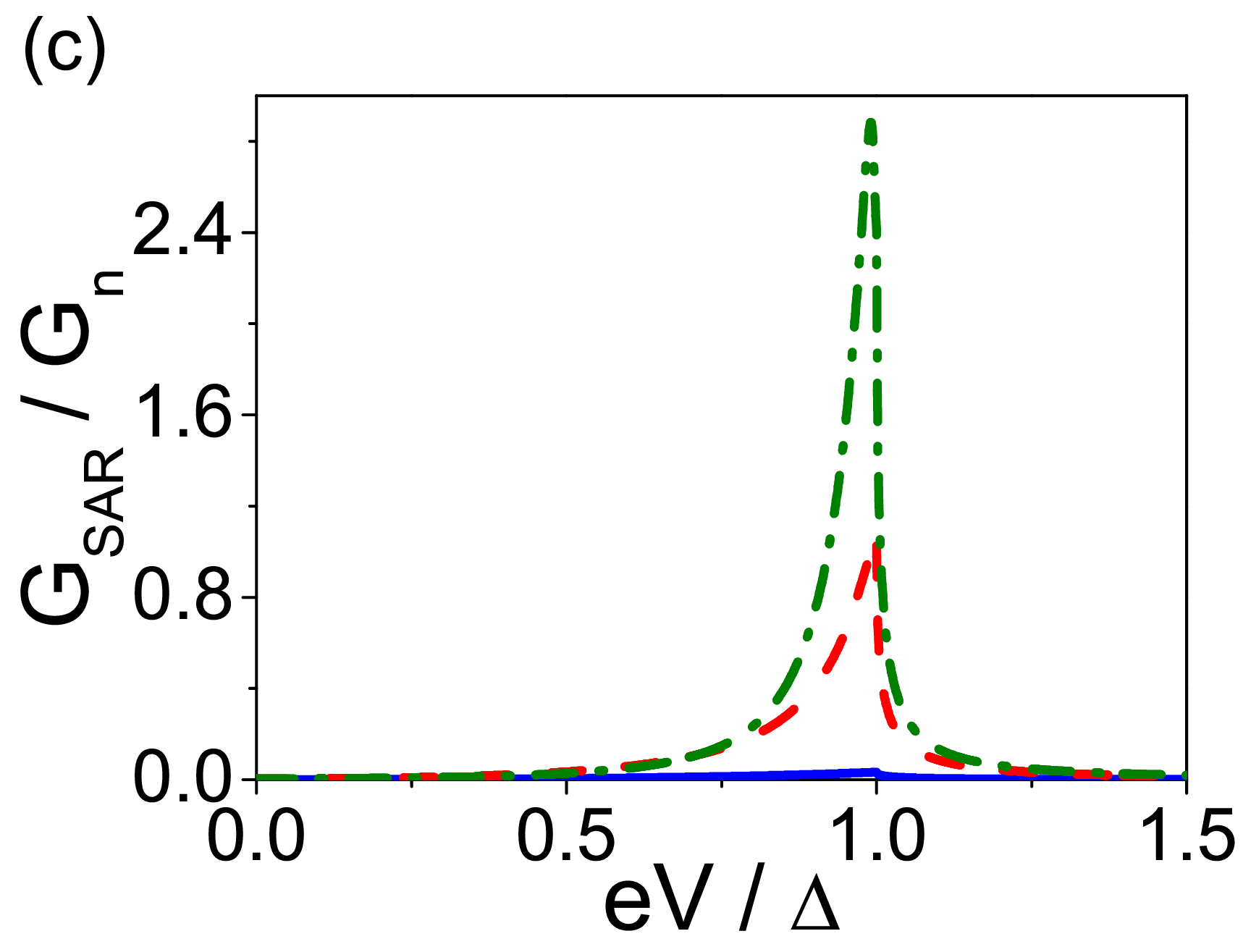}\includegraphics[width=0.49\columnwidth]{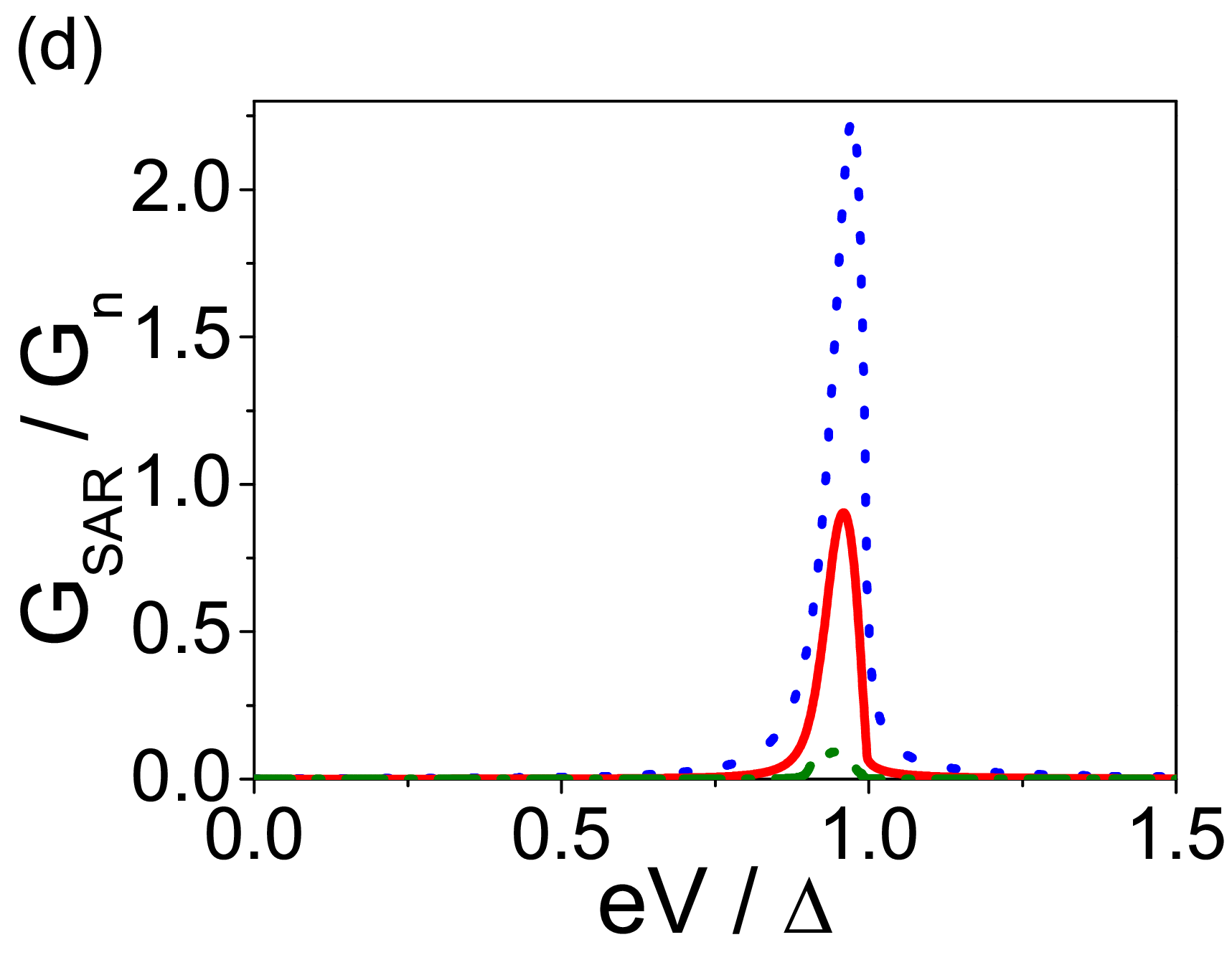}\\
\includegraphics[width=0.49\columnwidth]{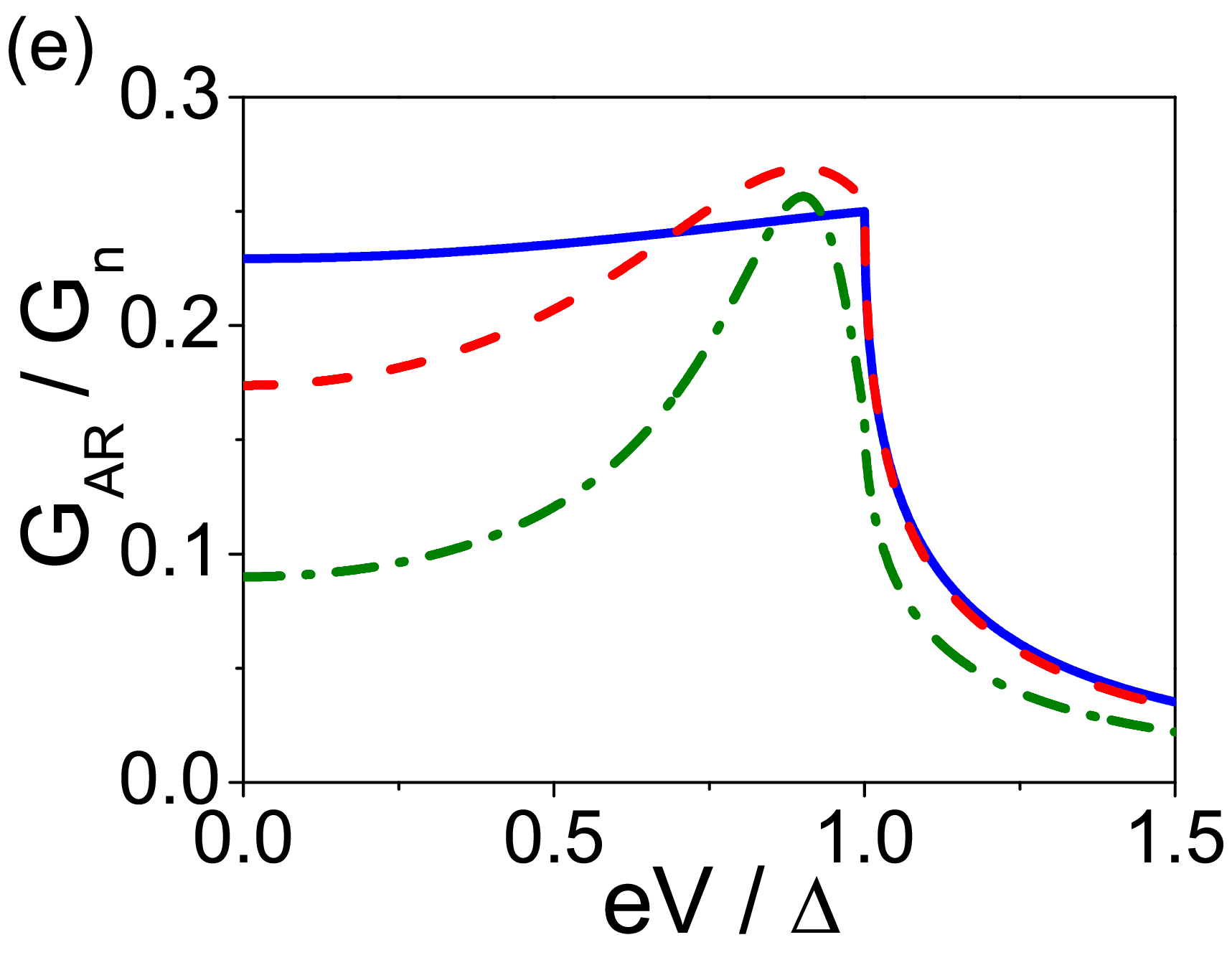}\includegraphics[width=0.49\columnwidth]{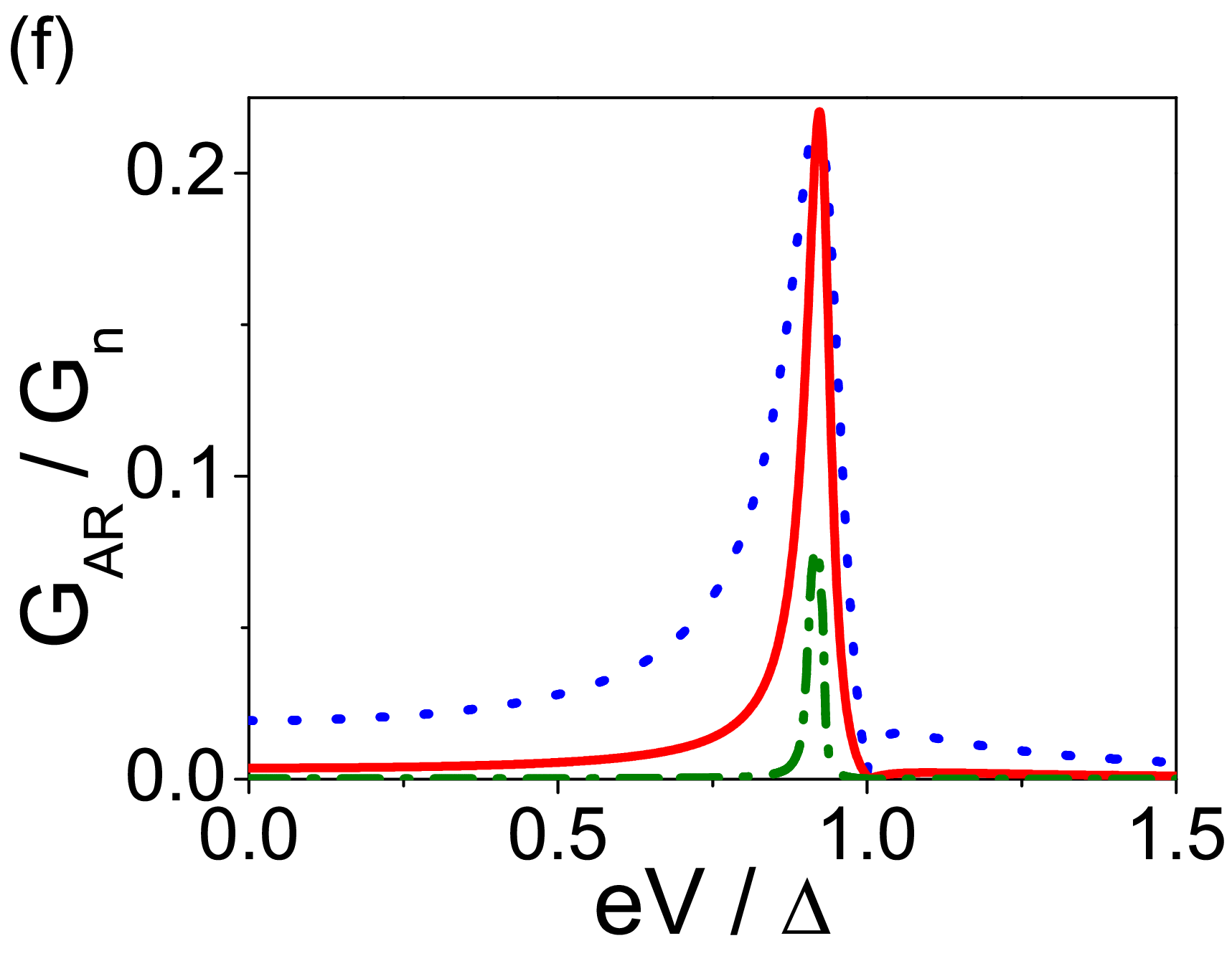}
\caption{\label{fig12}(color online)
The conductance $G$ of a point contact as function of contact voltage $V$ (first row),
and the corresponding SAR (second row) and AR (third row) contributions to it,
for $d=0.1,\ 0.5,\ 1.0\ \lambda_{\rm{F}}/2\pi$ (left column) and
$d=2.0, \ 3.0,\ 4.0,\ 5.0\ \lambda_{\rm{F}}/2\pi$ (right column).
The remaining parameters in all plots are $E_2=0.1 E_{\rm{F}}$, $E_3=0.9\ E_{\rm{F}}$, $U_+=1.1\ E_{\rm{F}}$, $U_-=1.9\ E_{\rm{F}}$, and $\alpha=0.5\ \pi$.}
\end{figure}
In Fig.~\ref{fig12} we discuss the results for the box potential using exactly the same parameters as in Fig.~\ref{fig4}.
This corresponds to a spin-polarization of the FM of $P=0.5$.
In Fig.~\ref{fig12}a,b we plot the total differential conductance. For thin interfaces we obtain spectra with a rather conventional shape. The solid line in Fig.~\ref{fig12}a corresponds to a highly transparent interface, but still the conductance does not rise to a value close to twice the normal state conductance as in the conventional BTK picture. This is a direct result of the FM spin-polarization. Looking at the same line in Fig.\ref{fig12}c,d, we see that the shape of $G_{\rm{AR}}$ actually follows the usual trend, albeit with reduced magnitude, while $G_{\rm{SAR}}$ gives almost no contribution in this case. The reduction of the Andreev conductance compared to the normal state is in this case due to the spin-polarization of the FM. Only a fraction of the quasiparticles impinging the interface can undergo AR due to the reduced density of states in the minority band.

As the thickness of the interface increases the conductance contribution of SAR is enhanced and even dominates the sub-gap conductance for tunneling interfaces (Fig.~\ref{fig12}d,f). This is because the magnitude of SAR is insensitive to the spin-polarization as it takes two quasiparticles with the same spin from the FM. On the other hand it is very sensitive to spin-active scattering, which is why it is reduced for thin interfaces. We also see that as the transparency of the interface decreases, a sub-gap peak develops, as discussed in the previous section (Fig.~\ref{fig12}b). However, the Andreev bound state stays close to the gap-edge in this scenario and thus smears out even for very small temperatures.

\begin{figure}
\includegraphics[width=0.49\columnwidth]{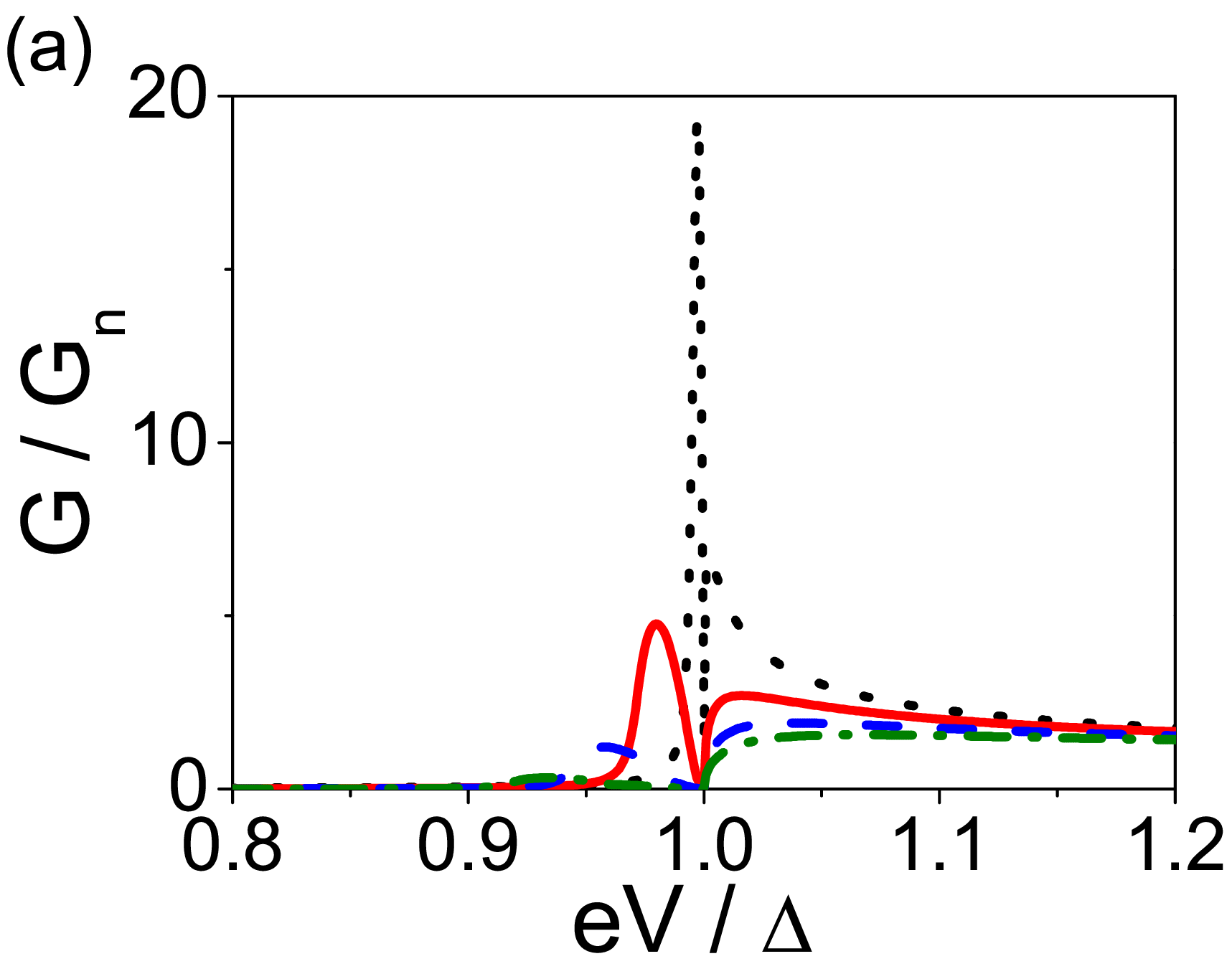}\includegraphics[width=0.49\columnwidth]{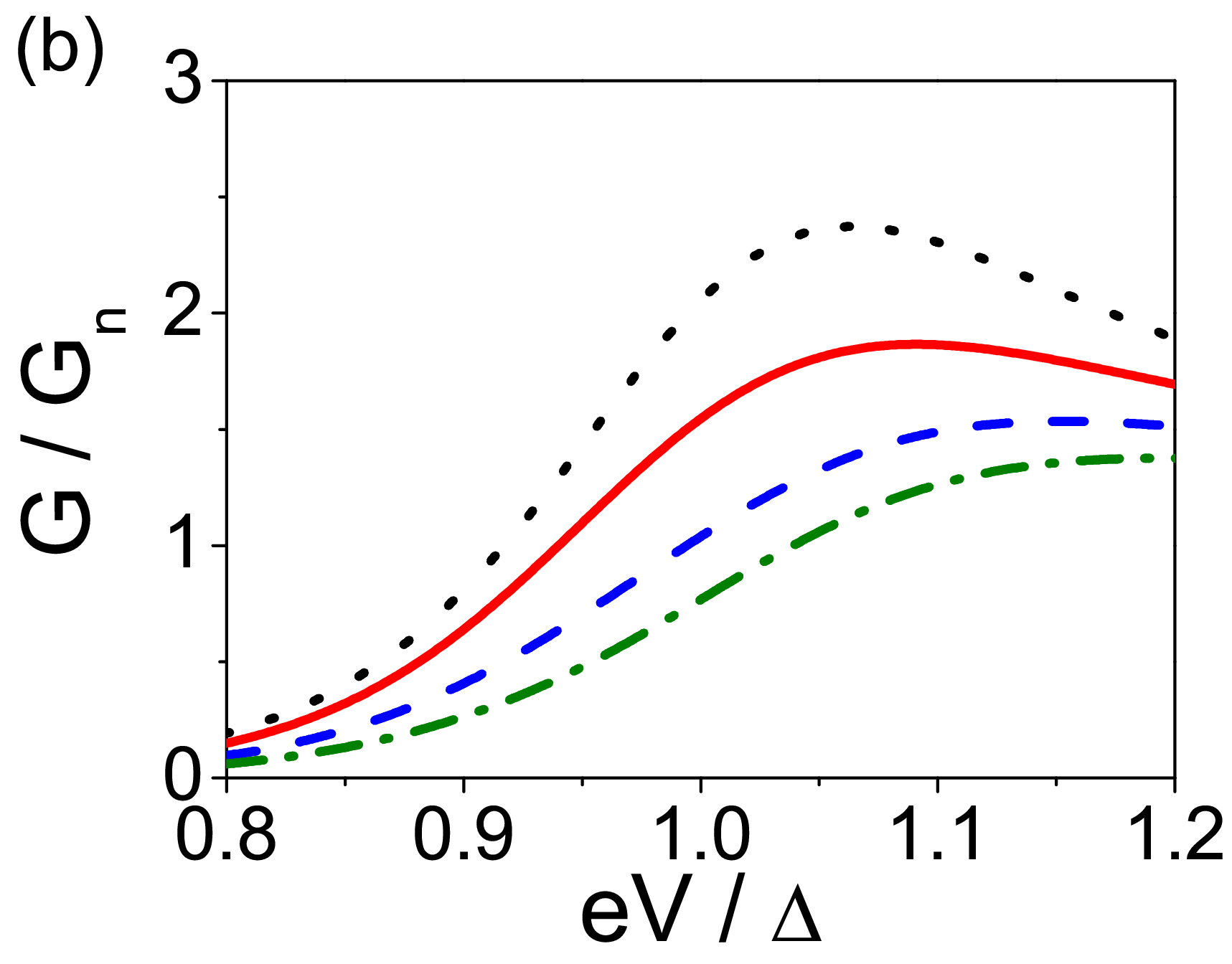}
\caption{\label{fig13}
The conductance $G$ of a point contact as function of contact voltage $V$, for
(a) $T=0$ and (b) $T=0.1\ T_c$.
In both cases, the values of
$E_3=0.2,0.4,0.6,0.9$, $U_-=E_-+E_{\rm{F}}$ increase from top to bottom.
The remaining parameters in
all plots are $E_2=0.1\ E_{\rm{F}}$, $U_+=1.1\ E_{\rm{F}}$, $\alpha=0.5\ \pi$, $d=5.0\ \lambda_{\rm{F}}/2\pi$.
}
\end{figure}
In Fig.~\ref{fig13}a,b we plot the spectrum around the gap energy for different polarizations, i.e. exchange fields, of the FM and a tunneling interface $d=5.0\ \lambda_{\rm{F}}/2\pi$. Apparently, the sub-gap-peak moves to lower energies as the exchange field increases but also decreases in magnitude. In any case, the peak is too small and too close to the gap-edge to be observable at finite temperatures (Fig.\ref{fig13}b). This situation cannot be circumvented in the frame of the box-potential model, the reason being that one cannot obtain high mixing angles for reasonable parameter ranges. Moreover, this situation is aggravated by the Fermi-surface average. As the mixing angle varies with the trajectory impact angle, the peak is broadened even at $T=0$. This points again to the crucial importance of the Fermi-surface geometry. If the Fermi-vector in the SC is considerably smaller than those of the FM bands, the scattering states which contribute to the current will be confined to a small range around perpendicular impact and hence a sharper peak structure can be expected.

\begin{figure}
\includegraphics[width=0.49\columnwidth]{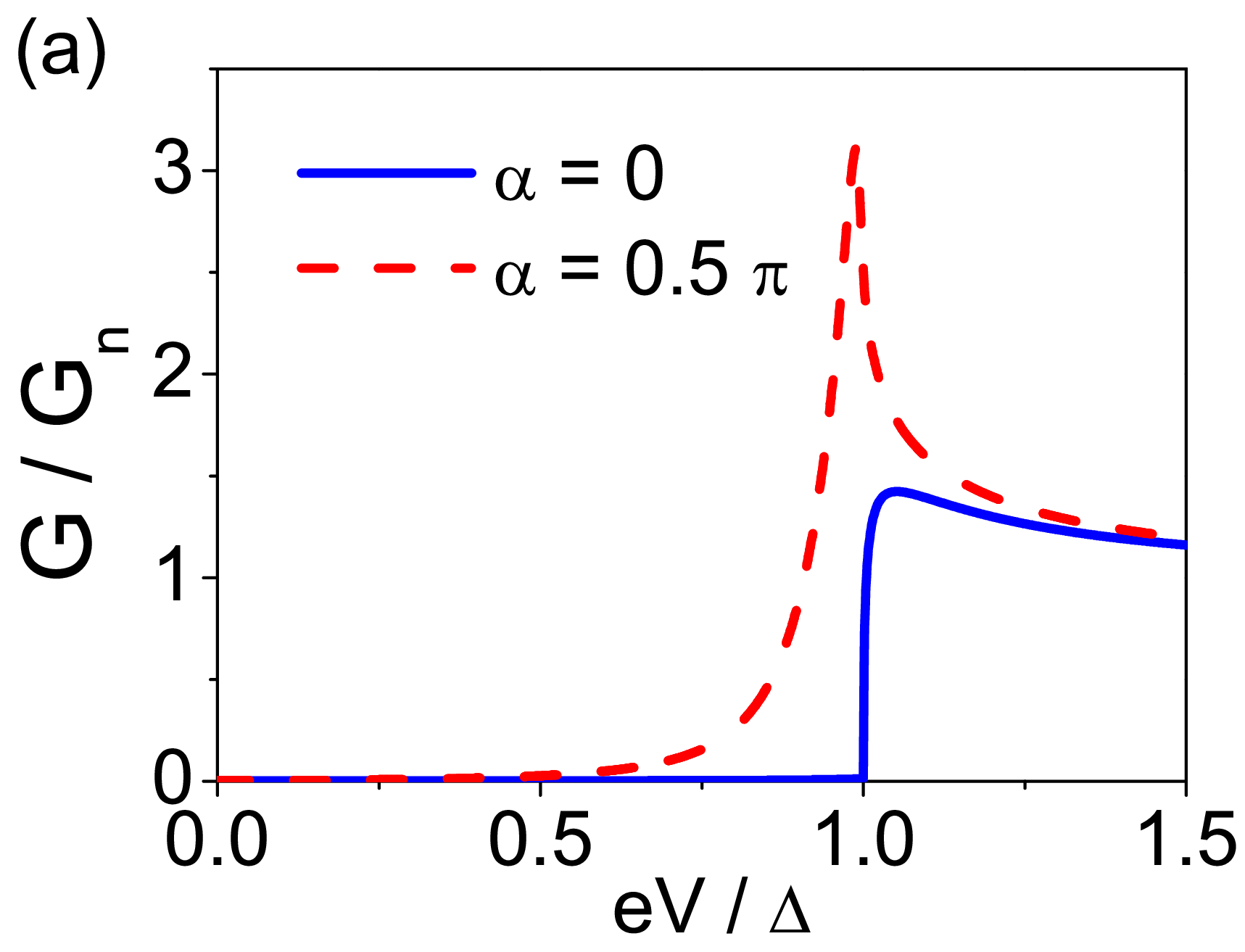}\includegraphics[width=0.49\columnwidth]{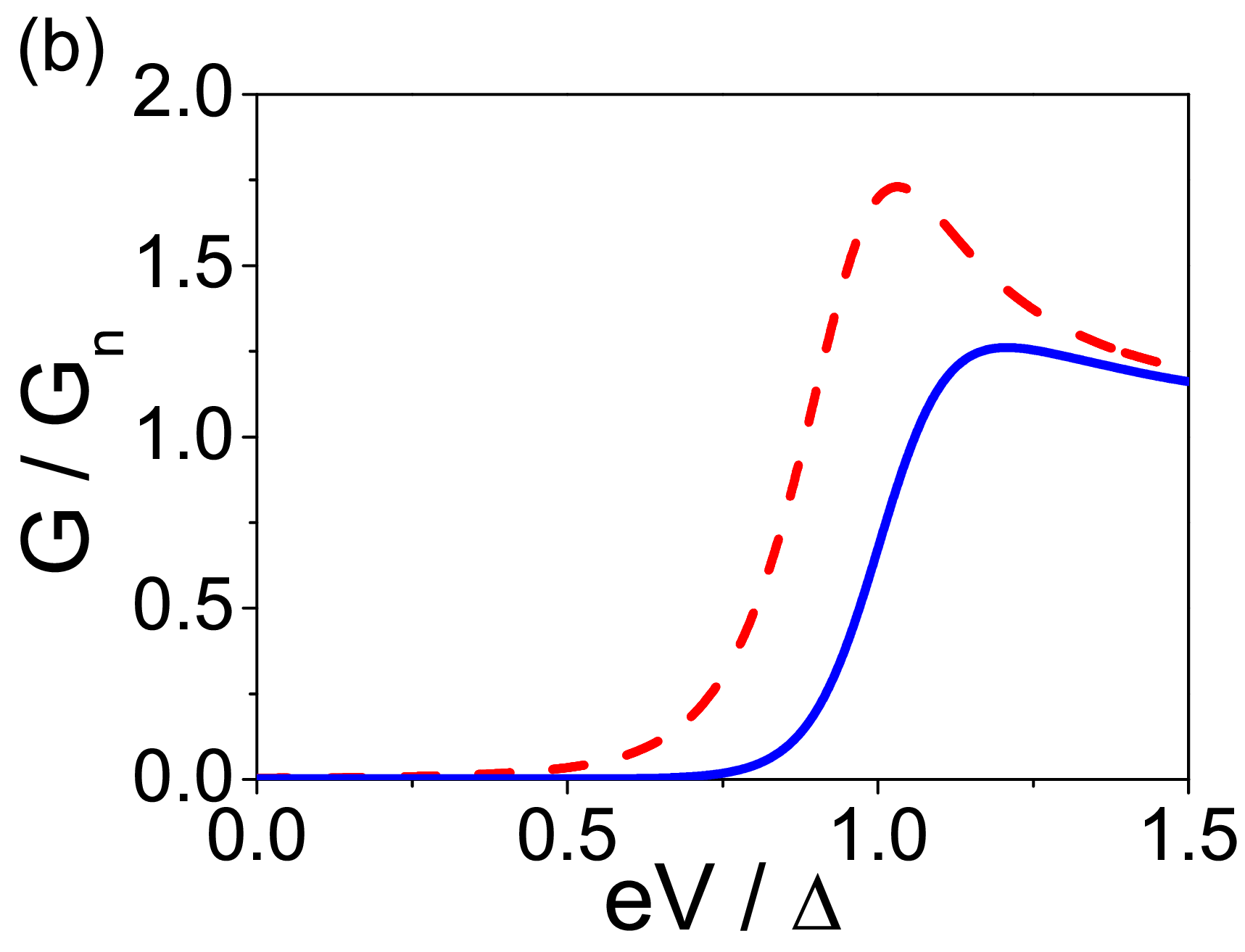}
\caption{\label{fig14}
The conductance $G$ of a point contact as function of contact voltage $V$, for
(a) $T=0$, and (b) $T=0.1\ T_c$, shown for two values of $\alpha $.
The remaining parameters
in all plots are $E_2=0.1 E_{\rm{F}}$, $E_3=0.99\ E_{\rm{F}}$, $U_+=1.1\ E_{\rm{F}}$,$U_-=1.99\ E_{\rm{F}}$ , $d=1.0\ \lambda_{\rm{F}}/2\pi$.}
\end{figure}

Finally, we show that even if this exotic feature in the conductance spectrum is not observable at finite temperatures, the impact of spin-active scattering can still be important. This holds in particular for FMs with high polarization, where SAR will naturally dominate the spectrum, if it is present. This can be seen in Fig.~\ref{fig14}, where we plot the conductance for a highly polarized ($P=0.8$) FM for $\alpha=0.5\ \pi$ and $\alpha=0$ respectively. In the latter case, SAR cannot occur. If $\alpha=0.5\ \pi$, the spectrum is largely enhanced around the gap energy. This is not surprising, since SAR is mainly contributing in this energy range. Even at finite temperatures an appreciable difference between the curves remains.
\begin{figure}[b]
\includegraphics[width=0.8\columnwidth]{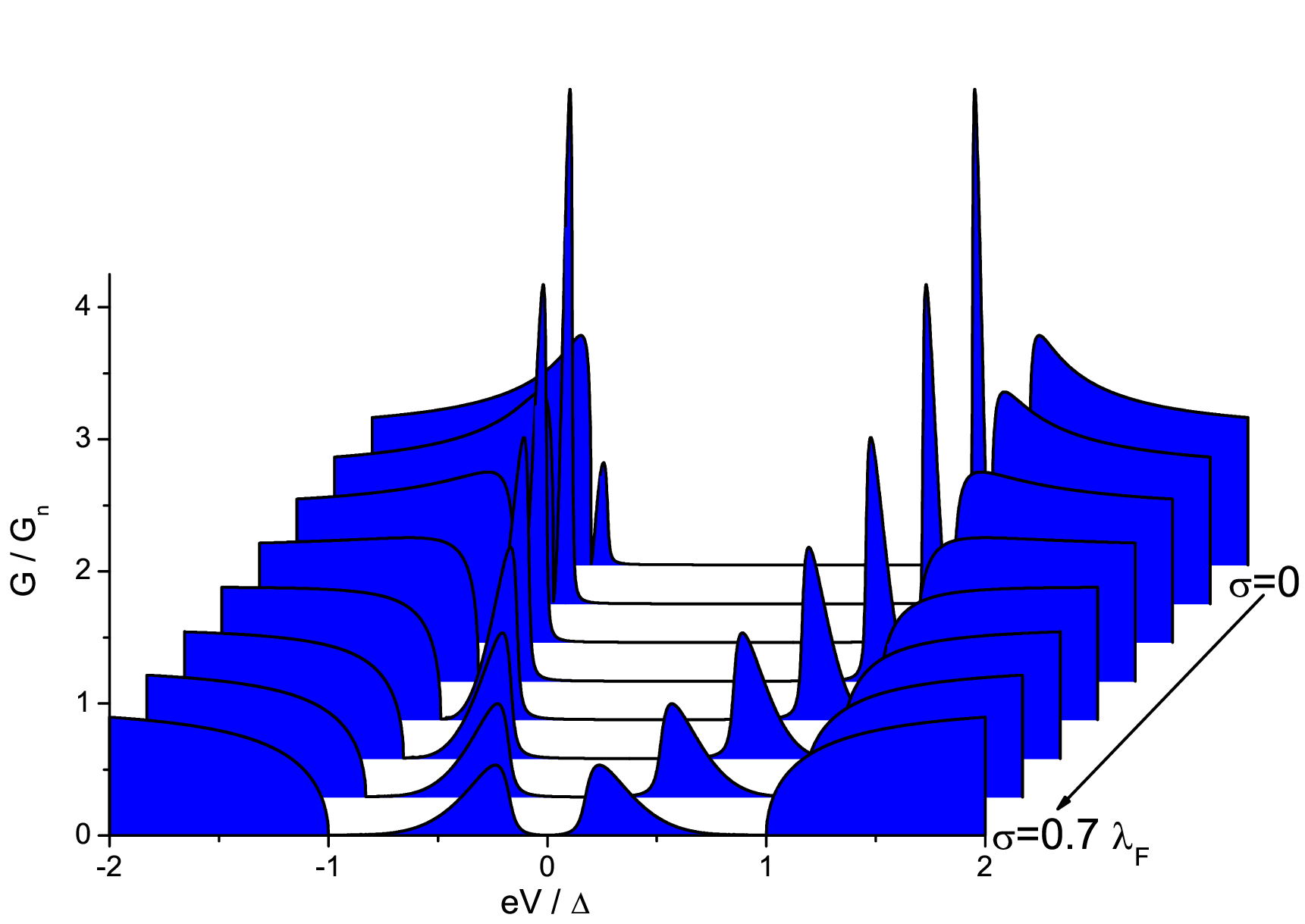}\\
\includegraphics[width=0.8\columnwidth]{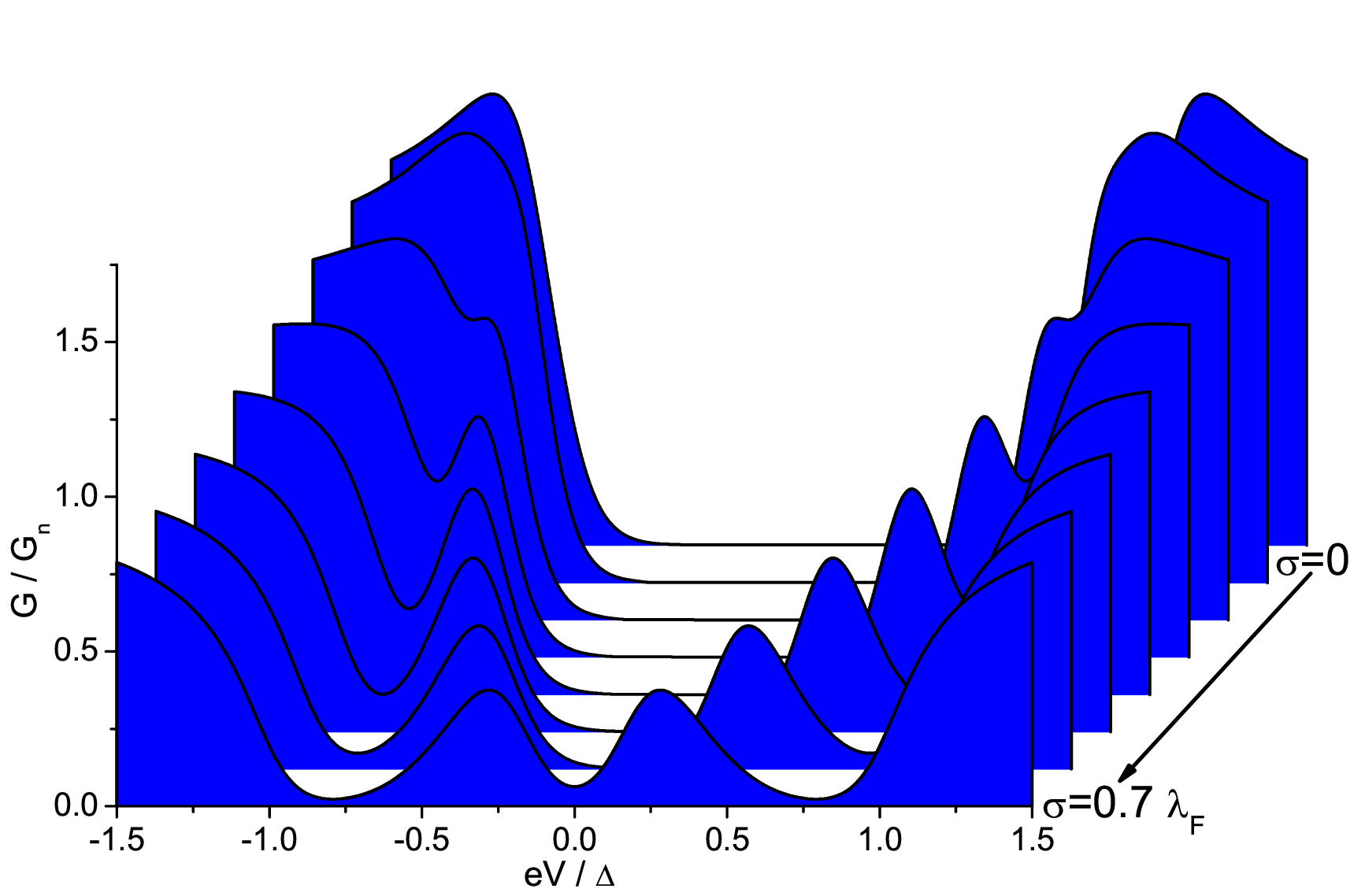}
\caption{\label{fig15} The differential conductance for the smooth potential model. The parameters are the same as in Fig.~\ref{fig6}. $\sigma$ increases from back to front by steps of $0.1\ \lambda_{\rm{F}}$.
top $T=0$,
bottom $T=0.1\ T_c$.}
\end{figure}

Turning to the smooth scattering potential, we see that the situation changes fundamentally. We calculate the spectrum for the same set of parameters as in Fig.~\ref{fig9}. These results are shown in Fig.~\ref{fig15}.
As we find a considerably enhanced mixing angle in this case, it is not surprising that the sub-gap peak is located far from the gap-edge if the potential is sufficiently smooth and may even be observed at finite temperatures.
The width of this peak is directly related to the Fermi-surface average.
The calculations in Fig.~\ref{fig15} are for
a tunneling limit situation ($t^2<0.01$), and formula \eqref{pole} holds approximately.
As one can see from Fig.~\ref{fig9}b, $\vartheta$ sweeps through the whole range from $0$
to its maximum value as a function of the trajectory impact angle.
This results in broadening and also implies that the Fermi-surface geometry may
have an important impact on the shape of this bound state peak. For the particular
geometry we consider here, with the Fermi surfaces of the FM-bands being both
smaller than that of the SC, the mixing angle reaches $0$ for grazing impact.
If however, the SC-band is smaller than at least one of the FM bands, this is no
longer true as it can be seen in Fig.~\ref{fig4}. Doing WKB calculations for
different geometries, we found that this may result in a kink at the tail of the
peak, if $\vartheta_{\rm{min}}$ is large enough.

\subsection{Connection to the extended BTK-model}
\label{BTK}

The extension of the BTK-model to ferromagnetic point contacts proposed in Ref.~\onlinecite{Beenakker} and further elaborated on in Ref.~\onlinecite{Mazin}, was first used in Refs.~\onlinecite{soulen98, Upad} to extract the FM spin-polarization from the spectra of such contacts. Here, we show how this model can be obtained from our theory. The extended BTK-model characterizes interfacial scattering by a single parameter $Z$, which controls the transparency of the interface. Z is assumed to be independent of the transport channel. The spin-polarization of the FM is then taken into account by noting that if $P$ is finite, the transport channels can be divided into ``non-magnetic'' and ``half-metallic'' channels\cite{Mazin, soulen98}, which is illustrated in Fig.~\ref{fig1} in this paper. This amounts to writing the conductance of the contact as a sum of the non-magnetic and half-metallic  contribution according to\cite{soulen98}:
\begin{equation}\label{GBTK} G=(1-P_C)G_N+P_C G_H. \end{equation}
Here, the \emph{transport} spin-polarization $P_C$ was introduced:
\begin{equation} P_C=\frac{\Nfz\vfz-\Nfd\vfd}{\Nfz\vfz+\Nfd\vfd}. \end{equation}
$G_H$ is zero below the gap, since spin-flip scattering cannot occur in this model. This means that for $|\mathrm{eV}|<\Delta$ one simply has the standard BTK formula reduced by a factor $(1-P_C)$. The connection to our model is now established by making corresponding assumptions for the normal state scattering matrix of the interface. Since there is no spin-flip scattering, the matrix is necessarily diagonal, spin-mixing effects are obviously also disregarded. Moreover, the fact that wavevector mismatches, let alone a spin-dependent interface potential, will introduce a spin-filtering effect is also not taken into account. Consequently, the whole scattering matrix is described by a single transmission parameter $T_N=t_2^2=t_3^2$. Evaluating the corresponding expressions for $\Gamma^{R,A}_{\eta}$ is straight forward and yields:
\begin{align}\label{speccurr} &j_{\eps}(\eps,\ \mathrm{V},\vec{p}_{\eta})=X_{\eta}-x_{\eta}-\Gamma^R_{\eta}x_{\eta}\tilde{\Gamma}^A_{\eta}\\\nonumber
&=-\frac{2\cdot T_N^2 \tilde{x}_{\eta}(\eps,\mathrm{eV})\Theta(p_{\mathrm{F}3}-|p_{\parallel}|)} {1+R_N^2-\frac{2R_N}{\Delta^2}(2\eps^2-\Delta^2)},\quad \eps<\Delta \end{align}
with $R_N=1-T_N$, and $\Theta(p_{\mathrm{F}3}-|p_{\parallel}|)$ is the kinematic constraint for trajectories to be ``non-magnetic''.
The total current density is:
\begin{equation}\label{BTKcurr1} \vec{j}(\mathrm{V})=-\sum_{\eta}\frac{{\rm e}N_{\mathrm{F}\eta}}{2}\int_{\eps}{\rm d}\eps \langle \vec{v}_{\eta} j_{\eps}\rangle_{\eta+},\end{equation}
where we sum over the contributions of both bands for which an FS-average is calculated independently.

Writing the FS-average explicitly, we get
\begin{eqnarray}
\vec{j}&=&-\frac{{\rm e}}{2}\int_{\eps}{\rm d}\eps\left[\int \frac{{\rm d}^2 p'_{\mathrm{F}2}}{(2\pi\hbar)^3}\frac{\vec{v}_{\mathrm{F}2}j_{\eps}(\vec{p}_{\mathrm{F}2})}{|\vec{v}_{\mathrm{F}2}(p'_{\mathrm{F}2})|}\right.+
\nonumber \\
&&\qquad \qquad \qquad +\left. \int \frac{{\rm d}^2 p'_{\mathrm{F}3}}{(2\pi\hbar)^3}\frac{\vec{v}_{\mathrm{F}3}j_{\eps}(\vec{p}_{\mathrm{F}3})}{|\vec{v}_{\mathrm{F}3}(p'_{\mathrm{F}3})|}\right].\end{eqnarray}
Note that
\begin{equation} S_{\eta}=\int_{v_{\mathrm{F}\eta,z}>0} {\rm d}^2 p'_{\mathrm{F}\eta}\frac{\vec{v}_{\mathrm{F}\eta}}{|\vec{v}_{\mathrm{F}\eta}(p'_{\mathrm{F}\eta})|}\cdot \vec{e}_z\end{equation}
is exactly the area of the projection of the Fermi-surface onto the contact plane. Due to the kinematic constraint we have  $j(\vec{p}_{\mathrm{F}2})=j(\vec{p}_{\mathrm{F}3})=j(p_{\parallel})$, if the parallel momentum components of $\vec{p}_{\mathrm{F}2}$ and $\vec{p}_{\mathrm{F}3}$ are identical, and $j(p_{\parallel})=0$ for $p_{\parallel}>\pfd$, which follows from equation \eqref{speccurr}. This together implies that both integrals give the same contribution to the current which is not surprising, since Andreev reflection induces the same current contribution in both bands. In the extended BTK-model case $j$ is not a function of $p_{\parallel}$, as $T_N$ is not trajectory dependent. Assuming spherical Fermi-surfaces, we can hence calculate the FS average explicitly:
\begin{equation} \vec{j}=j \vec{e}_z=-\frac{{\rm e} v_{\mathrm{F}3}N_{\mathrm{F}3}}{8}\int_{\eps}{\rm d}\eps j_{\eps}(\eps,\ \mathrm{V})\vec{e}_z.\end{equation}

The conductance is then given by:
\begin{equation} G^{\mathrm{quasi}}_N/G_{N,0}=A\partial_V j/G_{N,0}=\mathcal{T}_N \end{equation}
where $A$ is the contact area. Calculating $\mathcal{T}_N=\int_{\eps}{\rm d}\eps\ \partial_V j_{\eps}(\eps,\ \mathrm{V})/2{\rm e}$ at $T=0$ yields the BTK formula\cite{blonder82} with $T_N=1/[1+Z^2]$ (note that we used $\partial_V x=2{\rm e}\delta(\eps+\mathrm{eV})$ at $T=0$ at this point). $G_{N,0}=(2{\rm e}^2A v_{\mathrm{F}3}N_{\mathrm{F}3})/8$ is the contribution to the normal-state conductance of the non-magnetic trajectories. The corresponding term in \eqref{GBTK} reads:
\begin{equation} (1-P_C)G_N=\frac{2v_{\mathrm{F}3}N_{\mathrm{F}3}}{\Nfz\vfz+\Nfd\vfd} G_N. \end{equation} To obtain the correct contribution to the  normal-state conductance, $G_N$ must be related to the BTK-formula by $G_N=[{\rm e}^2A(\Nfz\vfz+\Nfd\vfd)/8] \mathcal{T}_N$. Hence we have exactly $G_N^{\mathrm{quasi}}=(1-P_C)G_N$.  Analogously we can derive $G_N^{\mathrm{quasi}}$ for $|\mathrm{eV}|>\Delta$ and recover the BTK result as well. We also obtain an expression for $G_H$ for $|\mathrm{eV}|>\Delta$ from our model by assuming a scattering matrix with $r_{1\u}=r\ e^{i\vartheta/2}$, $r_{1\d}=e^{-i\vartheta/2}$ which implies $T_{3}=0$ and $T_2=t_2^2=T_H\ e^{i\vartheta/2}$:
\begin{align} &G_H/G_{H,0}=\mathcal{T}_H=\\\notag&=\frac{4 T_H\beta}{2(\beta^2+1)-(\beta-1)^2T_H-2\cos\vartheta \sqrt{1-T_H}(\beta^2-1)} \end{align}
with $\beta=\mathrm{eV}/\sqrt{\mathrm{eV}^2-\Delta^2}$.
Comparison of this formula with that of Ref.~\onlinecite{Mazin} then shows that agreement requires
\begin{equation}\label{Tcos} \cos\vartheta=Z\sqrt{\frac{1}{1+Z^2}}\left(1-\frac{2(K/Z-1)}{(K-2Z)^2+1}\right) \end{equation}
with $T_H=1/[1+Z^2]$ and $K$ is a parameter introduced in Ref.~\onlinecite{Mazin} that we discuss in the following. For the sake of completeness, the corresponding contribution to the normal-state conductance is $G_{H,0}={\rm e}^2 A (\Nfz\vfz-\Nfd\vfd)/8$. Apparently, a spin-mixing phase is mandatory to reproduce the formula of Mazin \emph{et al}. This result is not surprising, since the model used in Ref.~\onlinecite{Mazin} to calculate $G_H$ necessarily introduces a spin-mixing effect, which is not true for the standard BTK model. The reason for this is that BTK assumes the same wavevectors in all channels and a non-spin-active interface. On the other hand Mazin \emph{et al.} introduce different wavevectors by assuming an evanescent mode in the minority band. This leads to the appearance of the quantity $K=\kappa/k$ in their formula, where $\kappa$ controls the attenuation of the evanescent mode ($e^{-\kappa z}$) and $k$ is the component normal to the interface of the wavevector in the propagating channel. From our point of view this is nothing but a manifestation of a spin-mixing phase, which is why we can only reach agreement by taking that into account. To make this point more convincing, we derived relation \eqref{Tcos} from an explicit calculation of the \emph{normal state} S-matrix using the same model as Ref.\onlinecite{Mazin}. We match plane waves with wavevector $\vec{k}$ in all propagating channels and the same $\kappa$ as above for the evanescent mode of the minority band in the FM. The interface is modeled by a spin-independent delta-function with a weight factor. This yields the reflection eigenvalues of the S-matrix on the SC-side $r_{1\u}$, $r_{1\d}$. By definition we have $\vartheta=\mathrm{arg}[r_{1\u}r_{1\d}^*]$ and find exactly Eq.~\eqref{Tcos}. In conclusion, we have shown here that earlier models for Andreev reflection in clean ferromagnetic heterostructures are contained as limiting cases in our theory. As already noted in Ref.~\onlinecite{Mazin}, the formula for $G_H$ used by Soulen \emph{et al.}\cite{soulen98} was not obtained from a rigorous calculation and is discontinuous at the gap-energy.

\section{Conclusions}
\label{Conc}

In summary, we have used an extension of the quasiclassical theory of superconductivity
to strongly spin-polarized ferromagnets to study the conductance of SC/FM point contacts
with a spin-active interface. We describe the interface by a microscopic model
that extends earlier models used in the description of Andreev reflection in such structures.
Our main result are:
(i) two types of Andreev reflection arise,
one of them being related to the creation of equal-spin triplet correlation.
These processes depend differently on various properties of the interface and
bulk materials involved.
(ii) the shape of the scattering potential has a pivotal impact in the magnitude of the
spin-mixing effect. The usually assumed box-like or delta-function-like
potential generically implies small mixing angles.
(iii)
we find spin-polarized Andreev bound state peaks in the conductance of a point contact
with a strong ferromagnet, that are more prominent for smooth interface potentials or
a finite magnetization near the interface in the superconductor. The latter effect could
be e.g. caused by the inverse proximity effect.
Lastly, we would like to stress that the feature $G(\rm{eV}=0)=0$ for $T=0$, which is universal for the spectra of half-metallic point contacts, may point to a criterion for identifying SAR in experiment at sufficiently low temperatures.

\acknowledgments
We would like to thank W. Belzig, M. Fogelstr\"om and G. Sch\"on for fruitful discussions, as well as D. Beckmann and F. H\"ubler for sharing their experimental results with us. In particular, we acknowledge helpful correspondence with I. I. Mazin.
\appendix

\section{Scattering matrix parameters}
\label{app}
The general boundary conditions for the scattering problem in quasiclassical theory of superconductivity have been derived in Ref.~\onlinecite{eschrig09}. These boundary conditions are formulated in terms of the normal state scattering matrix of the scattering region which has to be assumed, calculated from a microscopic model or fitted to experiment. In the case of a spin-active interface between a normal metal and a ferromagnet this matrix is a unitary $4\times4$ matrix and one may ask for a set of parameters that describes the most general matrix uniquely and still allows for an interpretation of these parameters with respect to the physical problem in question.

\subsection{Singular value decomposition}
\label{appsing}
Using a partial singular value decomposition and the spectral theorem one can arrive at a decomposition of $S$ that provides an appealing set of parameters. By partial we mean that a SVD is calculated for each block and not for the whole matrix, i.e. we have at the outset:
\begin{equation}\label{decomp} S=\left(\begin{array}{cc} URV^{\dagger} & WT\tilde{Z}^{\dagger} \\
                          \tilde{W}\tilde{T}Z^{\dagger} & -\tilde{U}\tilde{R}\tilde{V}^{\dagger}
                         \end{array}\right).
\end{equation}
$U,\ V,\ W,\ \tilde{Z},\ \tilde{W},\ Z,\ \tilde{U}$ and $\tilde{V}$ are unitary and independent $2\times2$-matrices, while $R,\ T,\ \tilde{T},\ \tilde{R}$ are diagonal and contain the corresponding singular values. Such a decomposition is possible for any $4\times4$ matrix, which means that we did not exploit the unitarity of $S$ so far. Exploiting unitarity we arrive at:
\begin{equation}\label{SVD} S=\left(\begin{array}{cc}\mathcal{U} & 0 \\ 0 & \tilde{\mathcal{U}}\end{array}\right)\left(\begin{array}{cc} R & T \\ T & -R \end{array}\right)\left(\begin{array}{cc}\mathcal{V}^{\dagger} & 0\\0 & \tilde{\mathcal{V}}^{\dagger}\end{array}\right).
\end{equation}
$\mathcal{U},\ \tilde{\mathcal{U}},\ \mathcal{V},\ \tilde{\mathcal{V}}$ are again unitary and independent. $R$ and $T$ contain the singular values of the composition and unitarity dictates $RR^{\dagger}+TT^{\dagger}=1$. To obtain a decomposition which allows for a clearcut interpretation in terms of scattering phases and spin-rotations, one has to continue decomposing $\mathcal{U},\ \tilde{\mathcal{U}},\ \mathcal{V}$ and $\tilde{\mathcal{V}}$ and eventually arrives at:
\begin{widetext}
\begin{equation}\label{final2} S=\left(\begin{array}{cc} Q & 0 \\ 0 & \tilde{Q}\end{array}\right)\left(\begin{array}{cc} \Phi^{\frac{1}{2}} & 0 \\ 0 & \tilde{\Phi}^{\frac{1}{2}}\end{array}\right)\left(\begin{array}{cc} Y & 0 \\ 0 & \tilde{Y}\end{array}\right)\left(\begin{array}{cc} \sqrt{1-TT} & \Psi T \\  \Psi^{\dagger}T & -\sqrt{1-TT} \end{array}\right)\left(\begin{array}{cc}Y^{\dagger} & 0 \\ 0 & \tilde{Y}^{\dagger}\end{array}\right)\left(\begin{array}{cc} \Phi^{\frac{1}{2}} & 0 \\ 0 & \tilde{\Phi}^{\frac{1}{2}}\end{array}\right)\left(\begin{array}{cc} Q^{\dagger} & 0 \\ 0 & \tilde{Q}^{\dagger}\end{array}\right).
\end{equation}
\end{widetext}
This decomposition is written in terms of $2\times2$ blocks which are related to reflection and transmission, these blocks are matrices in spin-space. The central matrix contains the singular values of the partial singular value decomposition. These singular values relate to the transmission and reflection amplitudes of the interface but not in a simple way, since the outer matrices contain several rotations in spin-space. The outer matrices come in two flavors. The matrices $Q$, $\tilde{Q}$, $Y$, $\tilde{Y}$ can be regarded as rotations of the quantization axis on either the left ($Q,\ Y$) or right ($\tilde{Q}, \ \tilde{Y}$) side of the interface. They have the structure:
\begin{equation} \mathrm{rot}(\alpha, \phi)=\left(\begin{array}{cc}\cos(\alpha/2) & \sin(\alpha/2)e^{i\phi} \\ -\sin(\alpha/2)e^{-i\phi} & \cos(\alpha/2). \end{array}\right)
\end{equation}
The matrices $\Phi^{\frac{1}{2}},\ \tilde{\Phi}^{\frac{1}{2}}$ and $\Psi^{\frac{1}{2}}$
are diagonal and contain complex phases ($\Psi=(\Psi^{\frac{1}{2}})^2$). Their structure is:
\begin{equation} \mathrm{phase}(\eta, \vartheta)=e^{i\eta}e^{i\sigma_z\vartheta/2}.\end{equation}
Apparently $\eta$ is a global phase and $\vartheta$ a relative phase.
The decomposition as it is presented here has 16 free parameters which agrees with the maximum number of free parameters a unitary $4\times4$ matrix can have. However, we can now identify parameters which will be irrelevant for our problem. First we use the freedom of choosing an arbitrary quantization axis in the SC and put $Q=1$. Secondly, we note that if the quantization axis of the interface does not rotate in the $x$-$y$-plane, we have $S=S^T$ and none of the rotation matrices defined above rotates in that plane. This implies $Y^{\dagger}=Y^T$, $\tilde{Y}^{\dagger}=\tilde{Y}^T$ and $\tilde{Q}^{\dagger}=Q^T$ and also that $\Psi$ is real. We may hence absorb $\Psi$ into $T$ which amounts to having transmission eigenvalues that can be negative. The transport properties, i.e. the current in this case, should also not depend on whether we extend the interface region arbitrarily further into the asymptotic region. This corresponds to the following transformation of the S-matrix:
\begin{equation} S'=\eta S\eta \end{equation}
with
\begin{equation}\eta=\left(\begin{array}{cc} e^{i\eta_1} & 0 \\ 0 & e^{i(\eta_2+\eta_3+(\eta_2-\eta_3)\sigma_z)/2}\end{array}\right). \end{equation}
Inspection of the boundary conditions shows in fact that both $X_{\eta}$ and $\Gamma^{R,A}_{\eta}$ are invariant under this transformation. Considering this as another gauge transformation, one can eliminate the global phase in $\Phi$ and use $\eta_2$, $\eta_3$ to obtain exactly the structure of \eqref{scatFM} in the transmission blocks. The reflection part on the SC-side reads:
\begin{equation}
\Phi^{\frac{1}{2}}
YRY^{\dagger}
\Phi^{\frac{1}{2}}
\end{equation}
from which we conclude that the relative phase $\vartheta_{\Phi}$ is what is usually referred to as the spin-mixing-angle:
\begin{equation}\label{thetaphi} \vartheta=\vartheta_{\Phi} \end{equation}
 This is also the quantity which we plot in Fig.~\ref{fig4},~\ref{fig9}. The necessity to have two additional mixing phases $\vartheta_2$ and $\vartheta_3$ comes about due to the additional rotations $\tilde{Q}$ and $\tilde{Y}$. They are a function of all parameters which enter the transmission part, thus a simple relation like \eqref{thetaphi} does not exist in this case.

The angle $\alpha_Y$ of which we make extensive use in the analytical discussion is associated to $Y$ by $Y=\mathrm{rot}(\alpha_Y, 0)$. In fact, the relations stated in \eqref{alphaU} are given by $YRY^{\dagger}$.

A fully analogous argumentation can be developed for the half-metallic case, however the corresponding scattering matrix is $3\times3$ and thus all tilde-quantities are scalar, making them irrelevant. Furthermore, one can show that the $\Psi$ matrix is also just a scalar phase in this case and hence $\Phi^{\frac{1}{2}}$ fully accounts for the spin-mixing effect. So we have in the half-metallic case:
\begin{equation} \vartheta=\vartheta_{\Phi} \quad \vartheta_2=\vartheta_{\Phi}/2. \end{equation}
This relation between $\vartheta_{\Phi} $ and $\vartheta_2 $ is due to the
fact that the evanescent solution in the ferromagnet is completely absorbed
in the scattering matrix.

\subsection{Box potential}
\label{appbox}
For the special case of a box shaped potential we obtain analytical solutions for
the scattering matrix, assuming for the normal metal (superconductor
in its normal state) a wave function of the form
\begin{equation}
\Psi_N = \frac{e^{i\vec{k}_{||} \vec{r}_{||}}}{\sqrt{v_1}}
\left[
\left(\begin{array}{c} s_{1+}\\s_{1-} \end{array}\right) e^{ik_1z}
+\left(\begin{array}{c} A_{1+}\\A_{1-} \end{array}\right) e^{-ik_1z}
\right],
\end{equation}
with $|\vec{k}_{||}|^2+k_1^2=2mE_F/\hbar^2$,
and in the barrier region
\begin{equation}
\Psi_B = e^{i\vec{k}_{||} \vec{r}_{||}}
{\cal U}({\alpha })^\dagger
\left(\begin{array}{c} B_+e^{\kappa_+z} +C_+ e^{-\kappa_+z}\\
B_- e^{\kappa_-z}+C_- e^{-\kappa_-z} \end{array}\right),
\end{equation}
with $|\vec{k}_{||}|^2-\kappa_\pm^2= -2m(U_\pm-E_F)/\hbar^2 $, and
with a certain spin rotation matrix ${\cal U}(\alpha )$
that represents the misalignment
of the barrier magnetic moment with the magnetization direction in the
ferromagnet by a misalignment angle $\alpha $.
The indices $\pm$ refer to spin-up and spin-down with respect to the
misaligned spin quantization axis in the barrier.
In the ferromagnet we can have, depending on the value of $\vec{k}_{||}$,
propagating or evanescent solutions in either of the two spin bands.
In the case of two propagating solutions they are
\begin{equation}
\Psi_F = e^{i\vec{k}_{||} \vec{r}_{||}}
\left[
\left(\begin{array}{c}
\frac{s_2}{\sqrt{v_2}} e^{-ik_{2} (z-a)}
\\
\frac{s_3}{\sqrt{v_3}} e^{-ik_{3} (z-a)}
\end{array}\right)
+
\left(\begin{array}{c}
\frac{A_2}{\sqrt{v_2}}  e^{ik_{2} (z-a)}
\\
\frac{A_3}{\sqrt{v_3}} e^{ik_{3} (z-a)}
\end{array}\right)
\right],
\end{equation}
where $|\vec{k}_{||}|^2+k_2^2= 2m(E_{\rm F}-E_2)/\hbar^2 $ and
$|\vec{k}_{||}|^2+k_3^2= 2m(E_{\rm F}-E_3)/\hbar^2 $ (in a more general
model the masses on the two sides of the interface could also differ;
we assumed them identical for definiteness).
In the case of one propagating and one evanescent solution,
\begin{equation}
\Psi_F = e^{i\vec{k}_{||} \vec{r}_{||}}
\left[
\left(\begin{array}{c}
\frac{s_2}{\sqrt{v_2}} e^{-ik_{2} (z-a)}
\\
0
\end{array}\right)
+
\left(\begin{array}{c}
\frac{A_2}{\sqrt{v_2}}  e^{ik_{2} (z-a)}
\\
D_3 e^{-\kappa_{3} (z-a)}
\end{array}\right)
\right],
\end{equation}
where $|\vec{k}_{||}|^2+k_2^2= 2m(E_{\rm F}-E_2)/\hbar^2 $ and
$|\vec{k}_{||}|^2-\kappa_3^2= 2m(E_{\rm F}-E_3)/\hbar^2 $,
and in the case of two evanescent solutions,
\begin{equation}
\Psi_F = e^{i\vec{k}_{||} \vec{r}_{||}}
\left(\begin{array}{c}
D_2 e^{-\kappa_{2} (z-a)}
\\
D_3 e^{-\kappa_{3} (z-a)}
\end{array}\right) ,
\end{equation}
where $|\vec{k}_{||}|^2-\kappa_2^2= 2m(E_{\rm F}-E_2)/\hbar^2 $ and
$|\vec{k}_{||}|^2-\kappa_3^2= 2m(E_{\rm F}-E_3)/\hbar^2 $.
We then match the wave functions and their derivatives at $z=0$ ($\Psi_N$ and
$\Psi_B$) and at
$z=a$ ($\Psi_B$ and $\Psi_F$), and eliminate the components $D_2$ and $D_3$. The scattering matrix
then is defined as the coefficient matrix in the relations
\begin{equation}
\left(\begin{array}{c} A_{1+} \\ A_{1-}\\A_2\\A_3 \end{array}\right)
= S
\left(\begin{array}{c} s_{1+} \\ s_{1-}\\s_2\\s_3 \end{array}\right)
\end{equation}
for the case of two propagating solutions in the ferromagnet,
\begin{equation}
\left(\begin{array}{c} A_{1+} \\ A_{1-}\\A_2 \end{array}\right)
= S
\left(\begin{array}{c} s_{1+} \\ s_{1-}\\s_2 \end{array}\right)
\end{equation}
for the case of one propagating and one evanescent solution, and
\begin{equation}
\left(\begin{array}{c} A_{1+} \\ A_{1-} \end{array}\right)
= S
\left(\begin{array}{c} s_{1+} \\ s_{1-} \end{array}\right)
\end{equation}
for the case of two evanescent solutions in the ferromagnet.


\vspace{-0.5cm}
\begin{thebibliography}{27}
\expandafter\ifx\csname natexlab\endcsname\relax\def\natexlab#1{#1}\fi
\expandafter\ifx\csname bibnamefont\endcsname\relax
  \def\bibnamefont#1{#1}\fi
\expandafter\ifx\csname bibfnamefont\endcsname\relax
  \def\bibfnamefont#1{#1}\fi
\expandafter\ifx\csname citenamefont\endcsname\relax
  \def\citenamefont#1{#1}\fi
\expandafter\ifx\csname url\endcsname\relax
  \def\url#1{\texttt{#1}}\fi
\expandafter\ifx\csname urlprefix\endcsname\relax\def\urlprefix{URL }\fi
\providecommand{\bibinfo}[2]{#2}
\providecommand{\eprint}[2][]{\url{#2}}


\bibitem{bergeret05}
F.~S. Bergeret, A.~F. Volkov, and K.~B. Efetov, Rev. Mod. Phys. {\bf 77}, 1321 (2005).

\bibitem[{\citenamefont{{A.~I. Buzdin}}(2005)}]{buzdin05}
\bibinfo{author}{\bibnamefont{{A.~I. Buzdin}}}, \bibinfo{journal}{Rev. Mod.
  Phys.} \textbf{\bibinfo{volume}{77}}, \bibinfo{pages}{935}
  (\bibinfo{year}{2005}).

\bibitem{eschrig08}
M. Eschrig and T. L\"ofwander, Nat. Phys. {\bf 4}, 138 (2008).

\bibitem{cuoco08}
M. Cuoco, A. Romano, C. Noce, and P. Gentile,
Phys. Rev. B {\bf 78}, 054503 (2008).

\bibitem{galak08}
A.V. Galaktionov,
M. S. Kalenkov and A. D. Zaikin, Phys. Rev. B {\bf 77}, 094520 (2008).

\bibitem{haltermann08}
K. Halterman, O.T. Valls, and P.H. Barsic, Phys. Rev. B {\bf 77}, 174511 (2008).

\bibitem{volkov08}
A.F. Volkov and K.B. Efetov, Phys. Rev. B {\bf 78}, 024519 (2008).

\bibitem[{\citenamefont{{P. M. R. Brydon} et~al.}(2008)\citenamefont{{P. M. R.
  Brydon}, {Boris Kastening}, {D. K. Morr}, and {D. Manske}}}]{brydon08}
\bibinfo{author}{\bibnamefont{{P. M. R. Brydon}}},
  \bibinfo{author}{\bibnamefont{{B. Kastening}}},
  \bibinfo{author}{\bibnamefont{{D. K. Morr}}}, \bibnamefont{and}
  \bibinfo{author}{\bibnamefont{{D. Manske}}}, \bibinfo{journal}{Phys. Rev.
  B} \textbf{\bibinfo{volume}{77}} (\bibinfo{year}{2008}).

\bibitem{zhao08}
E. Zhao and J.A. Sauls, Phys. Rev. B {\bf 78}, 174511 (2008).


\bibitem{linder09}
J. Linder, T. Yokoyama, A. Sudb{\o}, and M. Eschrig,
Phys. Rev. Lett. {\bf 102}, 107008 (2009).

\bibitem{grein09}
R. Grein, M. Eschrig, G. Metalidis, and G. Sch\"on,
Phys. Rev. Lett. {\bf 102}, 227005 (2009).

\bibitem{brydon09}
P. M. R. Brydon and D. Manske, Phys. Rev. Lett. {\bf 103}, 147001 (2009).

\bibitem{kalenkov09}
M.S. Kalenkov, A.V. Galaktionov, and A.D. Zaikin,
Phys. Rev. B {\bf 79}, 014521 (2009).

\bibitem{beri09}
B. B{\'e}ri, J.~N. Kupferschmidt, C.~W.~J. Beenakker, and P.W. Brouwer,
Phys. Rev. B {\bf 79}, 024517 (2009).

\bibitem{Barsic}
P. H. Barsic and O. T. Valls, Phys. Rev. B \textbf{79}, 014502 (2009)

\bibitem{eschrig09}
M. Eschrig, Phys. Rev. B {\bf 80}, 134511 (2009).

\bibitem{eschrig07}
M. Eschrig, T. L\"ofwander, T. Champel, J. C. Cuevas, J. Kopu, and G. Sch\"on,
J. Low. Temp. Phys. {\bf 147}, 457 (2007).

\bibitem[{\citenamefont{{Y. Tanaka} and {A. A. Golubov}}(2007)}]{tanaka07}
\bibinfo{author}{\bibnamefont{{Y. Tanaka}}} \bibnamefont{and}
  \bibinfo{author}{\bibnamefont{{A. A. Golubov}}}, \bibinfo{journal}{Phys. Rev.
  Lett.} \textbf{\bibinfo{volume}{98}} (\bibinfo{year}{2007}).

\bibitem{eschrig03}
M. Eschrig, J. Kopu, J. C Cuevas, and G. Sch\"on, Phys. Rev. Lett. {\bf 90},
137003 (2003).

\bibitem{volkov03}
A.F. Volkov, F.S. Bergeret, and K.B. Efetov, Phys. Rev. Lett. {\bf 90}
117006 (2003).

\bibitem{kopu04}
J. Kopu, M. Eschrig, J. C. Cuevas, and M. Fogelstr\"om, Phys. Rev. B {\bf 69}, 094501 (2004).

\bibitem{eschrig04}
M. Eschrig, J. Kopu, A. Konstandin, J. C. Cuevas, M. Fogelstr\"om, and G.
Sch\"on, Adv. in Sol. State Phys. {\bf 44}, pp. 533-546, (Springer Verlag, Heidelberg, 2004).

\bibitem{keizer06}
R.S. Keizer, S.T.B. Goennenwein, T.M. Klapwijk, G. Miao, G. Xiao, and A. Gupta,
Nature (London) {\bf 439}, 825 (2006).

\bibitem{braude07}
V. Braude and Y.V. Nazarov, Phys. Rev. Lett. {\bf 98}, 077003 (2007).

\bibitem{asano07}
Y. Asano, Y. Tanaka, and A.A. Golubov, Phys. Rev. Lett. {\bf 98}, 107002 (2007);Y. Asano, Y. Sawa, Y. Tanaka, and A.A. Golubov, Phys. Rev. B {\bf 76}, 224525 (2007).

\bibitem[{\citenamefont{Linder and Sudb\o}(2007)}]{linder}
\bibinfo{author}{\bibfnamefont{J.}~\bibnamefont{Linder}} \bibnamefont{and}
  \bibinfo{author}{\bibfnamefont{A.}~\bibnamefont{Sudb\o}},
  \bibinfo{journal}{Phys. Rev. B}
  \textbf{\bibinfo{volume}{75}}, \bibinfo{pages}{134509}
  (\bibinfo{year}{2007}).

\bibitem{linder07}
J. Linder and A. Sudb{\o}, Phys. Rev. B {\bf 76}, 064524 (2007).

\bibitem{takahashi07}
S. Takahashi, S. Hikino, M. Mori, J. Martinek, and S. Maekawa,
Phys. Rev. Lett. {\bf 99}, 057003 (2007).

\bibitem{lofwander05}
T. L\"ofwander, Th. Champel, J. Durst, and M. Eschrig,
Phys. Rev. Lett. {\bf 95}, 187003 (2005);
T. L\"ofwander, Th. Champel, and M. Eschrig,
Phys. Rev. B {\bf 75}, 014512 (2007).

\bibitem{pajovic06}
Z. Pajovi{\'c}, M. Bo\v{z}ovi{\'c}, Z. Radovi{\'c}, J. Cayssol, and A. Buzdin,
Phys. Rev. B {\bf 74}, 184509 (2006).

\bibitem{champel08}
T. Champel, T. L\"ofwander, and M. Eschrig,
Phys. Rev. Lett. {\bf 100}, 077003 (2008).

\bibitem[{\citenamefont{{M. Houzet} and {A. I. Buzdin}}(2007)}]{houzet1}
\bibinfo{author}{\bibnamefont{{M. Houzet}}} \bibnamefont{and}
  \bibinfo{author}{\bibnamefont{{A. I. Buzdin}}}, \bibinfo{journal}{Phys. Rev.
  B} \textbf{\bibinfo{volume}{76}} (\bibinfo{year}{2007}).

\bibitem{Valls}
K. Halterman, P. H. Barsic, and O. T. Valls, Phys. Rev. Lett. \textbf{99}, 127002 (2007)

\bibitem{cottet08}
A. Cottet and W. Belzig, Phys. Rev. B {\bf 77}, 064517 (2008).

\bibitem{cottet08b}
A. Cottet, B. Dou\c{c}ot, and W. Belzig,
Phys. Rev. Lett. {\bf 101}, 257001 (2008).

\bibitem{Beenakker}
M.~J.~M. de Jong, C.~W.~J. Beenakker, Phys. Rev. Lett. \textbf{74}, 1657 (1995)

\bibitem{Mazin}
I. I. Mazin, A. A. Golubov, and B. Nadgorny, J. Appl. Phys. \textbf{89}, 7576 (2001)
Unfortunately, the general formula for $G_H$ contains a typo in the published version
of this paper (private communication from I. I. Mazin). The formula for $\kappa\rightarrow\infty$
is correct.

\bibitem[{\citenamefont{{S. K. Upadhyay} et~al.}(1998)\citenamefont{{S.
  K. Upadhyay}, {A. Palanisami}, {R. N. Louie}, and {R. A.
  Buhrman}}}]{Upad}
\bibinfo{author}{\bibnamefont{{S. K. Upadhyay}}},
  \bibinfo{author}{\bibnamefont{{A. Palanisami}}},
  \bibinfo{author}{\bibnamefont{{R. N. Louie}}}, \bibnamefont{and}
  \bibinfo{author}{\bibnamefont{{R. A. Buhrman}}}, \bibinfo{journal}{Phys. Rev.
  Lett.} \textbf{\bibinfo{volume}{81}}, \bibinfo{pages}{3247}
  (\bibinfo{year}{1998}).

\bibitem{soulen98}
R. J. Soulen, Jr., J. M. Byers, M. S. Osofsky, B. Nadgorny, T. Ambrose, S. F. Cheng, P. R. Broussard, C. T. Tanaka, J. Nowak, J. S. Moodera, A. Barry, and J. M. D. Coey, Science {\bf 282}, 85 (1998).

\bibitem{desisto00}
W. J. DeSisto, P. R. Broussard, T. F. Ambrose, B. E. Nadgorny, and M. S. Osofsky, Appl. Phys. Lett. {\bf 76}, 3789 (2000).

\bibitem{ji01}
Y. Ji, G. J. Strijkers, F. Y. Yang, C. L. Chien, J. M. Byers, A. Anguelouch, G. Xiao, and A. Gupta, Phys. Rev. Lett. {\bf 86}, 5585 (2001).

\bibitem{angu01}
A. Anguelouch, A. Gupta, Gang Xiao, D. W. Abraham, Y. Ji, S. Ingvarsson, and C. L. Chien
Phys. Rev. B {\bf 64} 180408(R) (2001).


\bibitem{parker02}
J. S. Parker, S. M. Watts, P. G. Ivanov, and P. Xiong
Phys. Rev. Lett. {\bf 88} 196601 (2002).

\bibitem{woods04}
G. T. Woods, R. J. Soulen Jr, I.I. Mazin, B. Nadgorny, M. S. Osofsky, J. Sanders, H. Srikanth, W. F. Egelhoff, and R. Datla, Phys. Rev. B {\bf 70}, 054416 (2004).

\bibitem[{\citenamefont{{F. P\'erez-Willard} et~al.}(2004)\citenamefont{{F.
  P\'erez-Willard}, {J. C. Cuevas}, {C. S\"urgers}, {P. Pfundstein}, {J. Kopu},
  {M. Eschrig}, and {H. v. L\"ohneysen}}}]{Perez}
\bibinfo{author}{\bibnamefont{{F. P\'erez-Willard}}},
  \bibinfo{author}{\bibnamefont{{J. C. Cuevas}}},
  \bibinfo{author}{\bibnamefont{{C. S\"urgers}}},
  \bibinfo{author}{\bibnamefont{{P. Pfundstein}}},
  \bibinfo{author}{\bibnamefont{{J. Kopu}}}, \bibinfo{author}{\bibnamefont{{M.
  Eschrig}}}, \bibnamefont{and} \bibinfo{author}{\bibnamefont{{H. v.
  L\"ohneysen}}}, \bibinfo{journal}{Phys. Rev. B} \textbf{\bibinfo{volume}{69}}
  (\bibinfo{year}{2004}).

\bibitem{dyachenko06}
A.I. D'yachenko, V.N. Krivoruchko, and V.Yu. Tarenkov,
Low. Temp. Phys. {\bf 32}, 824 (2006).

\bibitem{yates07}
K.A. Yates, W.R. Branford, F. Magnus, Y. Miyoshi, B. Morris, L. F. Cohen,
P. M. Sousa, O. Conde, and A. J. Silvestre, Appl. Phys. Lett. {\bf 91}, 172504 (2007)

\bibitem{krivoruchko08}
V.N. Krivoruchko and V.Yu. Tarenkov,
Phys. Rev. B. {\bf 75}, 214508 (2007);
Phys. Rev. B {\bf 78}, 054522 (2008).

\bibitem{bocklage07}
L. Bocklage, J.M. Scholtyssek, U. Merkt, and G. Meier,
J. Appl. Phys. {\bf 101}, 09J512 (2007).

\bibitem[{\citenamefont{{K. Xia} et~al.}(2002)\citenamefont{{K. Xia}, {P. J.
  Kelly}, {G. E. W. Bauer}, and {I. Turek}}}]{Xia}
\bibinfo{author}{\bibnamefont{{K. Xia}}}, \bibinfo{author}{\bibnamefont{{P. J.
  Kelly}}}, \bibinfo{author}{\bibnamefont{{G. E. W. Bauer}}}, \bibnamefont{and}
  \bibinfo{author}{\bibnamefont{{I. Turek}}}, \bibinfo{journal}{Phys. Rev.
  Lett.} \textbf{\bibinfo{volume}{89}} (\bibinfo{year}{2002}).

\bibitem{blonder82}
G. E. Blonder, M. Tinkham, and T. M. Klapwijk,
Phys. Rev. B {\bf 25}, 4515 (1982).

\bibitem{meservey94}
R. Meservey and P. M. Tedrow, Phys. Rep. {\bf 238}, 173 (1994).

\bibitem{tokuyasu88}
T. Tokuyasu, J. A. Sauls, and D. Rainer,
Phys. Rev. B {\bf 38}, 8823 (1988).

\bibitem{fogelstrom00}
M. Fogelstr\"om, Phys. Rev. B {\bf 62}, 11812 (2000).

\bibitem{huertas02}
D. Huertas-Hernando, Yu. V. Nazarov, and W. Belzig, Phys. Rev. Lett. {\bf 88},
047003 (2002).

\bibitem{cottet05}
A. Cottet and W. Belzig, Phys. Rev. B {\bf 72}, 180503 (2005).

\bibitem{bobkova07}
I. V. Bobkova and A. M. Bobkov,
Phys Rev. B {\bf 76}, 094517 (2007).

\bibitem{Zaikin}
M. S. Kalenkov, A. D. Zaikin, Phys. Rev. B \textbf{76}, 224506 (2007)

\bibitem{saint64}
D. Saint-James, J. Phys. (Paris) {\bf 25}, 899 (1964).

\bibitem{deutscher05}
G. Deutscher, Rev. Mod. Phys. {77}, 109-135 (2005).

\bibitem{larkin68}
A.~I. Larkin and Y.~N. Ovchinnikov, Zh. Eksp. Teor. Fiz. {\bf 55},  2262
  (1968), [Sov. Phys. JETP {\bf28}, 1200 (1969)].

\bibitem{eilen}
G. Eilenberger, Z. Phys. {\bf 214},  195  (1968).


\bibitem{Serene}
J.~W. Serene and D. Rainer, Phys. Rep. {\bf 101},  221  (1983).

\bibitem{schmid75}
A. Schmid and G. Sch\"on, J. Low. Temp. Phys. {\bf 20}, 207 (1975).

\bibitem{schmid81}
A. Schmid,
in {\it Nonequilibrium Superconductivity, Phonons and Kapitza Boundaries},
Proceedings of NATO
Advanced Study Institute,
edited by K. E. Gray (Plenum Press, New York, 1981), Chapter 14.

\bibitem{rammer86}
J. Rammer and H. Smith, Rev. Mod. Phys. {\bf 58}, 323 (1986).

\bibitem{Larkin86}
A. I. Larkin and  Y. N. Ovchinnikov,
 in {\em Nonequilibrium Superconductivity},
edited by D. N. Langenberg and A. I. Larkin (Elsevier
Science Publishers, 1986),  p. 493.

\bibitem{FLT}
M. Eschrig, J. A. Sauls, H. Burkhardt, and D. Rainer,
in {\em High-$T_{\rm c}$ Superconductors and Related Materials, Fundamental
Properties, and Some Future Electronic Applications}, Proceedings
of the NATO Advanced Study Institute,
edited by S.-L. Drechsler and T. Mishonov,
pp. 413-446 (Kluwer Academic, Norwell, MA, 2001).

\bibitem{gorkov58}
L.~P. Gor'kov, Zh. Eksp. Teor. Fiz. {\bf 34},  735  (1958), [Sov.\ Phys.\ JETP
  {\bf7}, 505 (1958)],
L.~P. Gor'kov, Zh. Eksp. Teor. Fiz. {\bf 36},  1918  (1959), [Sov.\ Phys.\ JETP
  {\bf9}, 1364 (1959)].

\bibitem{keldysh64}
L.~V. Keldysh, Zh. Eksp. Teor. Fiz. {\bf 47},  1515  (1964), [Sov. Phys. JETP
  {\bf20}, 1018 (1965)].

\bibitem{shelankov85}
A.~L. Shelankov,
J. Low. Temp. Phys. {\bf 60}, 29 (1985).

\bibitem{zaitsev84}
A.~V. Zaitsev, Zh. Eksp. Teor. Fiz. {\bf 86}, 1742 (1984)
[Sov. Phys. JETP {\bf 59}, 1015 (1984)].
A. L. Shelankov, Sov. Phys. Solid State {\bf 26}, 981 (1984) [Fiz. Tved. Tela
{\bf 26}, 1615 (1984)]

\bibitem{zhao04}
E. Zhao, T. L\"ofwander, and J.A. Sauls, Phys. Rev. B {\bf 70}, 134510 (2004);
E. Zhao and J.A. Sauls, Phys. Rev. Lett. {\bf 98}, 206601 (2007).


\bibitem{barash02}
Yu. S. Barash and I. V. Bobkova,
Phys. Rev. B {\bf 65}, 144502 (2002);
Yu. S. Barash, I. V. Bobkova, and T. Kopp,
Phys. Rev. B {\bf 66}, 140503 (2002).

\bibitem{eschrig00}
M. Eschrig, Phys. Rev. B {\bf 61}, 9061 (2000).

\bibitem{cuevas06}
J. C. Cuevas, J. Hammer, J. Kopu, J. K. Viljas, and M. Eschrig,
Phys. Rev. B {\bf 73}, 184505 (2006).

\bibitem{nagato93}
Y. Nagato, K. Nagai, and J. Hara, J. Low Temp. Phys. {\bf 93},  33  (1993),
S. Higashitani and K. Nagai, J. Phys. Soc. Jpn. {\bf 64},  549  (1995),
Y. Nagato, S. Higashitani, K. Yamada, and K. Nagai, J. Low Temp. Phys. {\bf
  103},  1  (1996).

\bibitem{schopohl95}
N. Schopohl and K. Maki, Phys. Rev. B {\bf 52},  490  (1995),
N. Schopohl, cond-mat/9804064 (unpublished, 1998).

\bibitem{eschrig99}
M. Eschrig, J. A. Sauls, and D. Rainer, Phys. Rev. B {\bf 60}, 10447 (1999).

\bibitem{GolubovRMP} A. A. Golubov, M. Yu. Kupriyanov, and E. Il’ichev, Rev. Mod. Phys. \textbf{76}, 411 (2004)

\bibitem{smooth}
Instead of a smoothened potential, one could consider different barrier potential \emph{widths} for the two spin channels. However, this leads to exactly the same effects as we discuss, and a smoothened potential barrier (with differing
classical return points for the two spin channels, using a WKB language)
is a more realistic way of discussing these effects than two box potentials with differing widths.


\bibitem{WKB}
H. Jeffreys, Proceedings of the London Mathematical Society {\bf 23}, 428 (1924);
G. Wentzel, Zeitschrift der Physik {\bf 38}, 518 (1926);
H.A. Kramers, Zeitschrift der Physik {\bf 39}, 828 (1926);
L. Brillouin, Comptes Rendus de l'Academie des Sciences {\bf 183}, 24 (1926).

\bibitem{Georgo}
G. Metalidis and P. Bruno, Phys. Rev. B \textbf{72}, 235304 (2005).

\bibitem{Fisher}
D. S. Fisher and P. A. Lee, Phys. Rev. B \textbf{23}, 6851 (1981).

\end{thebibliography}
\end{document}